\documentstyle[emulateapj,psfig]{article}

\makeatletter

\newenvironment{inlinefigure}{%
\def\@captype{figure}%
\noindent\begin{minipage}{0.999\linewidth}\begin{center}}
{\end{center}\end{minipage}\smallskip}
\makeatother

\lefthead{Barger et al.}

\begin{document}
\title{The MicroJansky Radio Galaxy Population
\altaffilmark{1}}
\author{
A.~J.~Barger,$\!$\altaffilmark{2,3,4}
L.~L.~Cowie,$\!$\altaffilmark{4}
W.-H.~Wang$\!$\altaffilmark{4}
}

\altaffiltext{1}{Based in part on data obtained at the W. M. Keck
Observatory, which is operated as a scientific partnership among the
the California Institute of Technology, the University of
California, and NASA and was made possible by the generous financial
support of the W. M. Keck Foundation.}
\altaffiltext{2}{Department of Astronomy, University of
Wisconsin-Madison, 475 North Charter Street, Madison, WI 53706}
\altaffiltext{3}{Department of Physics and Astronomy,
University of Hawaii, 2505 Correa Road, Honolulu, HI 96822}
\altaffiltext{4}{Institute for Astronomy, University of Hawaii,
2680 Woodlawn Drive, Honolulu, HI 96822}

\slugcomment{Accepted by The Astrophysical Journal for January 1, 2007}

\begin{abstract}
We use highly spectroscopically complete observations of
the radio sources from the VLA 1.4~GHz survey of the Hubble 
Deep Field-North region to study the faint radio galaxy 
population and its evolution. The fraction of radio sources
that can be optically spectroscopically identified is fairly
independent of radio flux, with about $60-80$\% identified
at all fluxes. We spectrally classify the sources into four
spectral types: absorbers, star formers, Seyfert galaxies, 
and broad-line AGNs, and we analyze their properties by type. 
We supplement the spectroscopic redshifts with photometric 
redshifts measured from the rest-frame 
ultraviolet to mid-infrared spectral energy distributions.
Using deep X-ray observations of the field, we do not confirm 
the existence of an X-ray--radio correlation for star-forming 
galaxies. We also do not observe any
correlations between 1.4~GHz flux and $R$ magnitude or redshift.
We find that the radio powers of the host galaxies rise 
dramatically with increasing redshift, while the optical properties
of the host galaxies show at most small changes. Assuming that the locally 
determined far-infrared (FIR)--radio correlation holds at high 
redshifts, we estimate total FIR luminosities for the radio sources.
We note that the FIR luminosity estimates for any radio-loud 
AGNs, which we conservatively do not try to remove from the 
sample, will be overestimates. Considering only the radio 
sources with quasar-like bolometric luminosities, we find a maximum 
ratio of candidate highly-obscured AGNs to X-ray--luminous 
($L_{0.5-2~{\rm keV}}$ or 
$L_{2-8~{\rm keV}}\ge 10^{42}$~ergs~s$^{-1}$) 
sources of about 1.9. Finally, we use source-stacking analyses
to measure the X-ray surface brightnesses of various X-ray
and radio populations. We find the contributions to the
$4-8$~keV light from our candidate highly-obscured AGNs to be 
very small, and hence these sources are unable to account for the 
light that has been suggested may be missing at these energies.
\end{abstract}

\keywords{cosmology: observations --- galaxies: active --- galaxies: distances
          and redshifts --- galaxies: evolution --- galaxies: formation}

\section{Introduction}
\label{secintro}

In the local universe, Seyfert~2 galaxies outnumber Seyfert~1
galaxies by 4 to 1 (Maiolino \& Rieke 1995), and more than half
of Seyfert 2 nuclei are Compton-thick with obscuring column
densities $N_H>10^{24}$~cm$^{-2}$ (Maiolino et al.\ 1998;
Risalti et al.\ 1999; Guainazzi et al.\ 2005; Comastri 2004). 
In the high-redshift universe,
X-ray background population synthesis models (refined after the 
discovery of a lower redshift distribution than expected for 
the hard X-ray sources; e.g., Barger et al.\ 2002, 2003; 
Szokoly et al.\ 2004) similarly predict a population of 
highly-obscured active galactic nuclei (AGNs) that are missed 
in the deep {\em Chandra\/} and {\em XMM-Newton\/} hard X-ray 
surveys (e.g., Franceschini et al.\ 2002; Gandhi \& Fabian 2003; 
Ueda et al.\ 2003; Comastri 2004; Gilli 2004; Fabian \& Worsley 2004;
Treister \& Urry 2005; Ballantyne et al.\ 2006). 
Such theoretical predictions appear to be receiving some observational 
confirmation in the work of Worsley et al.\ (2005). Using a 
source-stacking analysis on the {\em Chandra} Deep Field-North (CDF-N) 
and {\em Chandra} Deep Field-South (CDF-S) fields, Worsley et al.\ (2005) 
found that the resolved fraction in the $6-8$~keV band was only about 60\%. 
(Worsley et al.\ 2004 did a similar analysis 
on the {\em XMM-Newton\/} Lockman Hole field and found a 50\% 
resolved fraction above 8~keV.) According to these authors,
the missing X-ray background 
(XRB) component has a spectral shape consistent with a population of 
highly-obscured AGNs at redshifts $z\sim 0.5-1.5$ having obscuring
column densities of $N_H\sim 10^{23}-10^{24}$~cm$^{-2}$.

Of course, to derive these results, Worsley et al.\ (2005) had to 
choose which estimate of the $2-8$~keV extragalactic XRB spectrum 
to use, and there is a great deal of uncertainty in these 
measurements (see Figure~15 of Hickox \& Markevitch 2006 for a summary; 
Worsley et al.\ 2005 used the XMM-Newton measurement made by
Deluca \& Molendi 2004, but Revnivtsev et al.\ 2005 found that
the intensity of the XRB measured by focusing telescopes is higher 
than that measured by collimated experiments by about $10-15$\%, 
possibly due to the extreme complexity of measuring effective solid 
angles of focusing telescopes).
Indeed, Barger et al.\ (2002) found 
that the issue of whether there is a need for a substantial 
population of as-yet undetected highly-obscured AGNs depends 
critically on how the low-energy and high-energy XRB measurements 
tie together. 

Regardless, the combination of the theoretical modelling predictions 
for a population of highly-obscured AGNs missing from the deep 
X-ray surveys and the possible observational evidence for a 
substantial unresolved fraction of the XRB has inspired observers 
to try to quantify the fraction of obscured AGNs that may be being missed. 
Deep mid-infrared (MIR) and radio images are an obvious avenue for 
searching for highly-obscured AGNs, since extinction in the MIR 
and radio is small. Various approaches using MIR or X-ray data
(e.g., Alonso-Herrero et al.\ 2006; Polletta et al.\ 2006), 
the combination of MIR and radio data 
(e.g., Mart\'{\i}nez-Sansigre et al.\ 2005), or the combination 
of MIR, radio, and X-ray data (e.g., Donley et al.\ 2005) have 
been adopted, and a candidate population of highly-obscured AGNs 
has been identified. However, the various selection effects in 
these approaches mean that a reliable upper limit on the number
of highly-obscured AGNs that are undetected even in 
the $2-8$~keV band has not yet been obtained.

In this paper, we take an alternative approach to look for
highly-obscured AGNs at high redshifts using a pure 
microJansky radio survey selection. Ultradeep radio surveys are 
very useful for tracing the evolution of dust-obscured galaxies 
and AGNs. In contrast to optical surveys,
which may omit dusty sources, and submillimeter surveys,
which currently have low spatial resolution and are limited to
small areas for any significant depth, the 1.4~GHz surveys do
not suffer from extinction, provide subarcsecond positional
accuracy, and cover large areas.

Although the Jansky and milliJansky radio
source populations are dominated by powerful AGNs, at the
microJansky level, the radio source population is
increasingly dominated by star-forming galaxies, which
produce non-thermal radio continuum at 1.4~GHz through
synchrotron emission from supernova remnants.
In local star-forming galaxies and radio-quiet AGNs, it is
well-known that the radio power is tightly correlated with
the far-infrared (FIR) luminosity (e.g., Helou et al.\ 1985;
Condon et al.\ 1991; Condon 1992), probably as a result of both
being linearly related to the massive star formation rate.
If we assume that this correlation continues to hold at
high redshifts (note that there is some observational
support that the correlation does extend to $z\ge 1$ from 
the work of Appleton et al.\ 2004, though the number of 
galaxies at the higher redshifts is relatively low),
then we can use this relation to estimate the total
FIR luminosity of a galaxy from its radio power alone.
Indeed, even with the advent of the {\em Spitzer Space Telescope\/},
this is still the most robust way to estimate a galaxy's
total FIR luminosity, due to the limited MIPS sensitivities at
70 and 160~$\mu$m and the uncertainties in the conversion from
24~$\mu$m luminosity to total FIR luminosity. Although estimating
total FIR luminosities from 24~$\mu$m data has been advocated
(e.g., Appleton et al.\ 2004; Marcillac et al.\ 2006),
the conversion depends on the assumed shape of the spectral energy
distribution (SED), which is expected to change as a function of
dust temperature and metallicity. 

Using their {\em Spitzer\/} Infrared 
Nearby Galaxies Survey (SINGS) data, Dale et al.\ (2005)
found a total range in rest-frame 8~$\mu$m (which corresponds to
observed-frame 24~$\mu$m at $z=2$) to total FIR luminosity
ratios of more than a factor of 20. When they excluded the metal-poor
dwarfs, the total range dropped to a factor of 10, which they
argued should be taken as a minimum systematic uncertainty
for total FIR luminosities inferred from rest-frame 8~$\mu$m
fluxes alone. They cautioned that if the metal abundance of 
the population cannot be inferred a priori, then the uncertainties
may be considerably higher. Likewise, they found that the rest-frame 
24~$\mu$m to total FIR ratio spans a factor of 5 in the SINGS sample.

In addition to our assumption that the FIR-radio correlation
continues to hold at high redshifts, the other possible concern
with our procedure of estimating the total FIR luminosities of 
the radio sources using the FIR-radio correlation 
is the presence of radio-loud AGNs. For these sources, the radio 
continuum will be in excess of the level expected from star 
formation, and we will overestimate their total FIR luminosities. 
Fortunately, however, since we are only interested in obtaining 
upper limits on the number of highly-obscured AGNs, 
this is not a major concern for our present analysis, and we
have conservatively not made any effort to remove such sources 
from our sample. 

Finally, we note that even in the case of the radio-quiet AGNs, we 
are not able to separate out the AGN contribution to the FIR luminosity 
from the star-forming contribution. However, since local radio-quiet 
AGNs are empirically observed to follow the FIR-radio correlation, 
we have an upper limit on the AGN FIR emission, which is
sufficient for our analysis.

In this paper, we use the ultradeep 1.4~GHz observations of the Hubble 
Deep Field-North (HDF-N; Richards 2000), for which we have obtained 
very complete spectroscopic identifications, to study the properties
of the radio sample. 
The HDF-N has a wealth of existing multiwavelength data, including 
the {\em HST\/} ACS Treasury Great Observatories Origins Deep 
Survey-North data (GOODS-N; Giavalisco et al.\ 2004) and the 
{\em Spitzer Space Telescope\/} Legacy GOODS-N data (P.I. M. Dickinson).
In addition, we have recently obtained deep wide-area $J-$ and $H-$band
observations using the University of Hawaii 2.2~m telescope 
(Trouille et al.\ 2006, in preparation), which enable
us to assign robust photometric redshifts to the spectroscopically
unidentified sources. 

The HDF-N is also the site of the deepest X-ray image of the sky, the 
2~Ms CDF-N. We use these data to examine the issue of whether there 
exists an X-ray--radio correlation for star-forming galaxies. In addition, 
we use these data to remove the X-ray--luminous ($L_{0.5-2~{\rm keV}}$ or
$L_{2-8~{\rm keV}}\ge 10^{42}$~ergs~s$^{-1}$) sources from our 
radio-identified candidate highly-obscured AGN population
for tighter upper limits. Finally, we measure the X-ray surface 
brightnesses for our candidate highly-obscured AGN population and 
others, which we compare with the shape and intensity of the XRB.

The structure of the paper is as follows. 
In \S\ref{secdata}, we describe the radio sample and the 
available multiwavelength imaging and spectroscopic data.
In \S\ref{secids}, we discuss the spectroscopic identifications 
of the radio sample.
In \S\ref{secphotz}, we briefly describe our photometric
redshift method, which uses the optical to MIR SEDs,
and we show our redshift distribution for the radio sources.
In \S\ref{sec24um}, we discuss the $24~\mu$m properties
of the radio sample and the difficulties with using these
data to estimate FIR luminosities.
In \S\ref{secspectral}, we describe our spectral classification scheme 
and examine the properties of the radio sample by spectral class.
In \S\ref{secxray}, we discuss the X-ray properties of the
radio sample and investigate whether there is an X-ray--radio correlation
for star-forming galaxies at these higher redshifts, like there is locally.
In \S\ref{secopt}, we describe the optical properties of the radio 
sample. In \S\ref{secuplim}, we determine the FIR luminosities of
the radio sources using our redshifts and the FIR-radio correlation.
We then put quantitative upper limits on the numbers of highly-obscured 
AGNs that could have gone undetected in this deepest of 
hard X-ray surveys, and we discuss the implications for the history 
of supermassive black hole growth. In \S\ref{secxrb}, we measure
the X-ray surface brightnesses of our radio-identified candidate 
highly-obscured AGN population, as well as of other radio and X-ray 
populations, and we compare them with the shape and intensity of the XRB. 
We summarize our results in \S\ref{secsummary}.

We assume $\Omega_M=0.3$, $\Omega_\Lambda=0.7$, and
$H_0=70$~km~s$^{-1}$~Mpc$^{-1}$. All magnitudes are in the AB 
magnitude system.

\section{The Data}
\label{secdata}

\subsection{Radio and X-ray Data}

Richards (2000) presented a catalog of 1.4~GHz sources detected 
in the Very Large Array (VLA)\footnote{The VLA is a facility of
the National Radio Astronomy Observatory (NRAO). The NRAO is a
facility of the National Science Foundation operated under
cooperative agreement by Associated Universities, Inc.} 
map of the HDF-N, which covers a $40'$ diameter region 
with an effective resolution of $1.8''$.
The absolute radio positions are known to $0.1''-0.2''$ rms.

Alexander et al.\ (2003) presented the 2~Ms X-ray image
of the CDF-N, which they aligned with the Richards (2000)
radio image. Near the aim point, the X-ray data reach
limiting fluxes of $f_{2-8~{\rm keV}} \approx1.4\times
10^{-16}$~ergs~cm$^{-2}$~s$^{-1}$ and
$f_{0.5-2~{\rm keV}} \approx 1.5 \times
10^{-17}$~ergs~cm$^{-2}$~s$^{-1}$.
We determined the maximal radius radio circle that fits within
the {\em Chandra\/} area to be a $10'$ radius circle centered
on the position R.A. = $12^h 36^m 54.29^s$,
decl. = $62^\circ 14' 16.0''$ (J2000.0).
This also roughly matches the highest sensitivity region of the
VLA observations, producing a relatively uniform radio map. Our
radio sample consists of the 207 radio sources contained within
this radius.

For sources within our $10'$ radius, matching X-ray counterparts
from the Alexander et al.\ (2003) catalogs to the radio sources is not
critically dependent on the choice of match radius. This can be seen
from Figure~7 of Alexander et al.\ (2003), which shows the positional
offset between the X-ray and radio sources versus off-axis angle.
We chose to use a $1.5''$ search radius as a reasonable compromise
between not pushing too hard on the accuracy of the data and not
introducing too much random error,
but only two additional sources would have been matched with a
$2''$ search radius, and only three of the current sources would
not have been matched with a $1''$ search radius.

Richards (2000) gave a $5\sigma$ completeness limit
of 40~$\mu$Jy for compact sources in the central region of the map.
The completeness limit is somewhat higher for extended radio sources.
We independently determined the completeness of the Richards (2000) 
1.4~GHz catalog by measuring the radio fluxes of a large,
optically-selected sample in $3''$ diameter apertures and
adjusting the normalization to match the measured fluxes
in the Richards (2000) catalog for the overlapping sources.
Figure~\ref{figcomplete}a shows that the Richards (2000) catalog 
is highly complete to 70~$\mu$Jy and then drops to about 50\% 
completeness at 60~$\mu$Jy.

%
% FIGURE 1 
%
\begin{inlinefigure}
\centerline{\psfig{figure=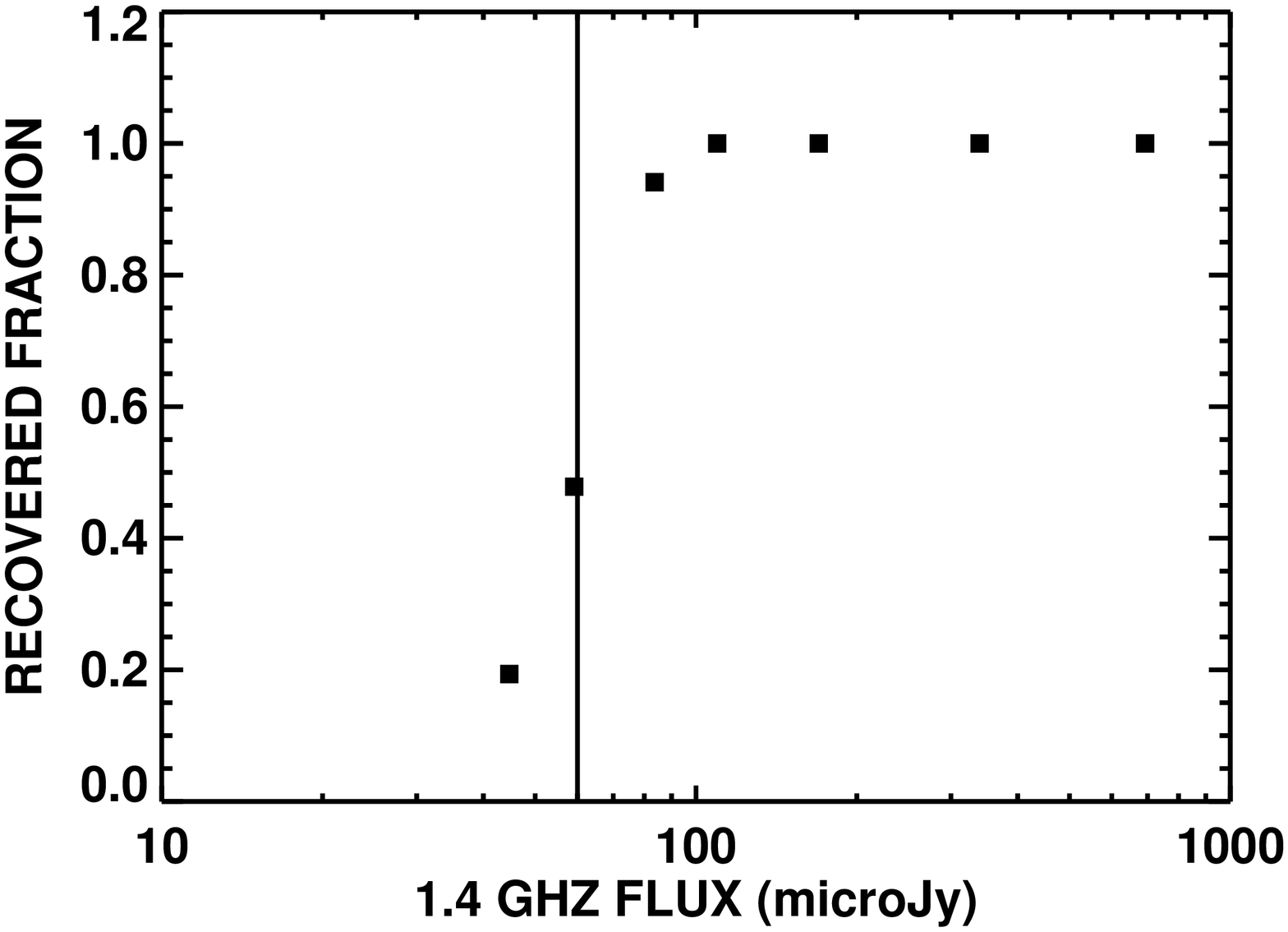,angle=90,width=3.5in}}
\centerline{\psfig{figure=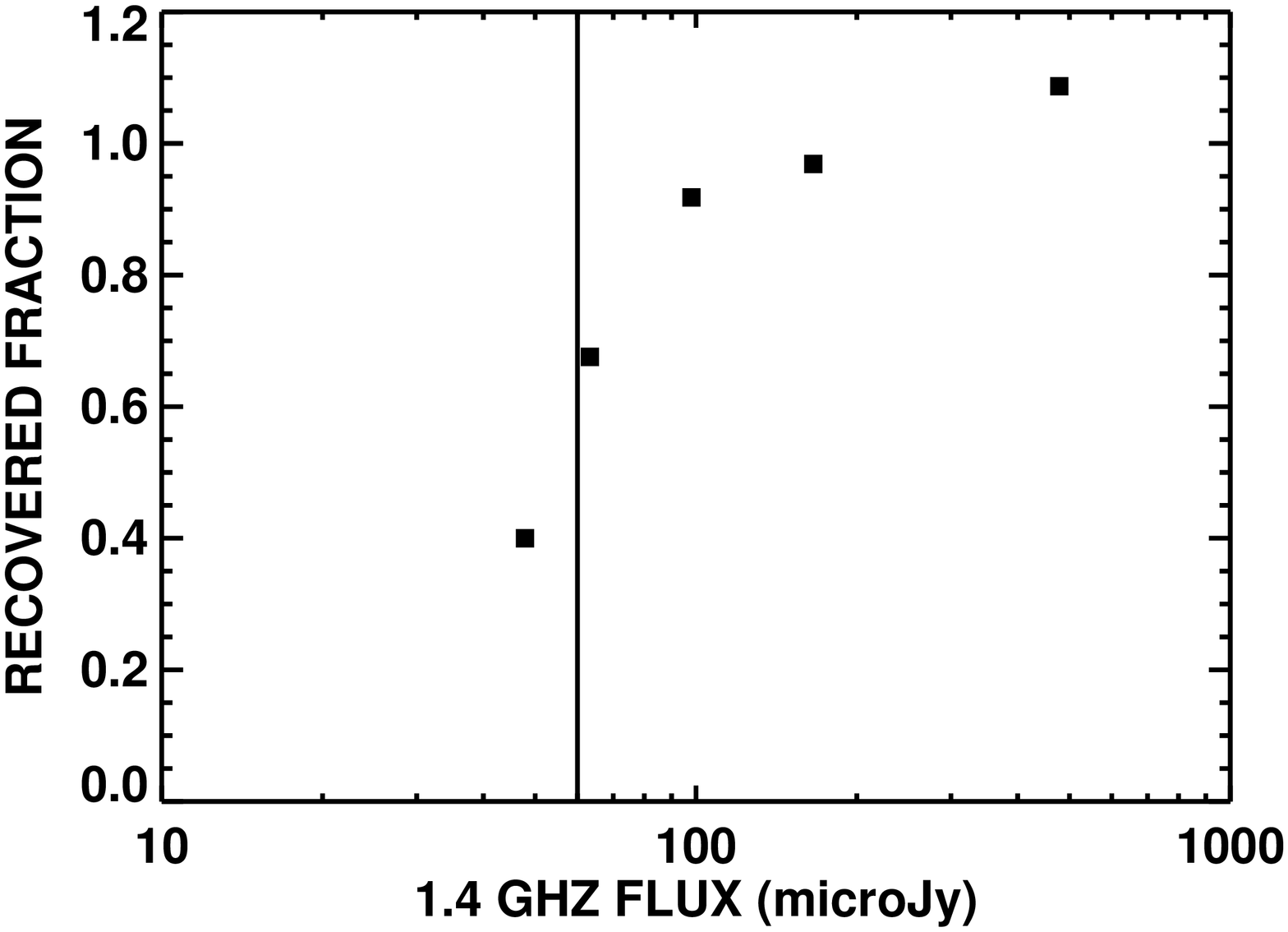,angle=90,width=3.5in}}
\figurenum{1}
\figcaption[]{
Completeness of the HDF-N radio catalog from Richards (2000)
as a function of 1.4~GHz flux based on (a) measuring the radio fluxes
of an optically-selected sample, and (b) comparing the Richards (2000)
catalog with the Biggs \& Ivison (2006) catalog. The solid lines
show our adopted 60~$\mu$Jy completeness limit. 
\label{figcomplete}
}
\end{inlinefigure}

Biggs \& Ivison (2006) have recently given a new catalog of 
1.4~GHz sources using all of the A-array data from the VLA. 
This catalog is slightly more sensitive than the Richards (2000)
catalog and may also be used to estimate the completeness of the 
Richards (2000) catalog. Within our $10'$ radius region, the new 
catalog contains 280 sources, of which 259 have fluxes greater
than 40~$\mu$Jy, as opposed to the 207 sources in the
Richards (2000) catalog. Of the 207 Richards (2000) sources, 188 
have direct counterparts in the Biggs \& Ivison (2006) catalog. 
For two other sources, the positions in the two catalogs
are offset due to 
different centerings on the radio sources. The fluxes of the 
overlapping sources are in reasonable agreement. Most of the 
17 Richards (2000) sources that are missing in the Biggs \& Ivison (2006)
catalog have optical counterparts and redshifts, so they are 
probably real and have just dropped below the flux limit of the 
Biggs \& Ivison (2006) catalog. In Figure~\ref{figcomplete}b, we 
show the completeness of the Richards (2000) catalog based on the 
Biggs \& Ivison (2006) catalog. The plot shows the ratio of the 
number of sources per flux bin detected in the Richards (2000) 
catalog to the number of sources per flux bin detected in the 
Biggs \& Ivison (2006) catalog. Consistent with our analysis of 
the completeness of the Richards (2000) catalog using the optical 
data, this shows that the Richards (2000) sample is substantially 
complete above 70~$\mu$Jy but becomes progressively more incomplete 
at lower fluxes. In our subsequent analysis, we adopt
60~$\mu$Jy as our effective completeness limit.

\subsection{Optical and Near-Infrared Data}

Giavalisco et al.\ (2004) presented {\em HST\/} ACS
images of the HDF-N region (the GOODS-N), and 
Capak et al.\ (2004) presented ground-based deep optical 
imaging of a very wide-field region encompassing the
GOODS-N. The ground-based data cover the whole MIPS and 
IRAC areas (\S\ref{secmir})
in the $U$, $B$, $V$, $R$, $I$, $z^{\prime}$,
and $HK^{\prime}$ bands. All of the images were registered
to the Richards (2000) radio catalog for accurate absolute 
astrometry.

Trouille et al.\ (2006, in preparation) 
carried out deep $J-$ and $H-$band imaging 
of the entire GOODS-N region using the Ultra-Low Background Camera 
(ULBCAM) on the University of Hawaii 2.2~m telescope during 2004 
and 2005. ULBCAM consists of four 2k$\times$2k 
HAWAII-2RG arrays (Loose et al.\ 2003) with a total 
$16\arcmin \times 16\arcmin$ field of view. The images were taken 
using a 13-point dither pattern with $\pm30\arcsec$ and $\pm60\arcsec$
dither steps in order to cover the chip gaps. 
The near-infrared (NIR) data were flattened 
using median sky flats from each dither pattern. The image distortion 
was corrected using the astrometry in the USNO-B1.0 catalog 
(Monet et al.\ 2003). The flattened, sky-subtracted, and warped
images (with typical seeing $0.7''$) were combined to form the 
final mosaic, which has a $20\arcmin \times 20\arcmin$ area that 
fully covers the GOODS-N region. The final image was registered to
the Richards (2000) radio catalog.
The integration times at each pixel are 9 hours in $J$ and 12.5 hours 
in $H$, respectively, and the $5\sigma$ sensitivities are
0.84~$\mu$Jy and 2.06~$\mu$Jy, corresponding to $5\sigma$ AB magnitude 
limits of 24.1 and 23.1, respectively. 

We measured all of the optical and NIR magnitudes at the 
positions of the radio sources using $3''$ diameter apertures 
and corrected them to approximate total magnitudes using an
average offset for each waveband (Cowie et al.\ 1994).

\subsection{Mid-Infrared Data}
\label{secmir}

Wang et al.\ (2006) combined the reduced DR1 and DR2 IRAC superdeep
images from the GOODS-N \emph{Spitzer} Legacy Science 
Program first, interim, and second data release products (DR1, DR1+, DR2; 
Dickinson et al.\ 2006, in preparation), 
weighted by exposure time, to form 3.6, 4.5, 5.8, 
and 8.0~$\mu$m images that fully cover the GOODS-N area. 
We measured the source fluxes at the radio positions using 
fixed $4\farcs8$ (3.6 and 4.5$~\mu$m) and 
$6\arcsec$ (5.8 and 8.0$~\mu$m) diameter apertures. 
These apertures are approximately three times the $\sim 1\farcs7$ 
(3.6$~\mu$m) to $\sim 2\arcsec$ (8.0$~\mu$m) FWHM of the point spread 
function (PSF) and 
are a good compromise between the PSF size and the source separation.  
We applied aperture corrections from the IRAC in-flight PSFs (January 2004),
which are consistent with the ones published in the IRAC Data Handbook, 
to the measured fluxes. The corrected IRAC fluxes 
should be reasonably close to the total fluxes of the sources, because 
the majority of the sources are point-like compared to the $\sim2\arcsec$ 
IRAC PSF. The primary errors in the photometry are caused by the high
density of sources, especially at 3.6 and 4.5$~\mu$m. In these two bands, 
the typical distance between sources is comparable to the PSF, and the maps 
are confusion-limited. Consequently, both the background estimate and the 
aperture photometry are highly subject to blending with nearby sources.

Also following Wang et al.\ (2006), we directly used the DR1+ MIPS 24$~\mu$m 
source list and version 0.36 MIPS 24$~\mu$m map provided by the 
\emph{Spitzer} Legacy Program. This source catalog is flux-limited at 
80$~\mu$Jy and is a subset of a more extensive catalog 
(Chary et al.\ 2006, in preparation). 
With the 0.065~deg$^2$ area coverage, the catalog contains 1199 24$~\mu$m 
sources and is highly complete at 80$~\mu$Jy. The source positions are based 
on sources detected in the deep IRAC images, and the fluxes are derived using 
PSF fitting. A $-0\farcs38$ offset in declination was applied to the source 
positions to match the radio-frame astrometry (Richards 2000).
We cross-identified the MIPS sources with the radio sources by
searching a $1.5''$ radius region around the radio positions.

\subsection{Spectroscopic Data}
\label{secz}

Cowie et al.\ (2004b) and Wirth et al.\ (2004) report
the results of an extensive spectroscopic survey of
galaxies in the ACS GOODS-N region. We supplemented
these data with spectral observations made with the Deep
Extragalactic Imaging Multi-Object Spectrograph (DEIMOS;
Faber et al.\ 2003) on the Keck~II 10~m
telescope subsequent to the publication of these papers
and with targeted observations of radio and X-ray sources
(Barger et al.\ 2002, 2003, 2005; Cowie et al.\ 2004a)
made with either DEIMOS or the Low-Resolution Imaging
Spectrograph (LRIS; Oke et al.\ 1995) on
the Keck~I telescope. We also use redshifts obtained
by Chapman et al.\ (2004, 2005) and Swinbank et al.\ (2004),
adopting the Swinbank et al.\ (2004) NIR redshifts over 
the Chapman et al.\ (2004, 2005) optical redshifts, where
available, because redshift measurements are generally 
more reliable when they are made from emission-line 
features. Where there are redshifts from both sources, 
the redshifts are within $<0.008$ of each other.

\section{Radio Sample Identifications}
\label{secids}

Cross-identification of the radio sources with 
their counterparts is non-trivial. 
Some radio sources have complex structures, and it
is only by inspection of overlays of the radio contours on
optical through MIR images that the likely counterparts
can be identified. In other cases, the counterparts are 
extremely red, only emerging in the NIR or MIR bands.
(Reassuringly, these red cores inevitably agree with the 
radio positions.) The most extreme of these sources can 
only be identified in the \emph{Spitzer\/} IRAC bands, 
which do not cover the entire radio field. 

There are also cases where previous spectroscopic 
identifications (found through a search of the
literature via the NASA/IPAC Extragalactic 
Database, NED) appear to be of neighboring, usually 
much bluer galaxies, rather than of the genuine counterparts. 
It is possible that some of these identifications
may, in fact, be correct if the bluer structures
are parts of a more extended galaxy centered
on the red core. However, it is very
hard to make a convincing argument that this
is indeed the case, rather than, for example,
that the blue source is a background galaxy
lensed by the foreground red source.

Here we try to take a conservative approach to the 
identifications. We first determined which sources
have unambiguous counterparts. These include
all of the sources which have either a $5\sigma$
ground-based $z'$- or $H$-band counterpart or 
an IRAC $5.4~\mu$m counterpart lying within 
$1''$ of the radio position. This immediately 
identifies 195 of the sources in the sample.
The dispersion of these sources around the radio 
positions is $0.3''$.

We then overlaid the radio contours of the remaining 
12 sources on the optical images of the field (Fig.~\ref{fignoid}).
One source, with a flux of 5.96~mJy, is clearly bi-lobal 
and centered on a small red galaxy. We take this red galaxy 
to be the counterpart to the radio source. The remaining 11 
sources have no obvious identifications, though three lie 
near bright galaxies (one of these could be an off-axis source;
see below), and a fourth may be associated with 
a pair of interacting galaxies at $z=0.642$. %Object 131
Nearly all of these are faint sources ($40-80~\mu$Jy), and 
some may be false. (Note that Richards 2000 found a $-9\sigma$
source within the HDF-N radio image, and since the probability
of finding such a source in the image is much less than
1\%, he concluded that the noise properties of the image
are not entirely Gaussian.) However, four of the remaining
11 sources have fluxes above 70~$\mu$Jy, and four (including 
only one of the sources with a flux above 70~$\mu$Jy) are in 
the deeper Biggs \& Ivison (2006) catalog (see Table~\ref{tab1}).
These may be real sources without counterparts
from the optical to the MIR. For example, they could be very 
obscured or perhaps high-redshift galaxies. 

%
% FIGURE 2
%
\begin{inlinefigure}
%\centerline{\psfig{figure=f2.ps,angle=-90,width=3.5in}}
\figurenum{2}
\figcaption[]{
Mosaic of the bi-lobal source and the 11 remaining sources in our
radio sample without
obvious counterparts. The sources are shown in the order of
Table~\ref{tab1}, starting from the bottom-left corner
and moving to the right, then looping back to the beginning
of the next row and again moving to the right.
The individual images are $12.5''$ on a side. There
are three colors in the images: blue is set to be the $B$ image,
green is set to be the $V$ image, and red is set to be the
average of $I+z'$, all from the ACS GOODS-N data. The contours
show the positions of the radio light.
\label{fignoid}
}
\end{inlinefigure}

%
% FIGURE 3 
%
\begin{inlinefigure} 
%\centerline{\psfig{figure=f3.ps,angle=-90,width=3.5in}}
\figurenum{3}
\figcaption[]{
Mosaic of the 12 sources whose redshifts we omitted due to
the uncertainty of whether these literature identifications are
directly associated with the radio sources. The sources are
shown in the order of Table~\ref{tab2}, starting from the bottom-left
corner and moving to the right, then looping back to the 
beginning of the next row and again moving to the right.
The individual images are $12.5''$ on a side. There
are three colors in the images: blue is set to be the $B$ image,
green is set to be the $V$ image, and red is set to be the
average of $I+z'$, all from the ACS GOODS-N data. The contours
show the positions of the radio light.
\label{figomit}
}
\end{inlinefigure} 

We summarize the radio and optical coordinates of the 
bi-lobal source and the radio coordinates of the 11 
sources without obvious conterparts in Table~\ref{tab1}.
Only one of these sources
has an X-ray or $24~\mu$m detection. This source lies 
in the outer regions of a bright $z=0.459$ galaxy and 
may be either an off-axis source in that galaxy or
a faint background source that is hidden
by the foreground galaxy and hence cannot be identified.

Next, we determined which of the sources with unambiguous
counterparts had spectroscopic redshifts. We omitted redshifts 
for 12 sources with identifications in the literature (from NED),
where inspection of the images suggested that the identifications
may not be directly associated with the radio sources.
As we have noted above, some of these identifications may 
be correct, but it appears safer to rely on the photometric 
redshifts for these sources. We summarize the 
omitted sources in Table~\ref{tab2}, and we show the radio
contours overlaid on the optical images for these sources
in Figure~\ref{figomit}.

%
% FIGURE 4 
%
\begin{inlinefigure}
\centerline{\psfig{figure=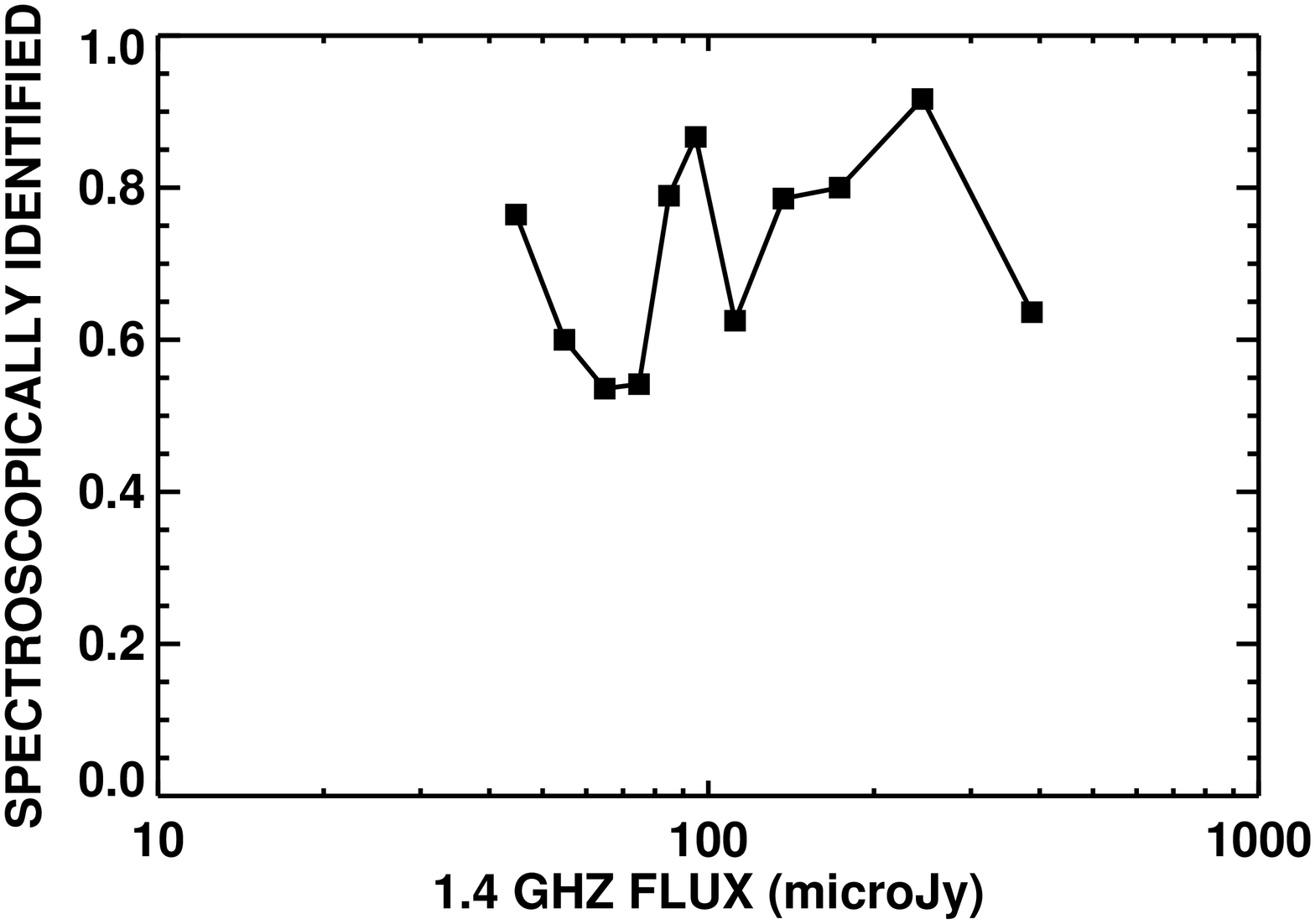,angle=90,width=3.5in}}
\figurenum{4}
\figcaption[]{
Spectroscopic completeness of our
radio sample. The sources are grouped into the
following flux bins: 40--50, 50--60, 60--70, 70--80, 80--90,
90--100, 100--125, 125--150, 150--200, 200-300, 300-500. All of
these bins contain 10 or more sources.
\label{figfraction}
}
\end{inlinefigure}

Our final compilation contains 143 spectroscopic redshifts, 
of which 101 are already in the literature, and 42 are new 
to this work. In our subsequent analysis,
we distinguish the 12 redshifts that we use from 
Chapman et al.\ (2004, 2005) and Swinbank et al.\ (2004)
from our other redshifts, referring to them as the CS sources.
In Figure~\ref{figfraction}, we show the fraction of
radio sources that are spectroscopically identified.
We see that the identified fraction is fairly independent
of radio flux, with about 60--80\% of the sources
identified at all fluxes.

%
% FIGURE 5
%
\begin{inlinefigure}
%\centerline{\psfig{figure=rhist.ps,angle=90,width=3.5in}}
\centerline{\psfig{figure=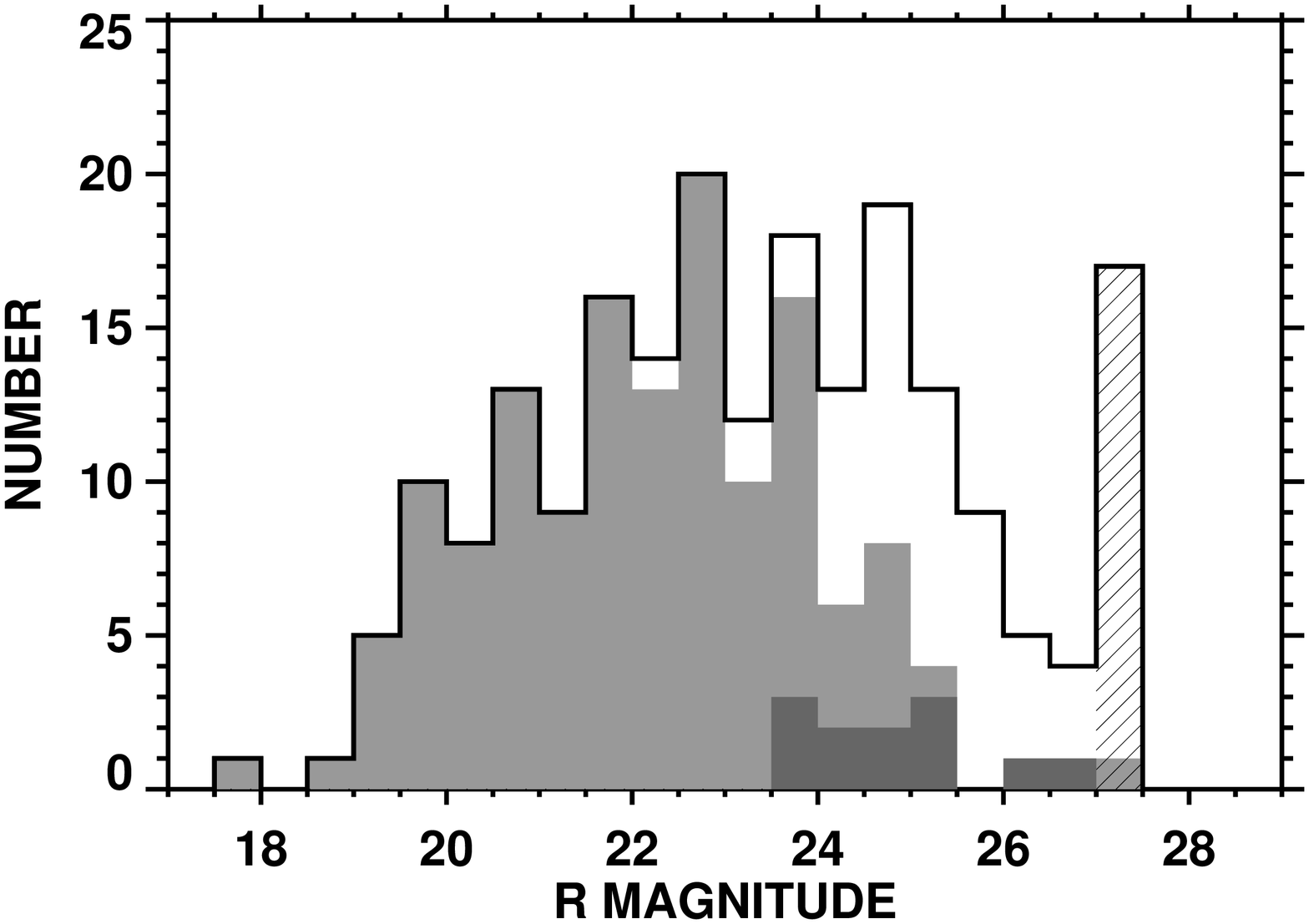,angle=90,width=3.5in}}
\figurenum{5}
\figcaption[]{
$R$ magnitude distribution for all of the sources
in our radio sample {\em (open)\/}.
Shading denotes the spectroscopically identified sample,
with darker shading for the CS sources.
All seventeen $R>27$ sources are placed in the final magnitude bin
{\em (hatched)\/}. Sources with $R<20$ suffer from saturation
problems and are likely to be brighter than measured.
\label{figrhist}
}
\end{inlinefigure}

In Figure~\ref{figrhist}, we show the $R$ magnitude 
distribution for all of the sources in our radio 
sample {\em (open)\/}. We denote the spectroscopic sample 
with light shading and the CS sources with dark shading. 
Beyond $R=24$, it becomes more difficult to
spectroscopically identify the radio sources, while at
brighter magnitudes, essentially all of the sources can
be identified.

\section{Photometric Redshifts}
\label{secphotz}

Wang et al.\ (2006), using the method of 
P{\'e}rez-Gonz{\'a}lez et al.\ (2005), built ``training-set'' 
SEDs from 1200 galaxies in the MIPS GOODS-N sample with 
known redshifts and spectral types. They built seven templates,
ranging from an elliptical galaxy spectrum to a very blue
star-forming galaxy spectrum, over the frequency range from
$6\times 10^{13}$~Hz to $4\times 10^{15}$~Hz. They then made
least-squares fits of their individual source SEDs to these 
templates to determine the 
photometric redshifts and spectral types for the galaxies
in their samples. They found that the method worked
extremely well over a wide range of redshifts and only
failed for a small number of sources. Their use of 
the deep $J$- and $H$-band magnitudes and the {\em Spitzer}
data, in addition to the optical data, substantially improved the 
high-redshift end of the photometric redshift determinations.

To derive photometric redshifts for our radio sources,
we similarly constructed SEDs using the optical
data, the {\em Spitzer\/} IRAC $3.6~\mu$m, $4.5~\mu$m, $5.8~\mu$m,
and $8.0~\mu$m data, the {\em Spitzer\/} MIPS $24~\mu$m data,
and the ULBCAM NIR $J$ and $H$-band data. 
We then made least-squares fits to the Wang et al.\ (2006)
templates to determine the photometric redshifts and
spectral types for the galaxies in our radio sample that
were detected in at least five bands.
This gives us an additional 47 redshifts. Thus, in total, 
only 17 of our radio sources (including the 11 without any
obvious counterparts)
have neither a spectroscopic nor a photometric redshift. 
We show a comparison of the photometrically identified 
sources {\em (open)\/} with the spectroscopically identified 
sources {\em (light shading)\/} in Figure~\ref{figzhist}.
The CS sources are denoted by dark shading. 
The gap in the spectroscopic
redshift distribution between $z\sim 1.4$ and $z\sim 1.9$, 
known as the ``redshift desert'', reflects the difficulty 
of identifying galaxies with redshifts in this range, where 
[OII] 3727~\AA\ has moved out of the optical window and
Ly$\alpha$~1216~\AA\ has not yet entered in.

%
% FIGURE 6
%
\begin{inlinefigure}
\centerline{\psfig{figure=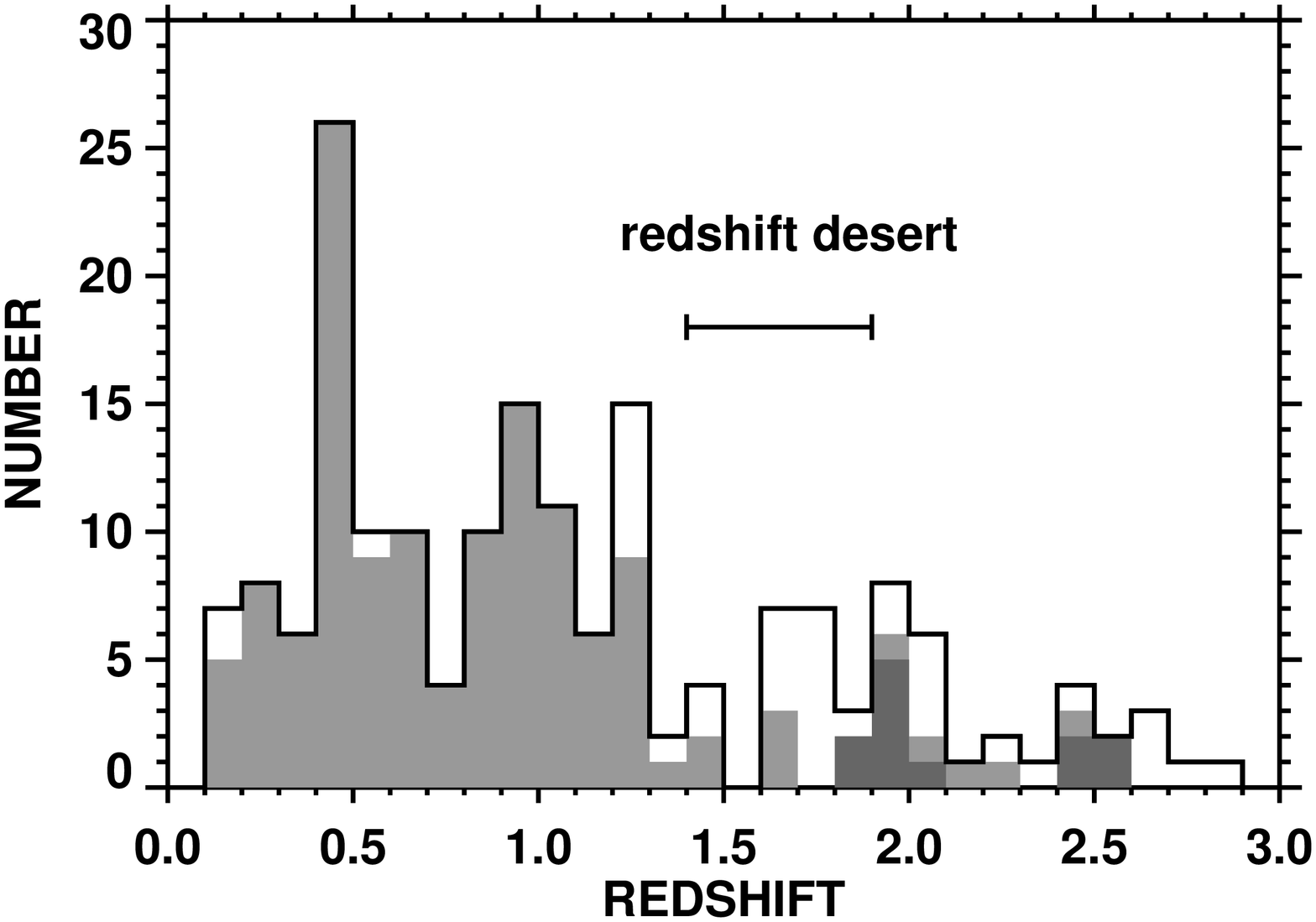,angle=90,width=3.5in}}
\figurenum{6}
\figcaption[]{
Number of spectroscopically ({\em shaded\/}) and
photometrically ({\em open\/}) identified sources vs.
redshift for the radio sample. The CS sources are denoted
by dark shading.
\label{figzhist}
}
\end{inlinefigure}

\section{24~micron Properties of the Radio Sample}
\label{sec24um}

Studies based on the {\em Infrared Space Observatory (ISO)\/} 
$15~\mu$m data found a loose correlation between the
MIR emission and the radio continuum
(Cohen et al.\ 2000; Elbaz et al.\ 2002; Garrett 2002;
Gruppioni et al.\ 2003). Appleton et al.\ (2004) confirmed
the existence of this correlation with \emph{Spitzer\/} data, 
and Marcillac et al.\ (2006) proposed using it to derive FIR 
luminosities. However, while obtaining the FIR luminosity from 
a radio measurement only depends on the FIR-radio correlation 
and the radio spectral index (and the assumption that the 
source is not a radio-loud AGN), obtaining the FIR luminosity 
from a MIR measurement strongly depends on the library of 
template SEDs that is used to $K$-correct the data.

%
% FIGURE 7 
%
\begin{inlinefigure} 
\centerline{\psfig{figure=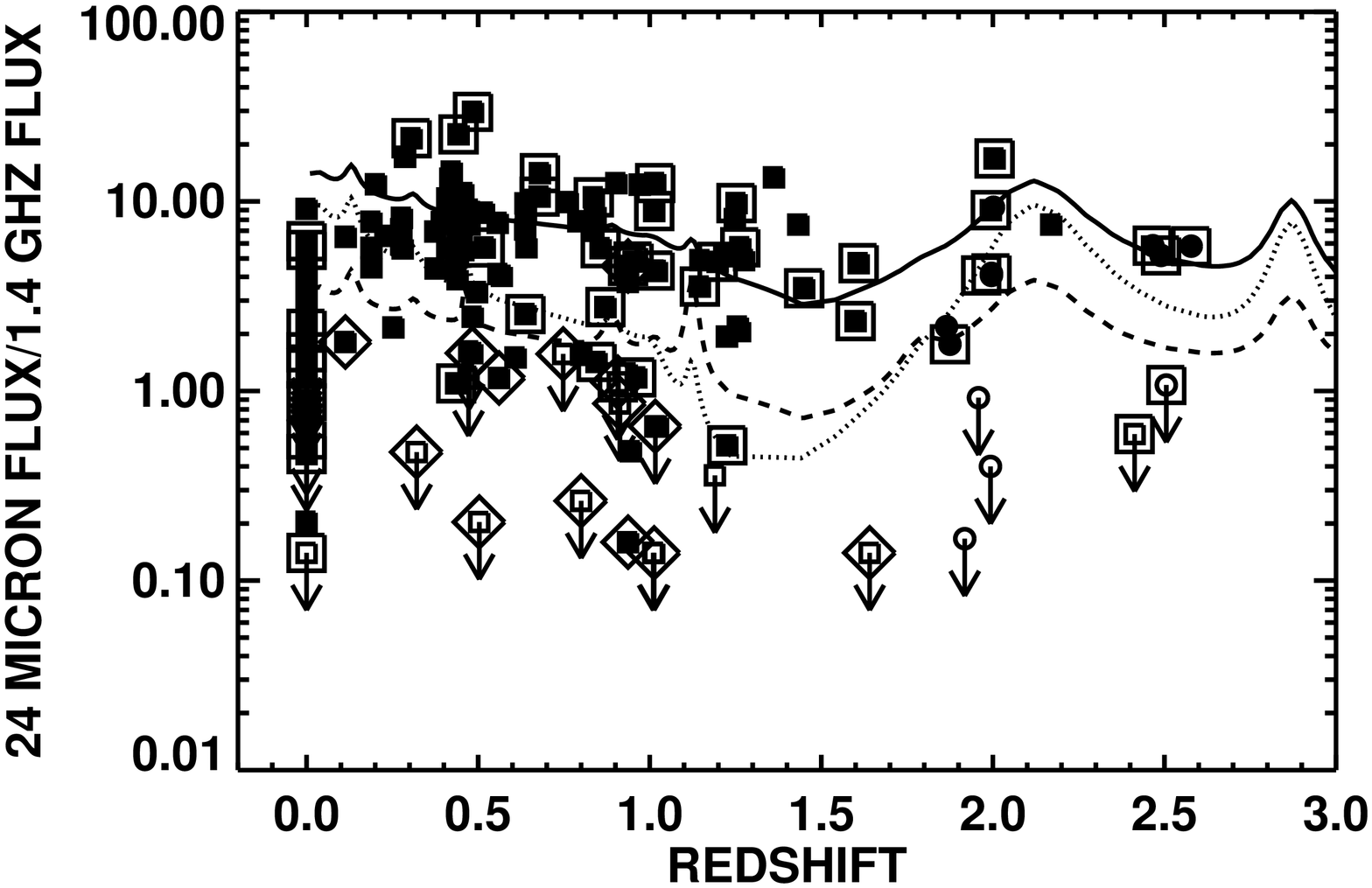,angle=90,width=3.5in}}
\figurenum{7}
\figcaption[]{
Ratio of 24~$\mu$m flux to 1.4~GHz flux vs. redshift for
the spectroscopically identified radio sources in the MIPS area.
Solid (open) symbols denote sources with (without) $24~\mu$m
detections. Circles denote the CS sources. 
Large open squares denote sources with either 
$L_{0.5-2~{\rm keV}}$ or $L_{2-8~{\rm keV}}\ge 10^{42}$~ergs~s$^{-1}$.
Large open diamonds denote spectroscopically classified absorbers.
Spectroscopically unidentified sources are nominally plotted
at $z=0$. The overplotted curves are the ratios from redshifted
template SEDs of M82 {\em (solid)\/}, Arp~220 {\em (dotted)\/},
and a spiral galaxy {\em (dashed)\/} (Silva et al.\ 1998).
\label{fig24um}
}
\end{inlinefigure}

We illustrate the difficulty with using MIR data to
estimate FIR luminosities in Figure~\ref{fig24um} 
(similar to Figure~1 of Donley et al.\ 2005), where 
we show the observed $24~\mu$m to 1.4~GHz flux ratio
versus redshift. Solid (open) symbols denote sources 
with (without) $24~\mu$m detections, and large open 
squares denote sources with soft or hard X-ray 
luminosities in excess of $10^{42}$~ergs~s$^{-1}$.
For the sources without $24~\mu$m detections, 
we have adopted the $80~\mu$Jy flux limit of the 
$24~\mu$m source catalog (see \S\ref{secmir}).
We have overplotted $24~\mu$m to 1.4~GHz flux 
ratios versus redshift for redshifted template SEDs 
(Silva et al.\ 1998). It is clear that there is 
substantial spread in the ratio, with the 24$~\mu$m 
measurements being very
sensitive both to redshift and to galaxy type. 
For the most part, the data points have values that
are within the range expected from the redshifted
template SEDs; however, there are also a number of 
sources not detected at 24~$\mu$m, including many of
the absorbers ({\em large open diamonds\/}; see \S\ref{secspectral}). 
The absorbers are most likely radio-loud AGNs,
which would not be expected to have SEDs like those shown.

We note that Donley et al.\ (2005), in their search 
for highly-obscured AGNs, defined a selection threshold 
of $q=\log(S_{24~\mu {\rm m}}/S_{1.4~{\rm GHz}})<0$
to classify a galaxy as probably having an AGN.
Based on the SEDs used in our Figure~\ref{fig24um}, 
this criterion might result in some contamination by
ultraluminous infrared galaxies, and, to a lesser extent, 
by spirals, as both of those SEDs dip below $q=0$. 
We do see sources below this value that do not have 
X-ray luminosities typical of an AGN and are not spectroscopically 
classified as absorbers, and some of these sources may be 
highly-obscured AGNs, but clearly we cannot rely on the 
24~$\mu$m data to determine their FIR luminosities.

\section{Properties of the Radio Sample by Spectral Class}
\label{secspectral}

For our spectroscopically identified sources, we classified
the optical spectra into four spectral classes, roughly
following the procedure used by Sadler et al.\ (2002) to
analyze low-redshift 1.4~GHz samples.
We classified sources without any strong
emission lines [EW([OII])$<3$~\AA\ or EW(H$\alpha+$NII)$<10$~\AA)]
as {\em absorbers\/}; sources with strong Balmer lines and
no broad or high-ionization lines as {\em star formers\/};
sources with [NeV] or CIV lines or strong
[OIII] [EW([OIII]~5007~\AA$)>3$~EW(H$\beta)$] as {\em Seyfert
galaxies\/,} and, finally, sources with optical lines having
FWHM line widths $>2000$~km~s$^{-1}$ as {\em broad-line AGNs\/}.
We have not classified one high-redshift source at $z=2.2032$. 
This source must be a strong
star former because it has UV absorption lines; however, its
[OII] emission is redshifted into the NIR and hence
cannot be measured with our spectra. We have also not
attempted to spectroscopically classify the CS sources,
which are generally of this type.

\subsection{Radio-to-Optical Ratios by Spectral Class}
\label{secradloud}

Some radio quasar studies have found bimodal distributions 
of rest-frame radio-to-optical ratios
(e.g., Kellermann et al.\ 1989; Stocke et al.\ 1992;
Ivezi{\'c} et al.\ 2002; however, see White et al.\ 2002;
Cirasuolo et al.\ 2003).
These have been interpreted as evidence that quasars come in 
two distinct populations, `radio-loud' and `radio-quiet'.
The boundary value between radio-loud and radio-quiet 
has been inferred to be 
$f_{5~GHz}/f_{2500~\AA}=10$ (Stocke et al.\ 1992). 

For our study, we define a rest-frame optical-to-radio
ratio of $R=f_{1.4~GHz}/f_{6500~\AA}$.
This is an approximately equivalent ratio, since the
$K$-correction to go from 5~GHz to 1.4~GHz is very similar
to the $K$-correction to go from 2500~\AA\ to 6500~\AA,
assuming spectral indices of $\alpha=0.8$
for $f_\nu\propto \nu^{-\alpha}$ in both cases
(Yun et al.\ 2001; Zheng et al.\ 1997).

In Figure~\ref{figsclass}, we show a histogram of 
logarithmic $R$ for our $z<1.6$ spectroscopically 
classified radio sample.
We have such uniform and complete wavelength coverage for all 
of the radio sources that we have just interpolated between 
our measurements to obtain rest-frame AB 6500~\AA\ magnitudes.
We have restricted the plot to sources with $z<1.6$ so that 
our interpolation remains valid. We have also $K$-corrected
the radio measurements to rest-frame 1.4~GHz.
The star formers clearly dominate the microJansky
sample, as was inferred from previous work
(e.g., Windhorst et al.\ 1995; Richards et al.\ 1998;
Richards 2000; Roche et al.\ 2002). 

Unfortunately, we do not have a large enough sample of Seyfert 
galaxies plus broad-line AGNs to investigate whether there is 
bimodality in the rest-frame radio-to-optical ratios. However,
we can see that all of the spectral classes show a wide range 
in radio-to-optical ratios. Thus, in contrast to the conclusions
of Machalski \& Condon (1999) from their study of
milliJansky radio sources in the Las Campanas Redshift Survey, 
we do not find this ratio to be a useful discriminator between 
star-forming galaxies and AGNs. Afonso et al.\ (2005) came to a 
similar conclusion using the Phoenix Deep Survey, which reaches 
into the sub-100~$\mu$Jy regime. They argued that the radio morphology
and polarimetry used by Machalski \& Condon (1999) to identify
AGNs results in the misclassification of many low-power AGNs,
and these sources would blur the distinction if properly
classified as AGNs. Since the radio-to-optical ratio does not 
appear to have much diagnostic power for our sample, we do not 
use it subsequently.

%
% FIGURE 8 
%
\begin{inlinefigure} 
\centerline{\psfig{figure=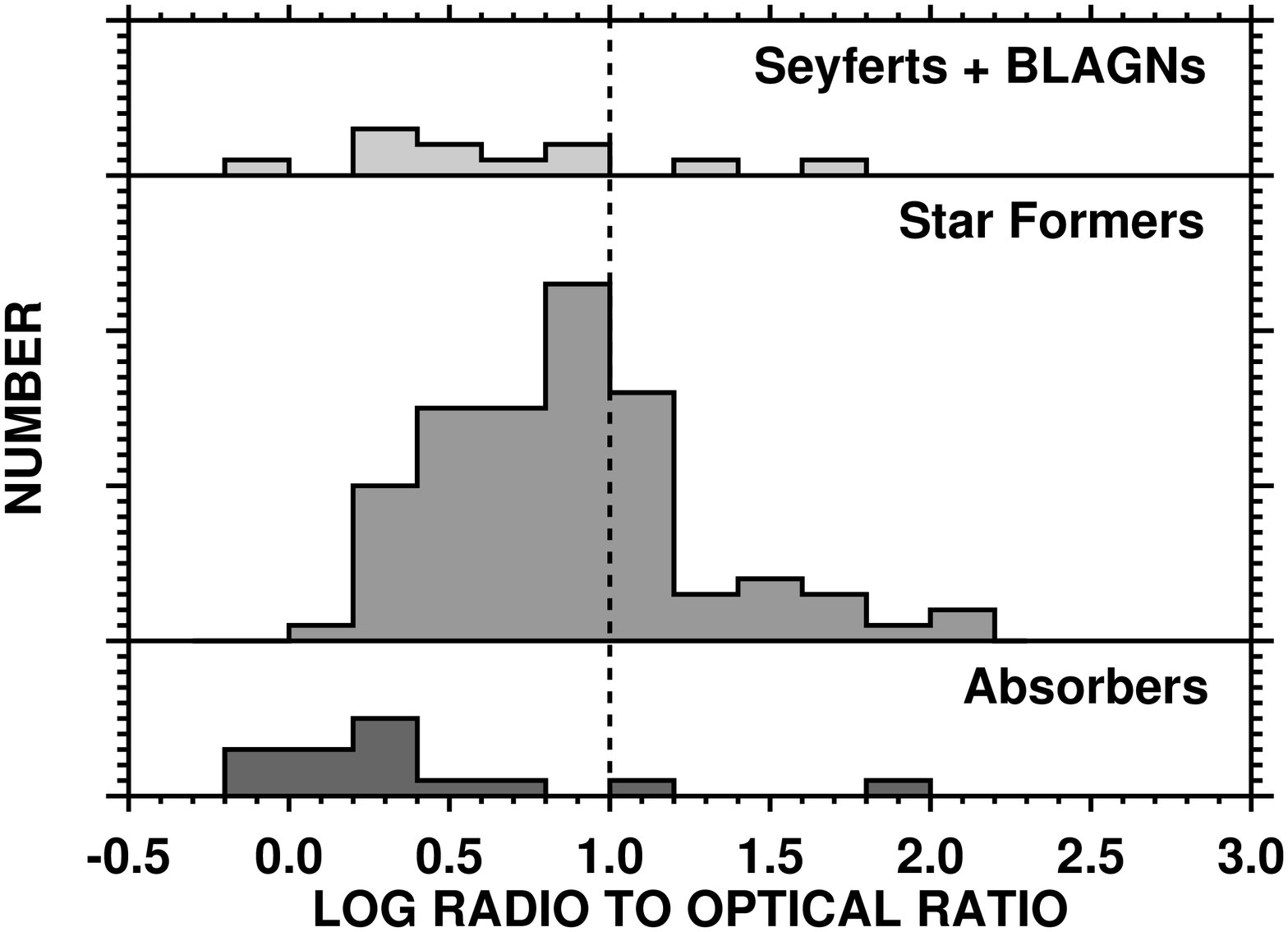,angle=90,width=3.5in}}
\figurenum{8}
\figcaption[]{
Histogram of the logarithmic rest-frame radio-to-optical
ratio, $R$, for our spectroscopically classified $z<1.6$
radio sample divided into absorbers, star formers, and
Seyfert galaxies plus broad-line AGNs (BLAGNs). Each
tickmark on the y-axis represents one source. The dashed
vertical line shows the classical boundary ($R=10$) between
radio-quiet and radio-loud.
\label{figsclass} 
}
\end{inlinefigure}

\subsection{Radio Power and Redshifts by Spectral Class}

In Figure~\ref{fighist}, we show histograms of the (a) logarithmic 
radio powers and (b) redshifts for the spectroscopically classified 
radio sample {\it (shaded)\/}. We calculate the rest-frame radio 
powers from the equation
\begin{equation}
P_{1.4~{\rm GHz}}=4\pi {d_L}^2 S_{1.4~{\rm GHz}} 10^{-29}
(1+z)^{\alpha - 1}~{\rm ergs~s^{-1}~Hz^{-1}} \,.
\label{eqradio}
\end{equation}
Here $d_L$ is the luminosity distance (cm), $S_{\rm 1.4~GHz}$
is the 1.4~GHz flux density ($\mu$Jy), and $\alpha$ is the
radio spectral index, which we take to be 0.8 (Yun et al.\ 2001).

The bulk of the star formers tend to lie at the lower radio powers.
The redshift distribution for the Seyfert galaxies plus broad-line
AGNs stretches to higher redshifts than the star formers and the
absorbers, though the absorbers in particular become more difficult
to identify spectroscopically at $z>1$. We show that this is a 
selection effect by including on Figure~\ref{fighist}b the redshift 
distribution for the photometrically identified absorbers
classified using their SEDs (see \S\ref{secphotz}) {\it (open)\/}.

\section{X-ray Properties of the Radio Sample}
\label{secxray}

An X-ray--radio correlation for star-forming galaxies has
been determined locally by a number of authors
(e.g., Bauer et al.\ 2002;
Ranalli et al.\ 2003; Grimm, Gilfanov, \& Sunyaev 2003;
Gilfanov, Grimm, \& Sunyaev 2004), suggesting that the
%X-ray and radio emission processes are associated and thus that
X-ray emission in star-forming galaxies
can be used as an indicator of the star formation rate.
These same authors, using a small sample of sources in the
CDF-N (at the time, only the 1~Ms {\em Chandra\/} data were
available; Brandt et al.\ 2001), also claim that the linear
relation extends to higher redshifts ($z<1.3$).
With our highly spectroscopically complete, cleanly
selected sample from just one flux band (1.4~GHz),
we are in an excellent position to test whether the local
X-ray--radio correlation does indeed continue to hold to higher
redshifts.

%
% FIGURE 9
%
\begin{inlinefigure}
\centerline{\psfig{figure=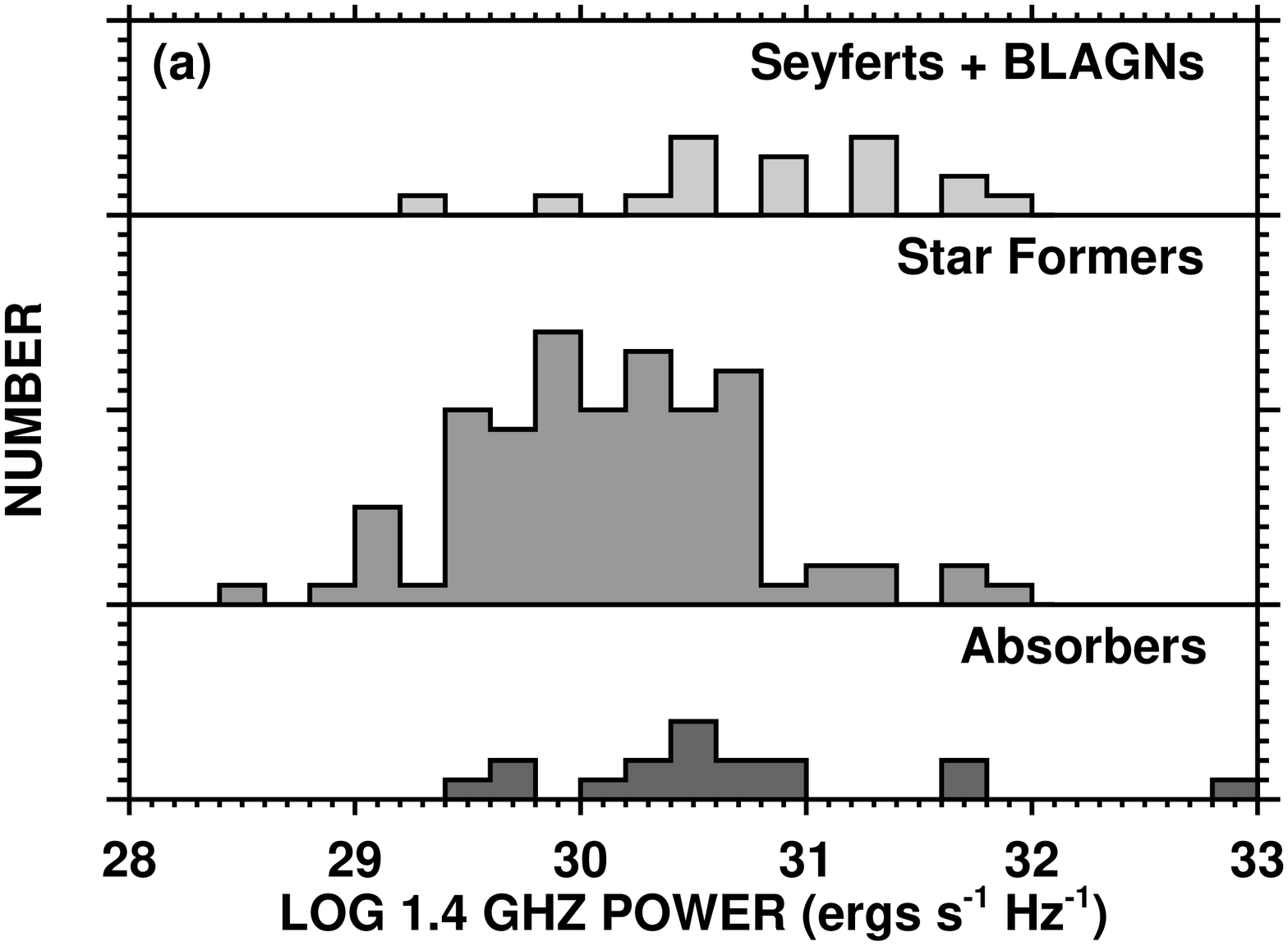,angle=90,width=3.5in}}
\centerline{\psfig{figure=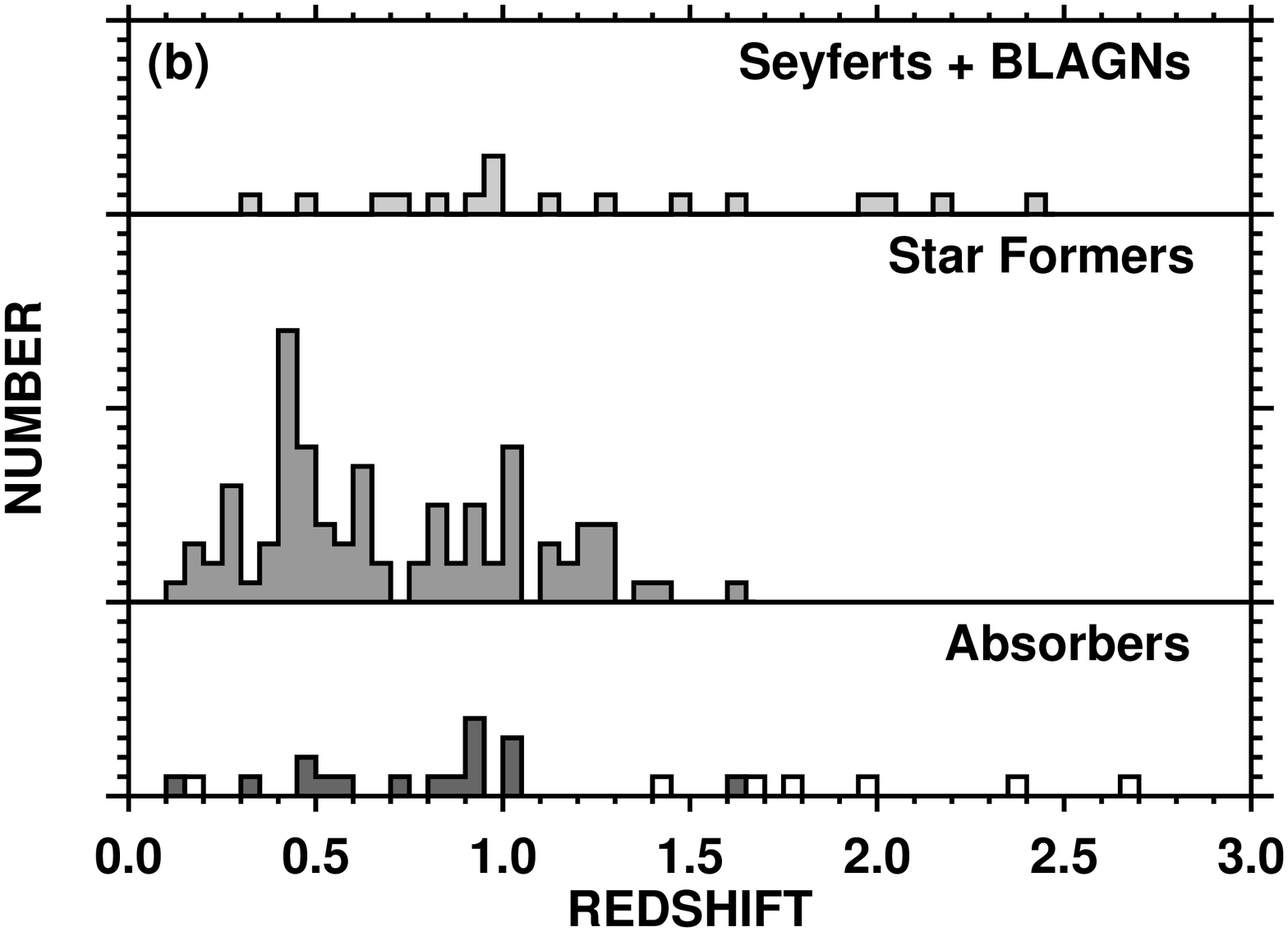,angle=90,width=3.5in}}
\figurenum{9}
\figcaption[]{
(a) Logarithmic power and (b) redshift distributions for our
spectroscopically classified radio sample {\it (shaded)\/} divided
into absorbers, star formers, and Seyfert galaxies plus broad-line
AGNs (BLAGNs). Each tickmark on the y-axis represents one source.
In (b), we also show the redshift distribution for photometrically
identified absorbers classified using their SEDs.
(see \S\ref{secphotz}) {\it (open)\/}.
\label{fighist}
}
\end{inlinefigure}

\begin{figure*}
\centerline{\psfig{figure=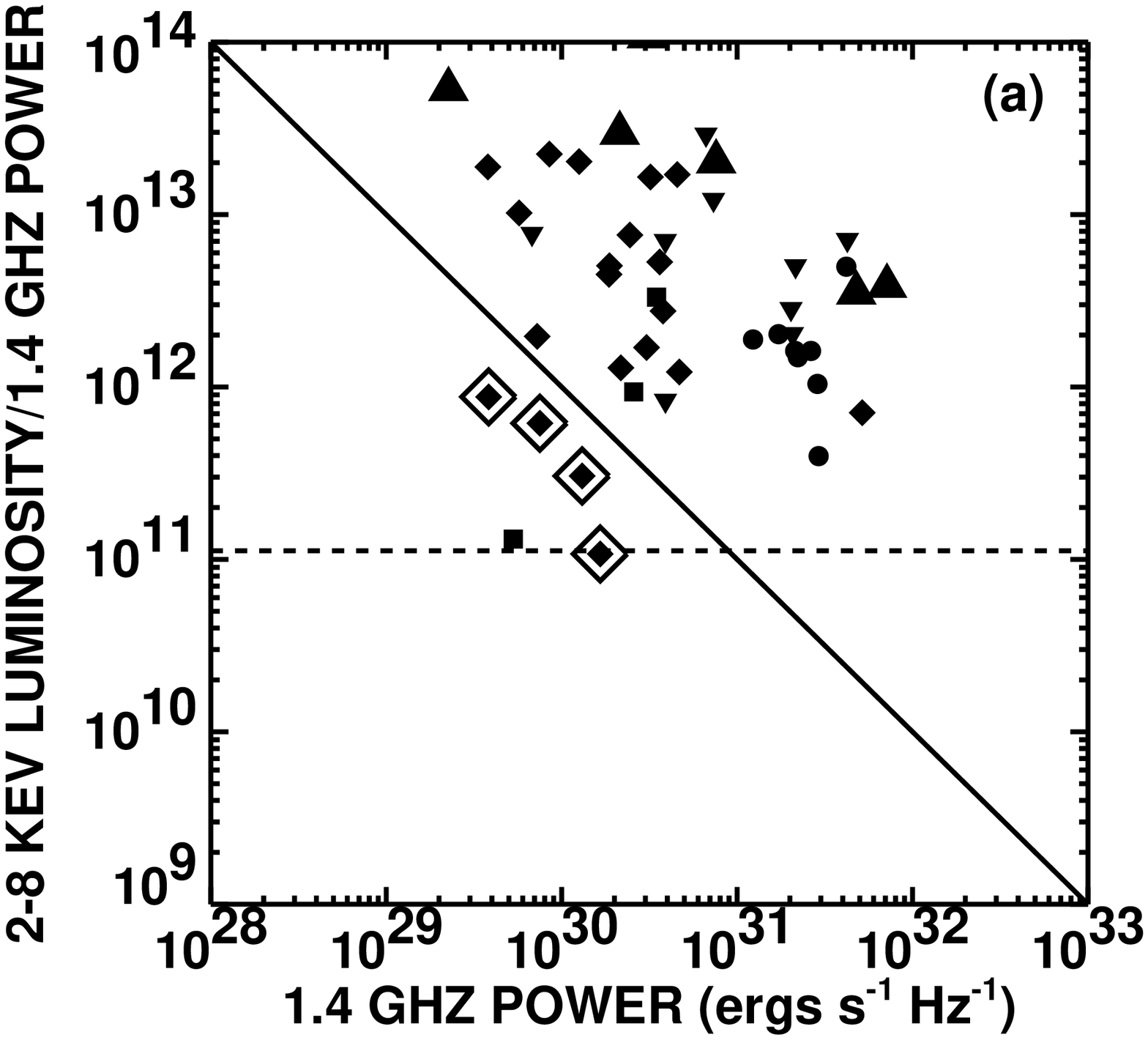,angle=90,width=3.5in}
\psfig{figure=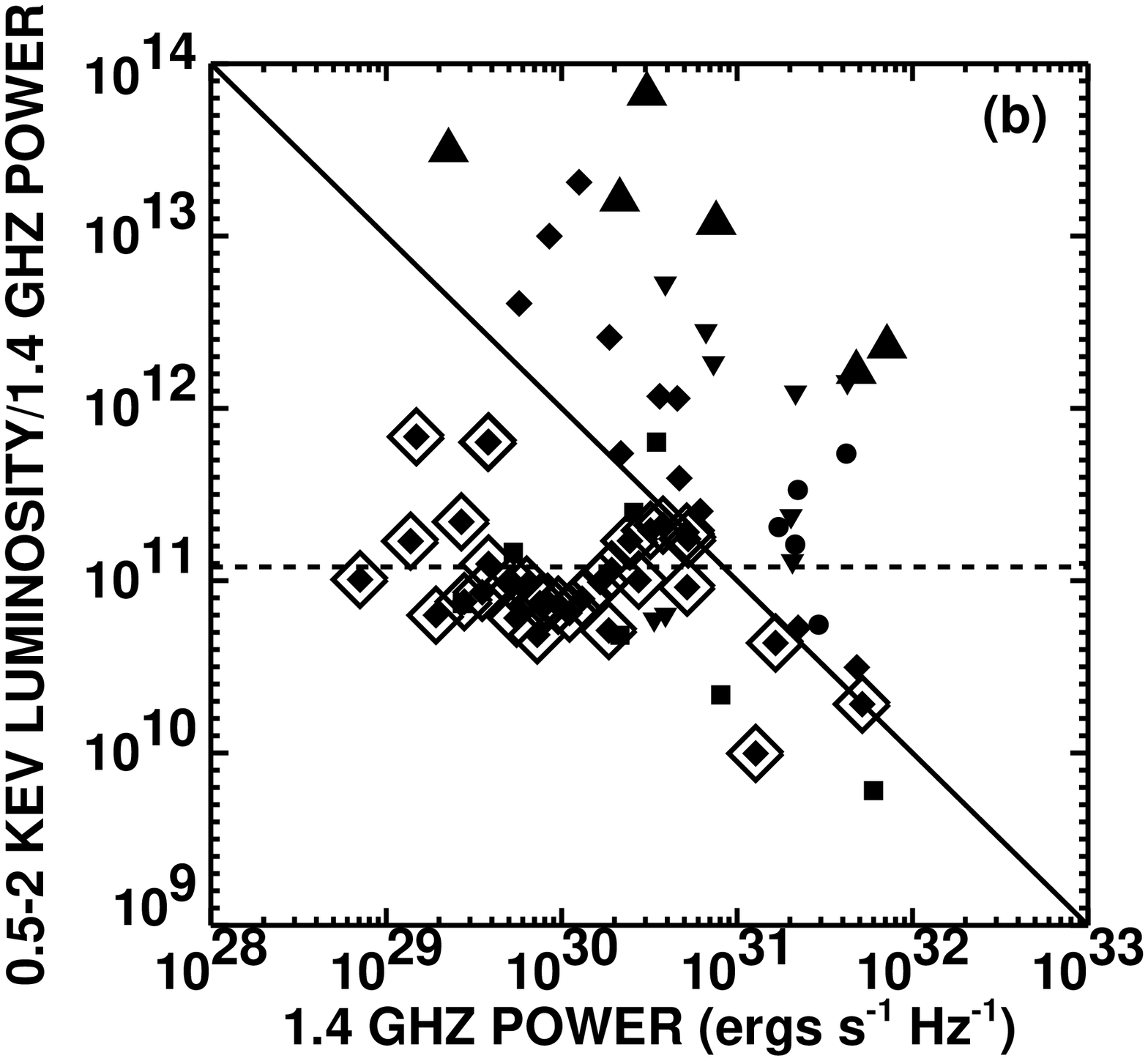,angle=90,width=3.5in}} 
\centerline{\psfig{figure=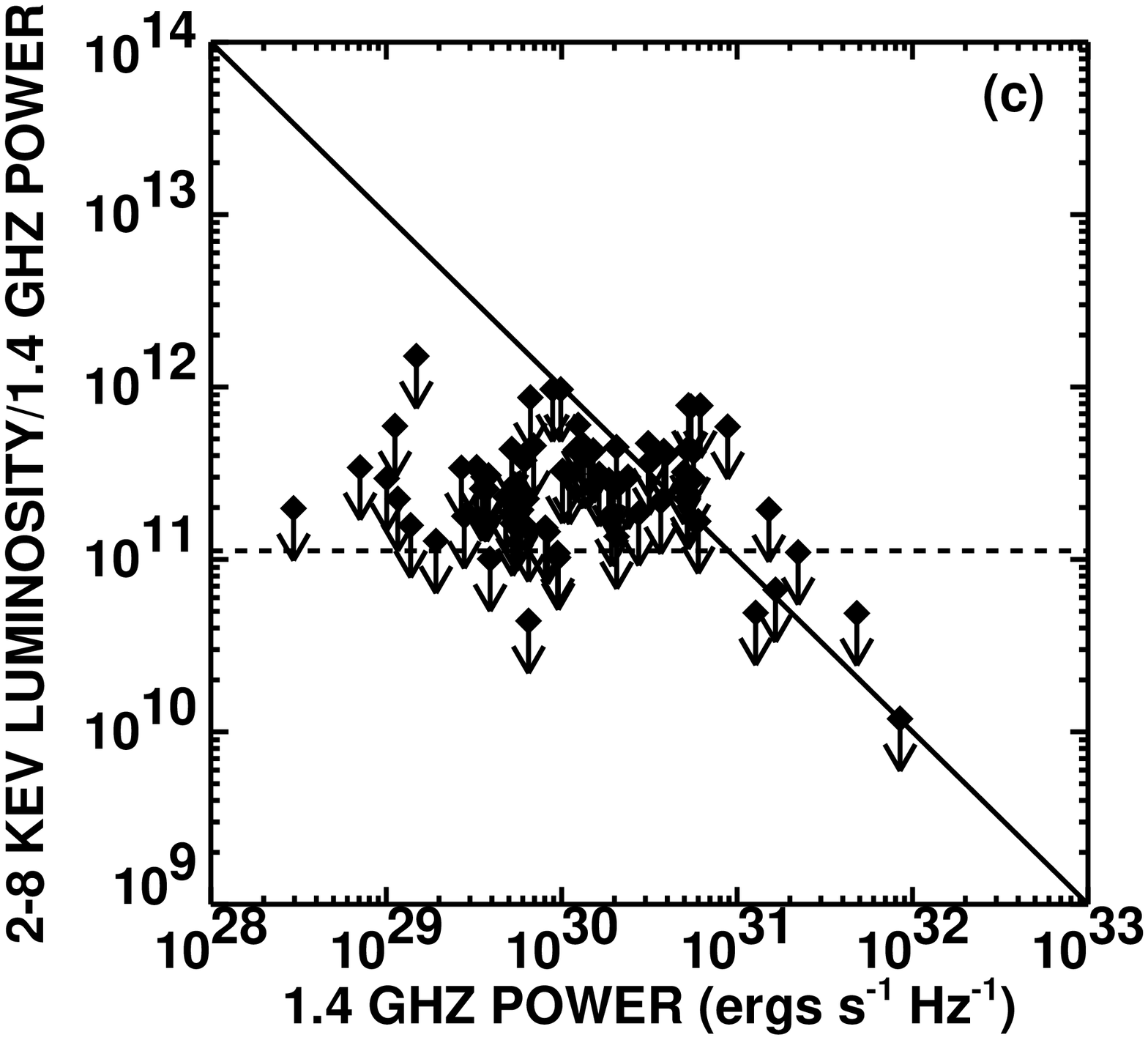,angle=90,width=3.5in}
\psfig{figure=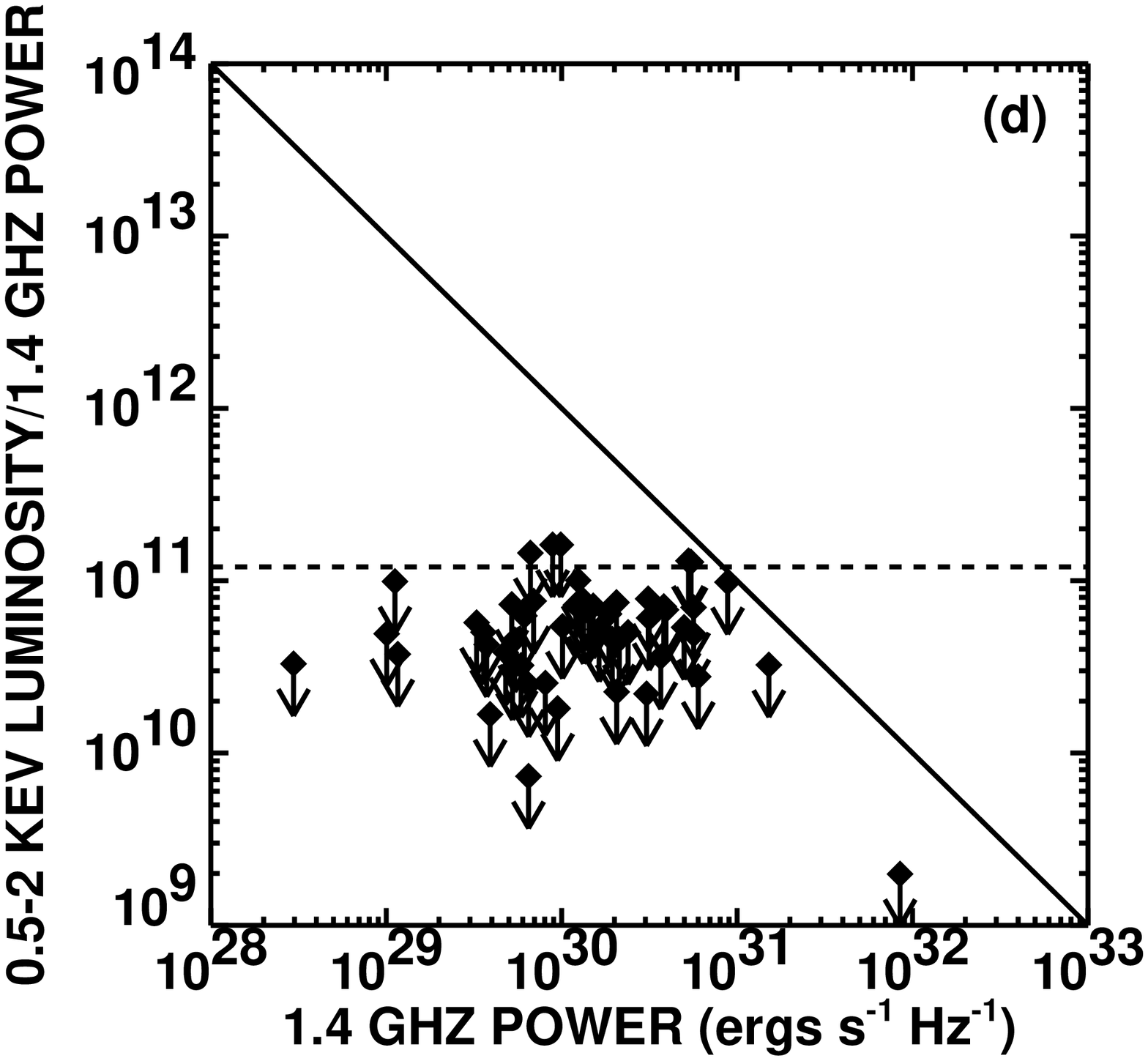,angle=90,width=3.5in}} 
\figurenum{10}
\figcaption[]{ 
Rest-frame $2-8$~keV or $0.5-2$~keV luminosity
over 1.4~GHz power vs. 1.4~GHz power for the spectroscopically identified
galaxies in the radio sample
(absorbers---{\em solid squares\/};
star formers---{\em solid diamonds\/};
Seyfert galaxies---{\em solid upside-down triangles\/};
broad-line AGNs---{\em large, solid triangles\/}; 
CS sources and our unclassified $z=2.2032$ source
[see \S\ref{secz}]---{\em solid circles\/})
that are (a) significantly hard X-ray detected, 
(b) significantly soft X-ray detected,
(c) star-forming galaxies only without significant hard X-ray detections,
(d) star-forming galaxies only without significant soft X-ray detections.
In (a) and (b), the large, open diamonds
denote significantly X-ray detected, star-forming galaxies with
either (a) $L_{\rm 2-8~keV}<10^{42}$~ergs~s$^{-1}$ or
(b) $L_{\rm 0.5-2~keV}<10^{42}$~ergs~s$^{-1}$.
In (c) and (d), the arrows show the hard and soft X-ray
detection limits from Alexander et al.\ (2003).
The dashed horizontal lines show the Ranalli et al.\ (2003) best-fit
relations for their local sample assuming a linear slope (their Eqs.~9 and
13 for the soft and hard X-ray bands, respectively; we have $K$-corrected
the latter to $2-8$~keV from $2-10$~keV assuming $\Gamma=1.8$).
The solid diagonal lines assume a fixed X-ray luminosity of
$10^{42}$~ergs~s$^{-1}$, the luminosity above which the sources are
most certainly powered by AGNs.
\label{figratiopower}
}
\end{figure*}

In Figure~\ref{figratiopower}a (\ref{figratiopower}b), 
using different symbols for the different spectral classes, 
we plot $2-8$~keV ($0.5-2$~keV) luminosity over 1.4~GHz power (we 
use the ratio to remove the redshift effect and improve the dynamic 
range) versus 1.4~GHz power for the spectroscopically
identified galaxies in our radio sample that were significantly
detected by Alexander et al.\ (2003) in the $2-8$~keV ($0.5-2$~keV) 
band. We calculated the rest-frame 1.4~GHz powers using
Eq.~\ref{eqradio}, and we calculated the rest-frame
hard (soft) X-ray luminosities from
\begin{equation}
L_X=4\pi d_L^2 f_X (1+z)^{\Gamma -2}~{\rm ergs~s^{-1}} \,.
\label{eqxray}
\end{equation}
Here $d_L$ is the luminosity distance (cm),
$f_X$ is the observed-frame hard (soft) X-ray flux
(ergs~cm$^{-2}$~s$^{-1}$), and $\Gamma$ is the photon index,
which we take to be 1.8 for all of the sources. Note that
using the individual photon indices (rather than the universal
power-law index of $\Gamma=1.8$) to calculate the
$K$-corrections would result in only a small difference
in the rest-frame luminosities (Barger et al.\ 2002).

We assume that any source more X-ray luminous than 
$10^{42}$~ergs~s$^{-1}$ is very likely to be an AGN on 
energetic grounds (Zezas et al.\ 1998; Moran et al.\ 1999), 
so we illustrate the ratio of this fixed hard (soft) 
X-ray luminosity to the 1.4~GHz power versus the 1.4~GHz 
power with a diagonal line in the figures. We note that 
since none of the star-forming galaxies {\em (solid diamonds)\/} 
show any obvious AGN signatures in their optical spectra, 
those with $L_X\ge 10^{42}$~ergs~s$^{-1}$ are most likely 
to be obscured AGNs. 

With a dashed horizontal line, we show the local hard 
(soft) X-ray/radio relation found by Ranalli et al.\ (2003; 
their Eqs. 13 and 9, respectively). We have converted their 
$2-10$~keV relation to $2-8$~keV, assuming a photon index 
of $\Gamma=1.8$. Although in Figure~\ref{figratiopower}a
we have too few significantly hard X-ray detected, 
star-forming galaxies with $L_{\rm 2-8~keV}<10^{42}$~ergs~s$^{-1}$ 
{\em (large open diamonds)} to identify any kind of 
correlation, in Figure~\ref{figratiopower}b, we can 
see the correlation that previous authors had identified.
However, several radio-powerful star formers significantly detected 
in soft X-rays lie well below the apparent correlation
and may be the luminous edge of the X-ray undetected
population. In fact, it is important to stress that the 
above figures only include X-ray detected
sources, not upper limits. Including the upper limits
would be the true test of whether a correlation actually
exists or whether one is just observing the upper end of 
the distribution and hence being fooled into thinking 
that there is a correlation when, in actuality, it is 
merely a selection effect.

Unfortunately, from Figures~\ref{figratiopower}c and d,
we can see that the apparent correlation is indeed merely 
a selection effect. Here we plot only the spectroscopically 
identified star formers that are not significantly detected 
in the $2-8$~keV and $0.5-2$~keV bands, respectively, using
the hard and soft X-ray detection limits from Alexander et al.\ (2003)
{\em (downward-pointing arrows)\/}.
The diagonal and horizontal lines are as before. 
The first thing to note is that even with the 2~Ms
exposure, we do not have the sensitivity to measure the 
bulk of the star-forming galaxy population in X-rays.
The second thing to note is that---again seen most
clearly in the soft X-ray figure---the upper limits are
not consistent with the Ranalli et al.\ (2003) relation. 

We note that some of the sources in Figure~\ref{figratiopower}
may be radio-loud AGNs, but anyone using the supposed correlation 
with their X-ray data is going to be faced with that uncertainty. 
Thus, the fact that we do not confirm the existence of an X-ray--radio
correlation at these redshifts, even with our knowledge of the 
optical spectra of the sources, is a major concern for those who 
want to infer star formation rates for their galaxies.  Although 
there are attractions to measuring a star formation history based 
on X-ray surveys because it avoids problems of obscuration 
(e.g., Norman et al.\ 2004), we are forced to conclude that it is 
not easy to do so.

%
% FIGURE 10 
%

\section{Optical Properties of the Radio Sample}
\label{secopt}

In Figure~\ref{figfradior}a, we plot 1.4~GHz flux versus
$R$ magnitude for the radio sample. Interestingly, there 
is no observed
correlation between 1.4~GHz flux and $R$ magnitude, which
helps to explain the uniformity of the spectroscopically
identified fraction with radio flux (see Fig.~\ref{figfraction}).
We illustrate this in histogram form in Figure~\ref{figfradior}b,
where we plot number versus $R$ magnitude for all of the radio
sources in our sample, divided into five radio flux bins.
The $R$ magnitude distribution is consistent with being drawn
from an invariant population. This result differs from that of
Georgakakis et al.\ (1999) and Afonso et al.\ (2005),
who found fainter median $R$ magnitudes with decreasing
radio flux density. The most likely reason for the discrepancy
is that the optical observations used by those authors
were much shallower, probing only to $R=22.5$.

%
% FIGURE 11
%
\begin{inlinefigure}
\centerline{\psfig{figure=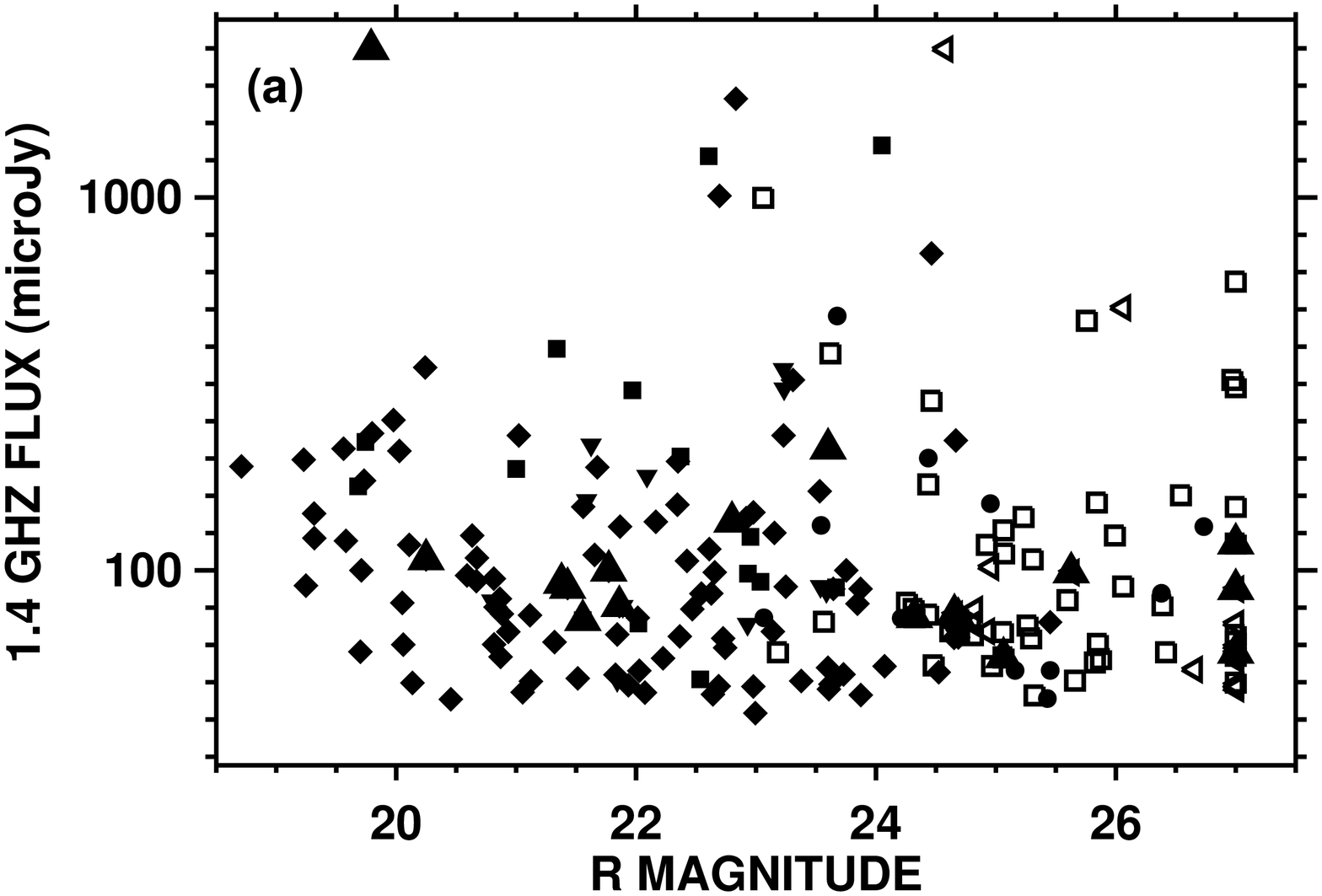,angle=90,width=3.5in}}
\centerline{\psfig{figure=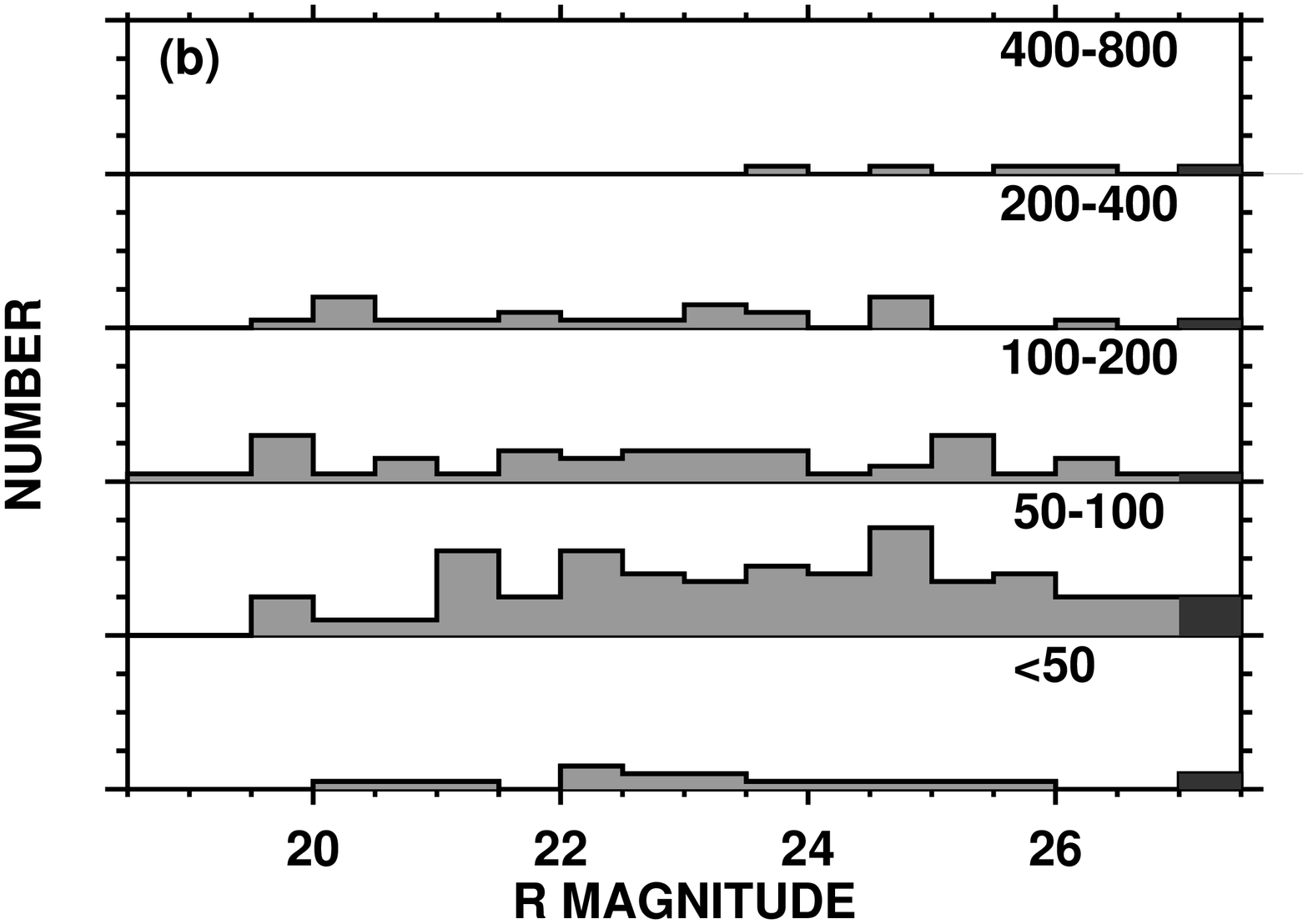,angle=90,width=3.5in}}
\figurenum{11}
\figcaption[]{
(a) 1.4~GHz flux vs. $R$ magnitude for the radio sample
(absorbers---{\em solid squares\/};
star formers---{\em solid diamonds\/};
Seyfert galaxies---{\em solid upside-down triangles\/};
broad-line AGNs---{\em large, solid triangles\/};
CS sources and our unclassified z=2.2032 source
[see \S\ref{secz}]---{\em solid circles\/};
photometric redshifts--{\em open squares\/};
unidentified sources---{\em open leftward-pointing triangles\/}).
Sources with radio fluxes greater than
2500~$\mu$Jy are shown at that flux, and sources with
$R>27$ are shown at that magnitude.
Sources with $R<20$ suffer from saturation
problems and are likely to be brighter than measured.
Radio flux ranges in microJanskys
are given for each bin. Each tickmark on the y-axis represents
five sources. Darker shading denotes sources with $R>27$.
\label{figfradior}
}
\end{inlinefigure}

In Figure~\ref{figzfradio}a, we show radio flux versus redshift
for the spectroscopically ({\em solid squares\/}) and 
photometrically ({\em open squares\/}) identified sources in 
the radio sample. We plot the unidentified sources at $z<0$.
Again we see no correlation between the two variables. 
In Figure~\ref{figzfradio}b we show this in histogram form, 
plotting number versus redshift for all of the radio sources
with either spectroscopic or photometric redshifts, divided
into five radio flux bins.
The redshift distribution is consistent with being invariant
with radio flux, although the selection of optically bright 
sources through the spectroscopic identification process may 
mean that we are missing the higher redshift tail of the 
redshift distribution function. Thus, probing to fainter radio 
fluxes does not sample to higher redshifts (Condon 1989).

%
% FIGURE 12
%
\begin{inlinefigure}
\centerline{\psfig{figure=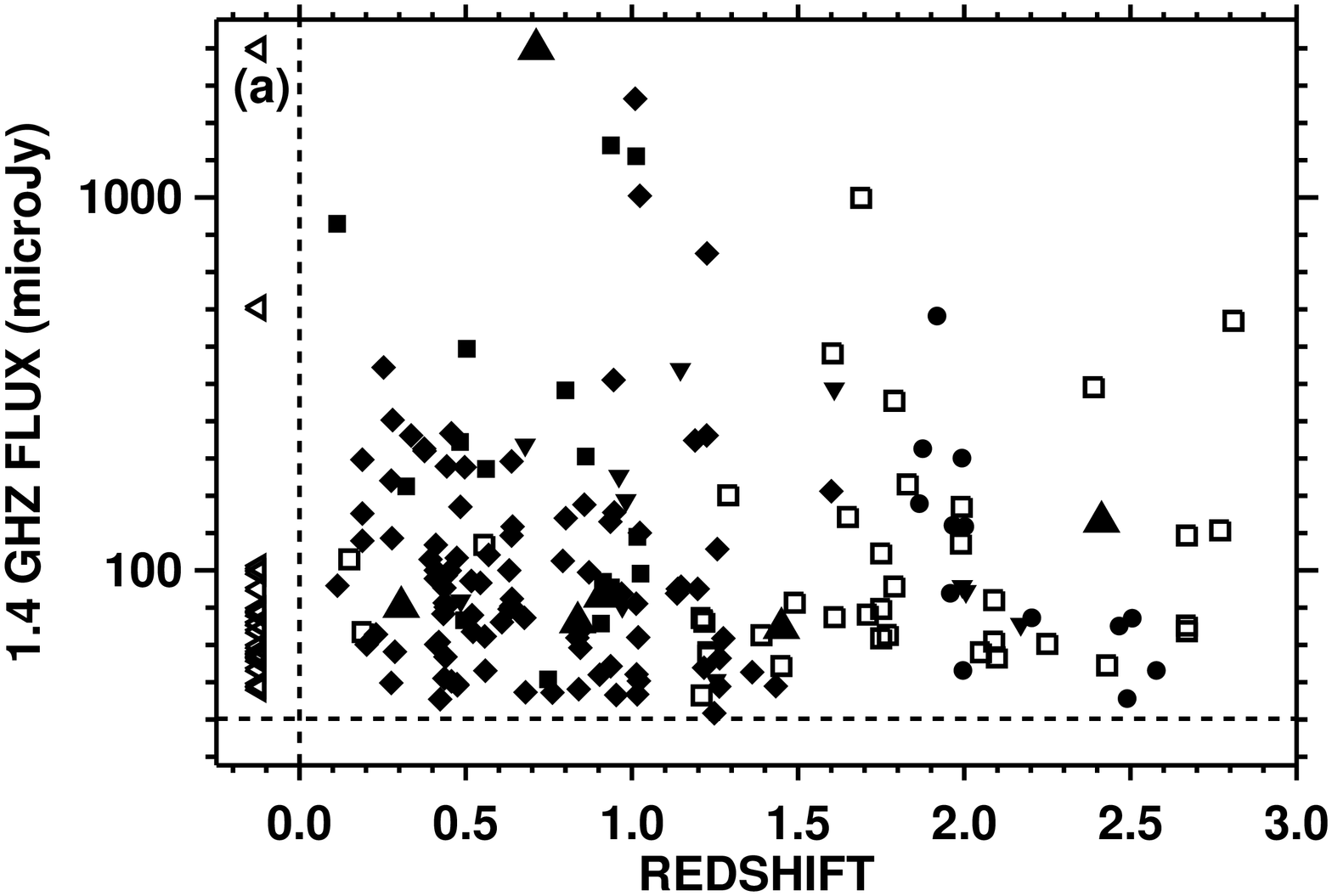,angle=90,width=3.5in}}
\centerline{\psfig{figure=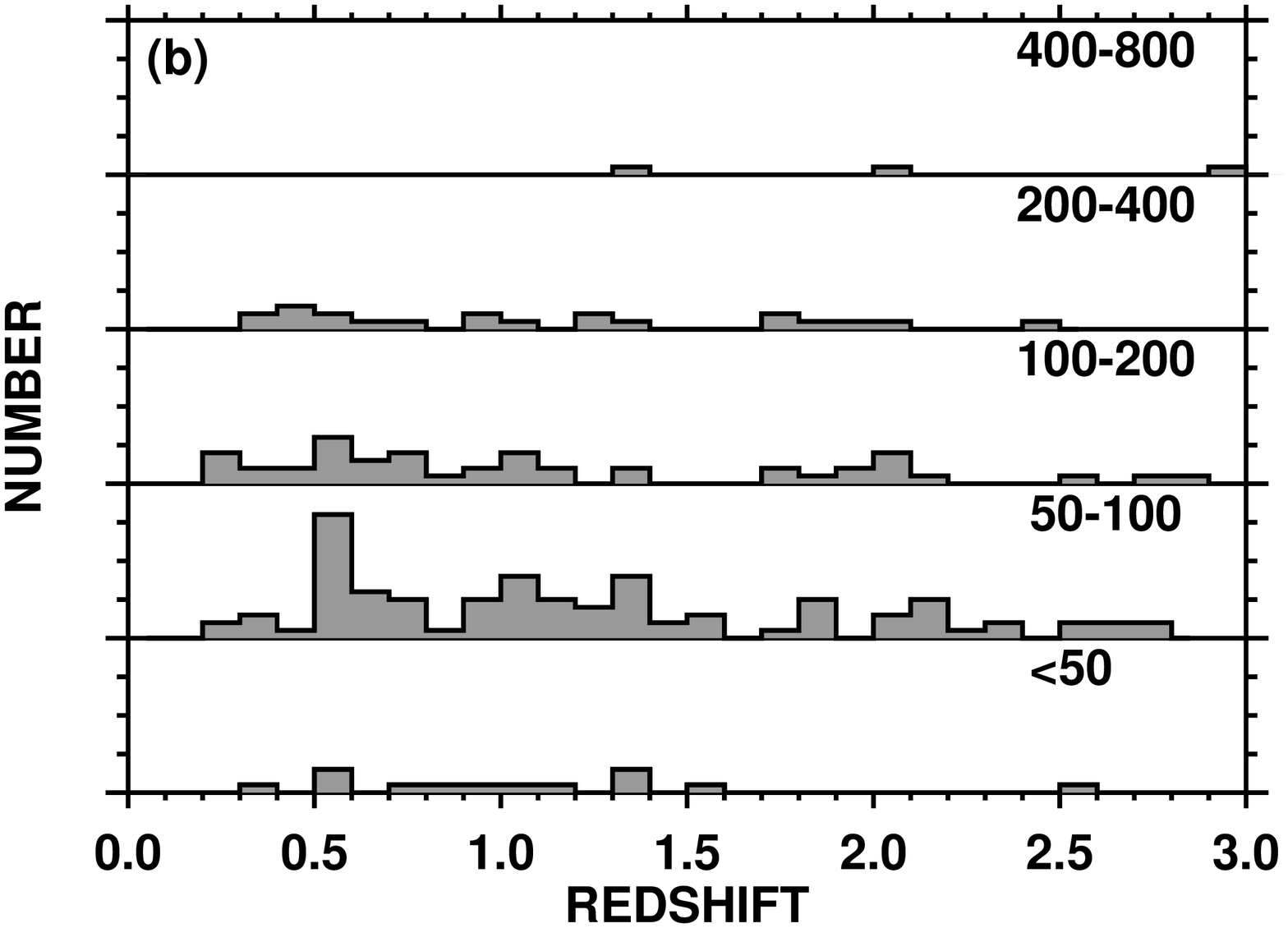,angle=90,width=3.5in}}
\figurenum{12}
\figcaption[]{
(a) 1.4~GHz flux vs. redshift for the radio sample
(absorbers---{\em solid squares\/};
star formers---{\em solid diamonds\/};
Seyfert galaxies---{\em solid upside-down triangles\/};
broad-line AGNs---{\em large, solid triangles\/}; 
CS sources and our unclassified z=2.2032 source
[see \S\ref{secz}]---{\em solid circles\/};
photometric redshifts--{\em open squares\/};
unidentified sources---{\em open leftward-pointing triangles at $z<0$\/}).
Sources with radio fluxes greater than 2500~$\mu$Jy are shown
at that flux.
(b) Redshift distribution for the radio sample with
spectroscopic or photometric redshifts, divided into
five flux bins. Radio flux ranges in microJanskys are
given for each bin. Each tickmark on the y-axis represents
five sources.
\label{figzfradio}
}
\end{inlinefigure}

This invariance in both the optical apparent magnitudes and the
redshifts of the host galaxies is a striking result. 
In Figure~\ref{figcolor}, 
we plot rest-frame AB 4500~\AA\ $-$ 8500~\AA\ color versus
(a) redshift and (b) 1.4~GHz flux for the spectroscopically
{\em (solid symbols)\/} and photometrically {\em (open squares)\/}
identified radio sources. Again, our uniform and complete
wavelength coverage for all of the radio sources means that
we can just interpolate between our measurements to obtain the
rest-frame colors. We only use redshifts up to $z=1.2$ so that
our interpolation remains valid.
The solid and dotted lines show the rest-frame AB 4500~\AA\ $-$ 8500~\AA\
color of an elliptical and irregular galaxy, respectively, using
Coleman et al.\ (1980). In both plots, there is very little color 
variation with redshift. The radio
sources are clearly drawn from the same host galaxy population,
and the host galaxy properties are not evolving much with redshift.

%
% FIGURE 13
%
\begin{inlinefigure}
\centerline{\psfig{figure=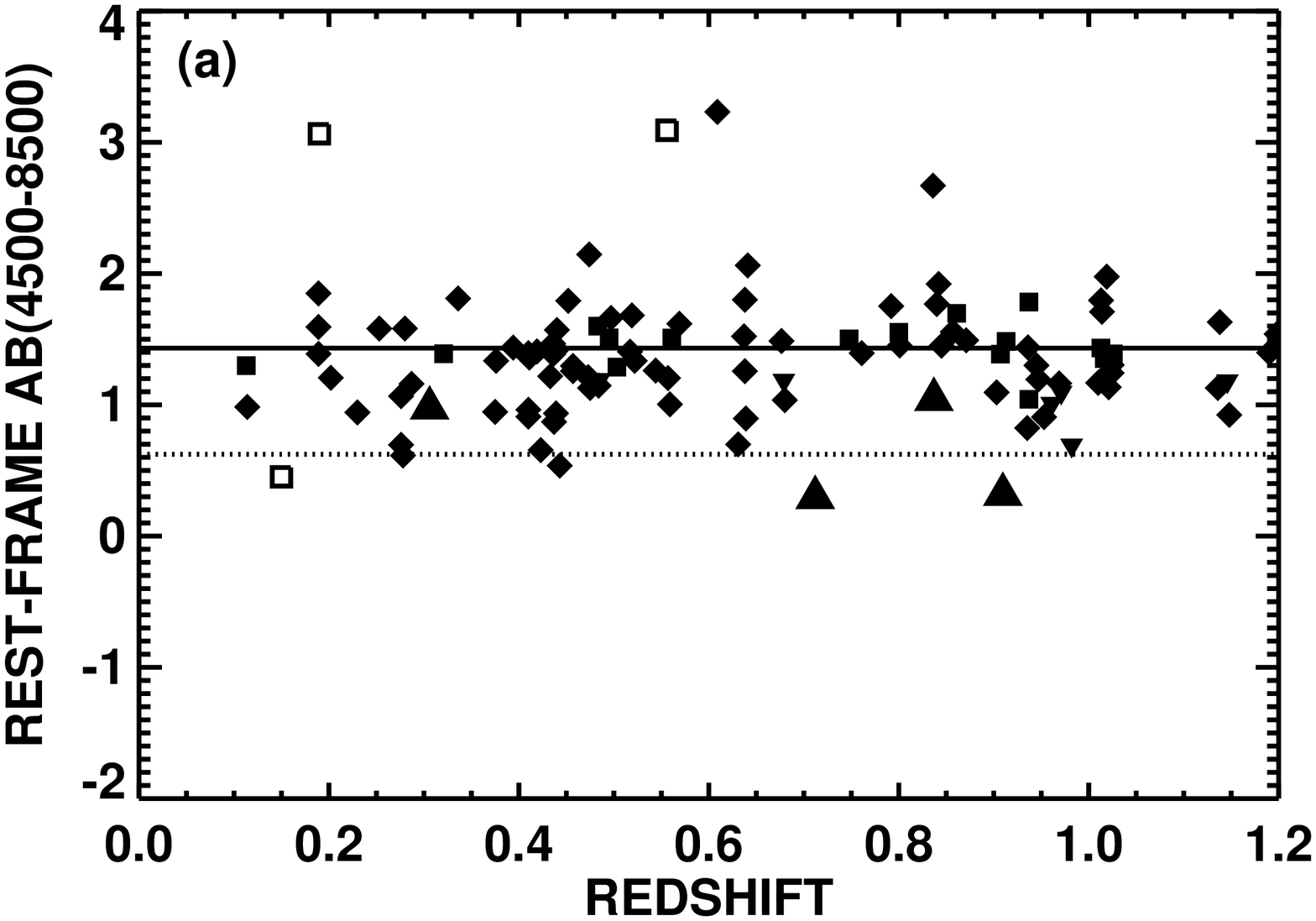,angle=90,width=3.5in}}
\centerline{\psfig{figure=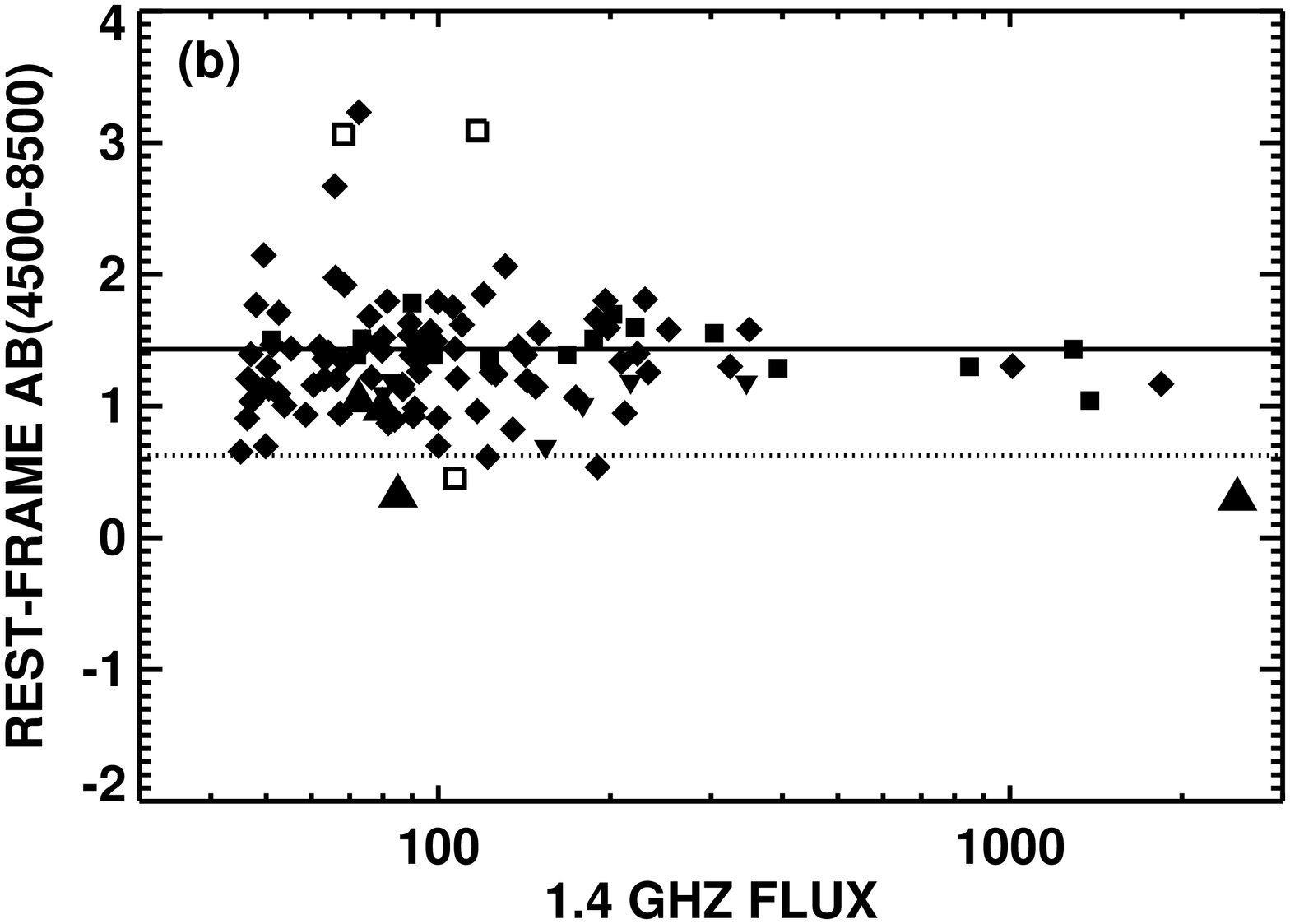,angle=90,width=3.5in}}
\figurenum{13}
\figcaption[]{
Rest-frame AB 4500~\AA\ $-$ 8500~\AA\ color vs. (a) redshift
and (b) 1.4~GHz flux for the $z<1.2$ spectroscopically
(absorbers---{\em solid squares\/};
star formers---{\em solid diamonds\/};
Seyfert galaxies---{\em solid upside-down triangles\/};
broad-line AGNs---{\em large, solid triangles\/})
and photometrically {\em (open squares)\/}
identified radio sample.
The solid (dotted) line shows the rest-frame color of
an elliptical (irregular) galaxy from Coleman et al.\ (1980).
\label{figcolor}
}
\end{inlinefigure}

In Figure~\ref{figabsr}a, we plot the absolute rest-frame $R$
magnitude, $M_R$, versus redshift for the spectroscopically
{\em (solid symbols)\/} and photometrically {\em (open squares)\/}
identified radio sources, interpolating between our 
measurements to obtain rest-frame AB 6500~\AA, and 
restricting to $z<1.6$ so that our interpolation remains valid. 
The approximate effect of our $R\sim 24$ spectroscopic selection 
limit on $M_R$ is shown by the dotted curve. The median absolute
magnitude at $z<0.4$ is $M_R=-21.4$, while that between $z=0.4$ and
$z=0.8$ is $M_R=-21.5$, so there may be a weak evolution to brighter
magnitudes at higher redshift. However, this evolution is small,
particularly when passive evolution of the galaxies is taken 
into account. At no redshift do the host galaxies become much
brighter than $M_R=-23$. Thus, we see a 
narrow range in $M_R$ with redshift ({\em dashed lines\/}), which 
suggests that the radio sources are chosen from approximately $L^*$
optical galaxies at all redshifts. At a very crude level,
the radio hosts are standard candles in their optical properties.

%
% FIGURE 14
%
\begin{inlinefigure}
\centerline{\psfig{figure=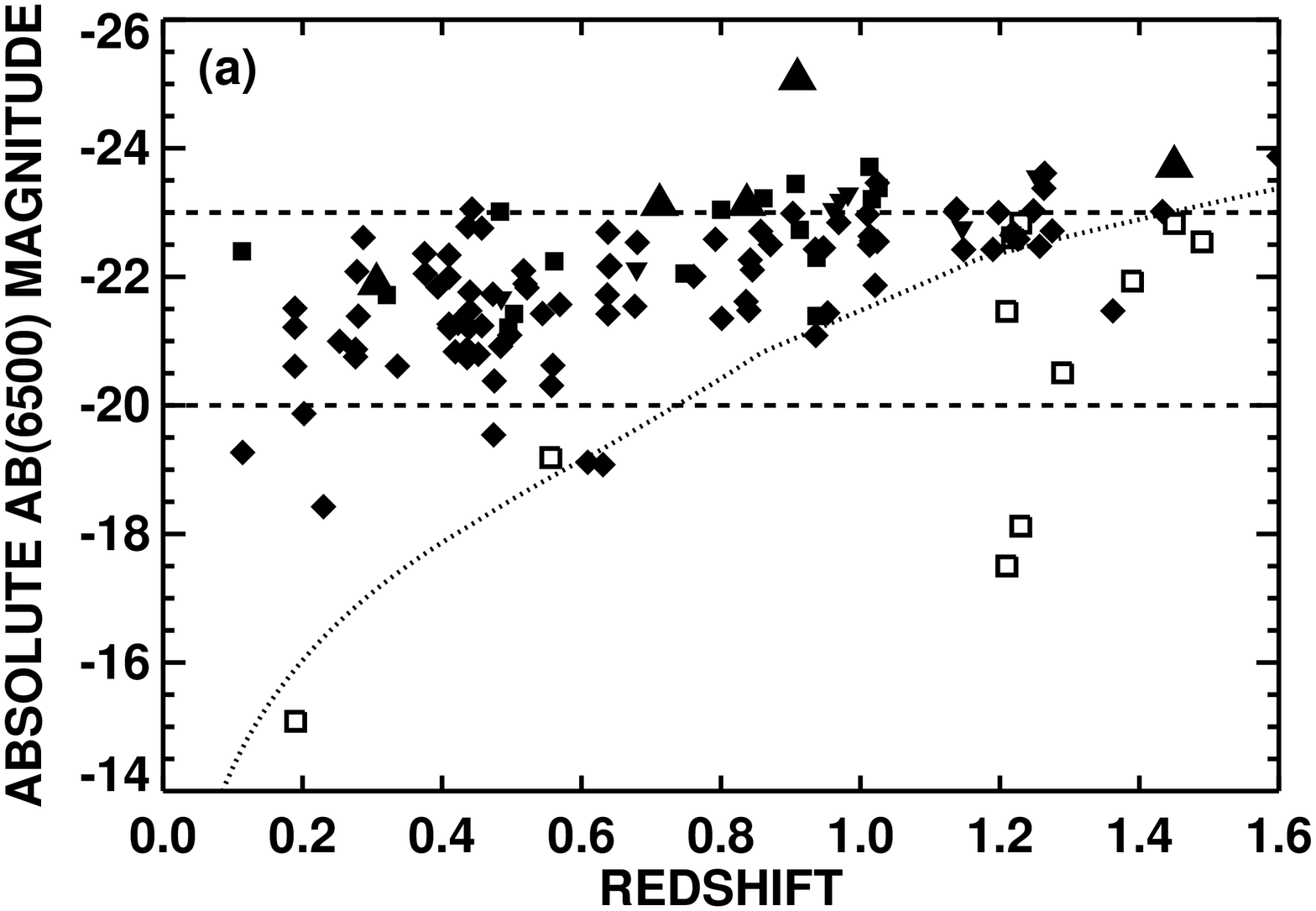,angle=90,width=3.5in}}
\centerline{\psfig{figure=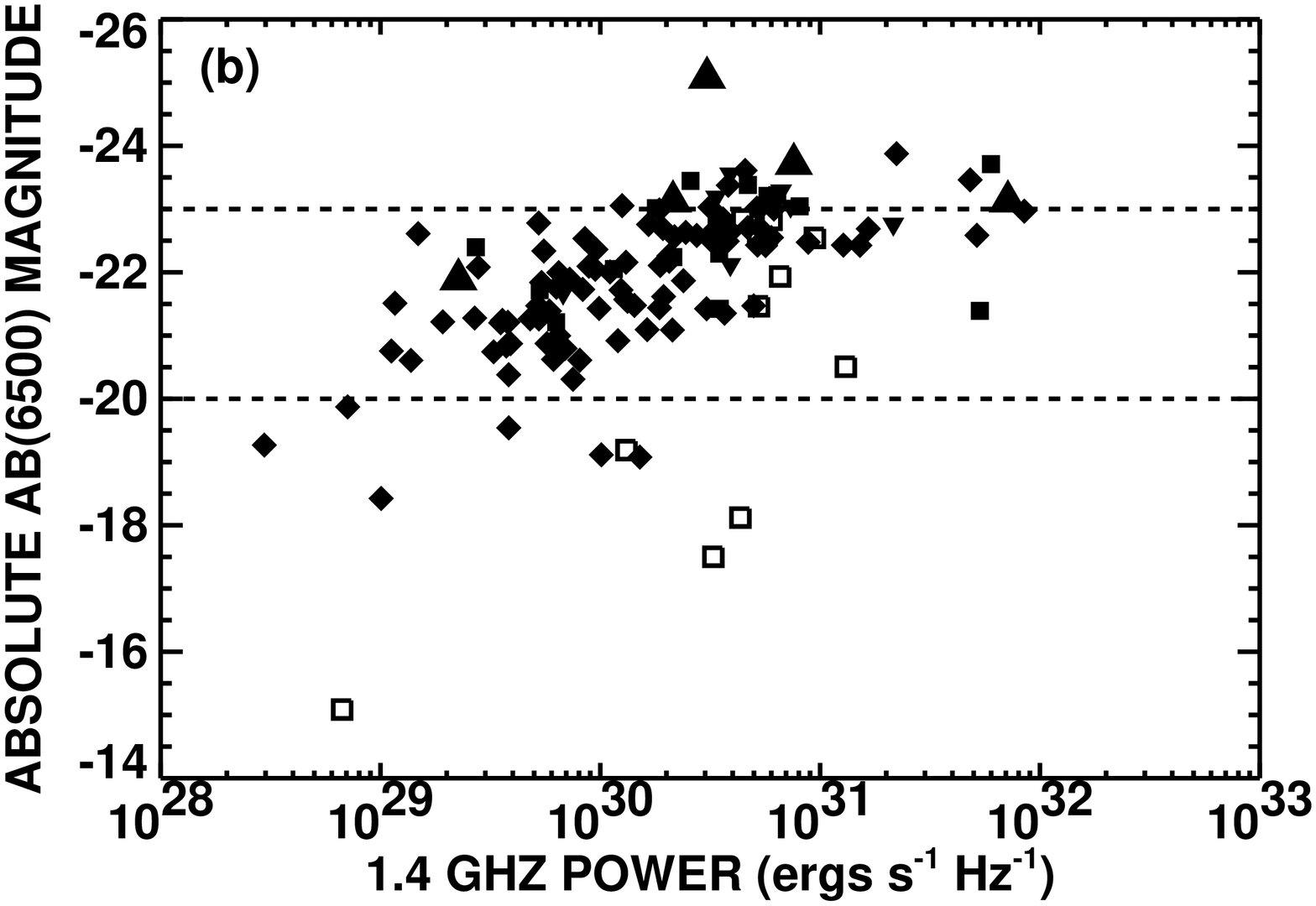,angle=90,width=3.5in}}
\figurenum{14}
\figcaption[]{
Absolute AB 6500~\AA\ magnitude versus (a) redshift and
(b) 1.4~GHz power for the $z<1.6$ spectroscopically
(absorbers---{\em solid squares\/};
star formers---{\em solid diamonds\/};
Seyfert galaxies---{\em solid upside-down triangles\/};
broad-line AGNs---{\em large, solid triangles\/})
and photometrically ({\em open squares\/}) identified radio sample.
Dashed lines show the approximate range in absolute magnitude,
from $-20$ to $-23$, where the majority of the radio sources
lie. Dotted curve in (a) illustrates the effect of an $R\sim 24$
spectroscopic incompleteness limit on the absolute magnitudes.
\label{figabsr}
}
\end{inlinefigure}

In Figure~\ref{figabsr}b, we plot $M_R$ versus 1.4~GHz power.
Even though the galaxy hosts have similar optical luminosities
at all redshifts, we can see from the figure that they have a wide 
range of radio powers (see also Cirasuolo et al.\ 2003). 
Thus, the radio powers of the host galaxies must be rising 
dramatically with increasing redshift, while the optical properties 
of the host galaxies are not changing.

We can see this dramatic evolution in Figure~\ref{figradiopower},
which shows the spectroscopically ({\em solid symbols\/})
or photometrically ({\em open squares\/}) identified radio 
sample in a redshift versus 1.4~GHz power plot. The vertical
dashed and solid lines show the radio powers 
(see \S\ref{secuplim} for how we determined these) corresponding
to a luminous infrared galaxy (LIRG;
$10^{11}~L_\odot \le L_{FIR}< 10^{12}~L_\odot$)
and an ultraluminous infrared galaxy
(ULIRG; $L_{FIR}\ge 10^{12}~L_\odot$), respectively. 
All of the ULIRGs in the 1.4~GHz sample to a completeness limit of 
$60~\mu$Jy {\em (dotted curve)\/} should be detected to $z\sim 1.3$.

Interestingly, we see from Figure~\ref{figradiopower} that only
a narrow range of luminosities is observed at each redshift
and that very high-luminosity sources are primarily found at
high redshifts ($z\gtrsim 1$). We therefore conclude that 
similar optical galaxies (see Fig.~\ref{figabsr}a) are
hosting more powerful radio galaxies at higher redshifts.
(See Cowie et al.\ 2004a for a quantitative description of
the evolution of the 1.4~GHz luminosity function.)

%
% FIGURE 15
%
\begin{inlinefigure}
\centerline{\psfig{figure=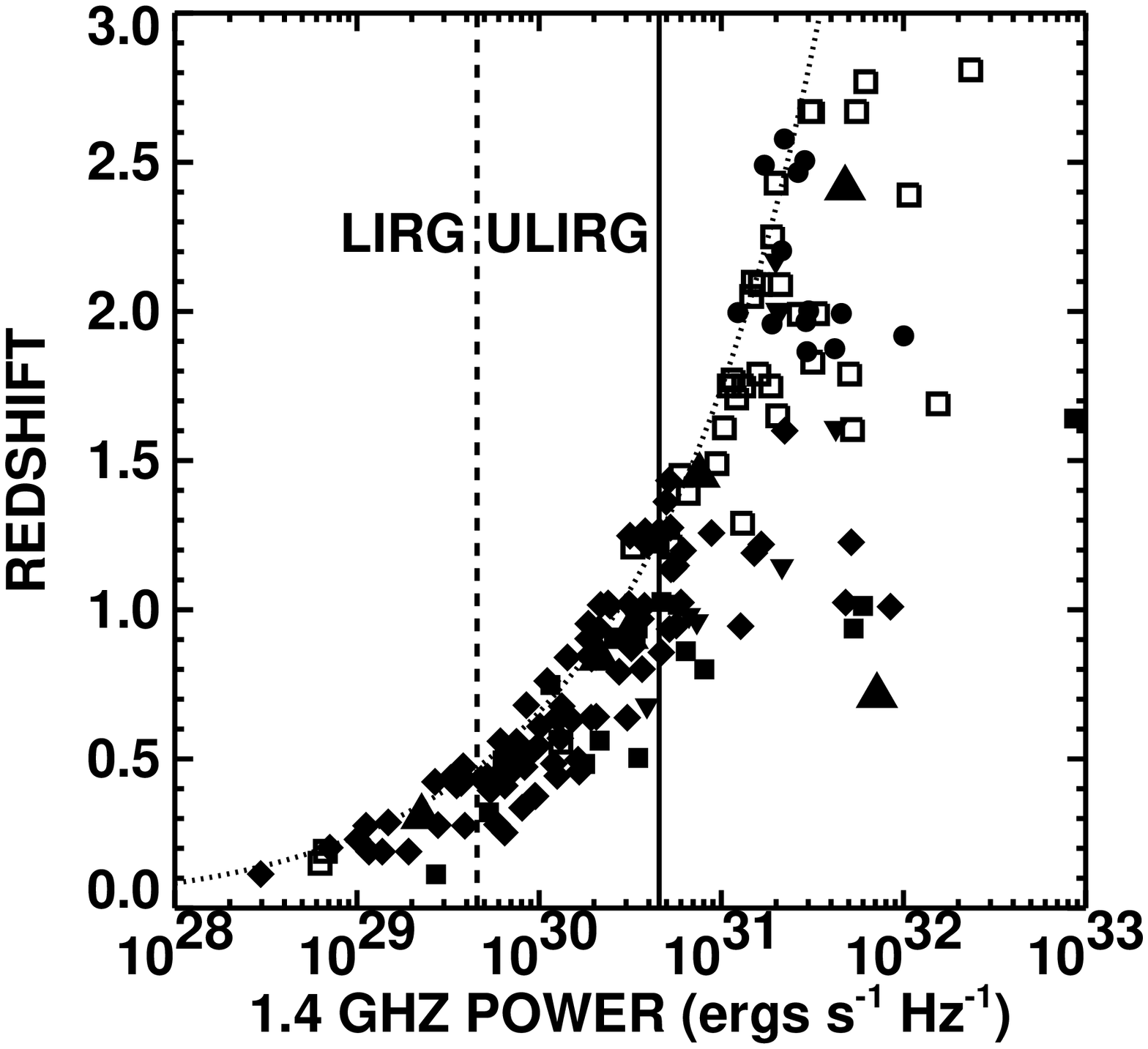,angle=90,width=3.5in}}
\figurenum{15}
\figcaption[]{
Redshift vs. 1.4~GHz luminosity for the
spectroscopically (absorbers---{\em solid squares\/};
star formers---{\em solid diamonds\/};
Seyfert galaxies---{\em solid upside-down triangles\/};
broad-line AGNs---{\em large, solid triangles\/};
CS sources and our unclassified $z=2.2032$ source
[see \S\ref{secz}]---{\em solid circles\/})
and photometrically ({\em open squares\/}) identified
radio sample.
The dashed and solid vertical lines show the equivalent radio
powers of a LIRG and a ULIRG, respectively.
The dotted curve shows the radio powers corresponding to the
60~$\mu$Jy completeness limit of the radio sample.
\label{figradiopower}
}
\end{inlinefigure}

%
% FIGURE 16
%
\begin{inlinefigure}
\centerline{\psfig{figure=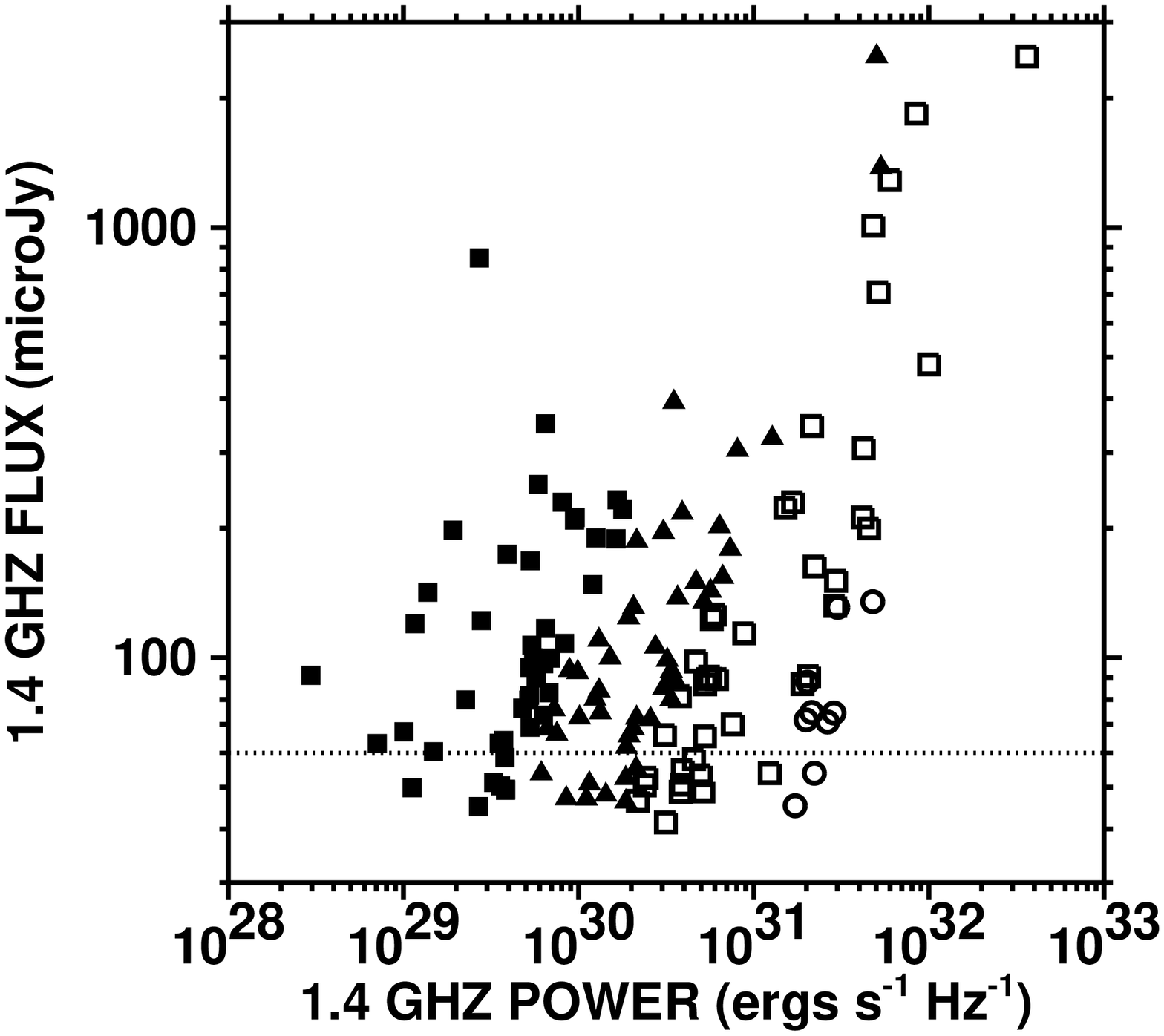,angle=90,width=3.5in}}
\figurenum{16}
\figcaption[]{
1.4~GHz flux versus 1.4~GHz power for the
spectroscopically identified radio sample, including the CS
sources ($z<0.5$---{\em solid squares\/};
$0.5\le z < 1$---{\em solid triangles\/};
$1\le z < 2$---{\em open squares\/};
$z\ge 2$---{\em open circles\/}).
The dotted line shows the 60~$\mu$Jy completeness limit.
Sources with radio fluxes greater than 2500~$\mu$Jy are shown
at that flux.
\label{figshells}
}
\end{inlinefigure}

We expand on this result in Figure~\ref{figshells}, where we
show 1.4~GHz flux versus 1.4~GHz power. Here we denote the different
redshift intervals by different symbols
($z<0.5$, {\em solid squares\/}; $0.5\le z <1$,
{\em solid triangles\/}; $1\le z <2$, {\em open squares\/};
$z\ge 2$, {\em open circles\/}). We see a series of
increasing luminosity slices with increasing redshift.
The reason for this effect was first suggested by Condon (1989)
based on the radio source counts. At low redshifts ($z\lesssim1$),
the radio surveys are dominated by the most luminous sources
in each redshift shell because the radio population is
experiencing such a dramatic luminosity evolution.
This is unlike any other wavelength survey, where sources are
generally picked up throughout the whole volume distribution.
However, at higher redshifts, the evolution slows, and a wider
range of luminosities is observed. 

Figure~\ref{figshells} contradicts the suggestion by
Ciliegi et al.\ (2005) that most
of the faintest radio sources may be associated with relatively
low radio luminosity sources at relatively modest redshifts
rather than with high radio luminosity AGNs at high redshifts.
In fact, the faintest radio sources have a wide range of
radio powers and redshifts, with the higher radio powers
corresponding to the higher redshifts.

\section{Upper Limits on Highly-Obscured AGNs}
\label{secuplim}

An important goal of the present work is to determine an upper limit 
on the number of AGNs with quasar-like bolometric luminosities 
($\sim 4\times 10^{45}$~ergs~s$^{-1}$; see below) that 
could be highly obscured---such that the light emerges in the FIR---and
hence not be detected in current deep X-ray 
surveys. This determination will be an upper limit, since for any 
of the radio sources, it is likely that much of the FIR light will 
be due to star formation rather than to AGN activity. Moreover,
such a determination will contain radio-loud sources, where the 
FIR-radio correlation substantially overestimates the FIR.

We calculate total FIR luminosities for the radio sources,
assuming that the FIR-radio correlation for star formers
and radio-quiet AGNs holds at high redshifts, and assuming that 
the sources are well described by this correlation.
Following Barger et al.\ (2001), we use the FIR-radio
correlation given in Sanders \& Mirabel (1996) with $q=2.35$
that they find holds for sources covering several orders of
magnitude in FIR luminosity.
Then $f_{FIR}=8.4\times 10^{14}~f^{rf}_{1.4}$, 
where $f^{rf}_{1.4}$ is the rest-frame
flux at 1.4~GHz. Assuming a synchrotron spectrum with a spectral index 
of $0.8$ (Yun et al.\ 2001), we can calculate the total FIR luminosity
from the observed 1.4~GHz flux using
\begin{equation}
L_{FIR}=4\pi d_L^2 (8.4\times 10^{14})f_{1.4}(1+z)^{-0.2} \,,
\end{equation}
where $d_L$ is the luminosity distance in cm and $f_{1.4}$ has
units of ergs~cm$^{-2}$~s$^{-1}$~Hz$^{-1}$.

Of particular interest for our analysis are the sources
that do not also have high ($\ge 10^{42}$~ergs~s$^{-1}$) X-ray 
luminosities. Quasars are typically defined as having $2-8$~keV
X-ray luminosities $\ge 10^{44}$~ergs~s$^{-1}$,
which, with a factor of 35 bolometric correction
(e.g., Elvis et al.\ 1994; Kuraszkiewicz et al.\ 2003),
would roughly correspond to ULIRG luminosities.
Thus, we consider any ULIRGs that do not have high X-ray 
luminosities to be our primary candidates to contain 
highly-obscured AGNs.

In Figure~\ref{figfirpower}, we plot redshift versus total
FIR luminosity for the sources with ULIRG {\em (vertical
line)\/} luminosities $L_{FIR}\ge 4\times 10^{45}$~ergs~s$^{-1}$.
We denote sources with spectroscopic redshifts
by solid symbols and sources with photometric redshifts by open
squares. We circle the sources that have either hard ($2-8$~keV)
or soft ($0.5-2$~keV) X-ray luminosities 
$\ge 10^{42}$~ergs~s$^{-1}$. The dotted curve shows the 60~$\mu$Jy
completeness limit for the radio observations. The dashed
horizontal line shows an adopted upper redshift cut-off of 
$z=2$, which is a compromise to keep our ULIRG selection
as uniform as possible (since the radio completeness limit begins
to cut out the lower luminosity ULIRGs at $z\gtrsim 1.3$),
while still retaining a statistically significant sample of sources.

%
% FIGURE 17
%
\begin{inlinefigure}
\centerline{\psfig{figure=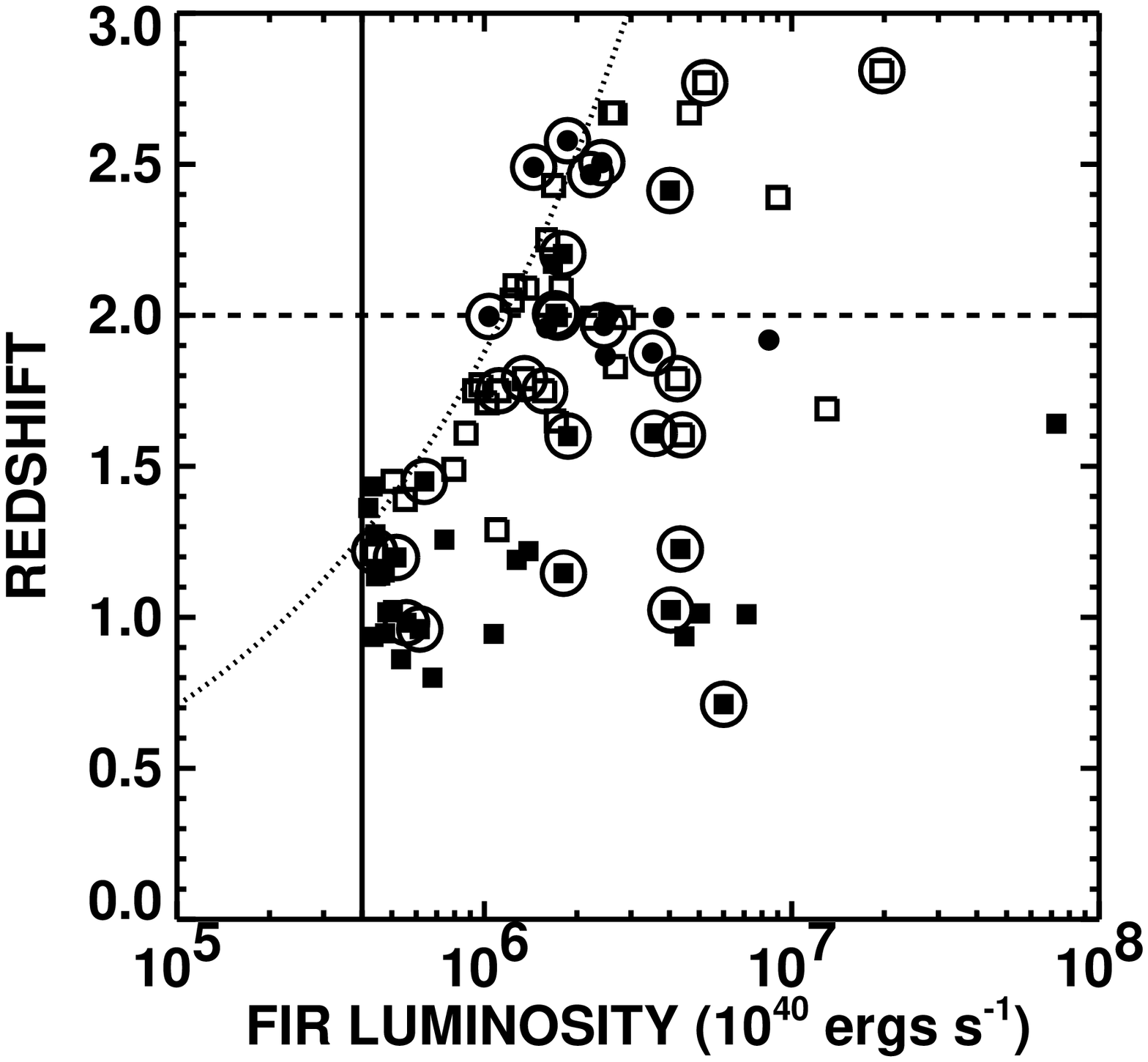,angle=90,width=3.5in}}
\figurenum{17}
\figcaption[]{
Redshift vs. total FIR luminosity for the radio sample with
ULIRG luminosities. Solid squares and solid circles
(the latter are CS sources)
denote spectroscopic redshifts, and open squares denote
photometric redshifts. Large, open circles denote sources that
have either $L_{0.5-2~{\rm keV}}$ or
$L_{2-8~{\rm keV}}\ge 10^{42}$~ergs~s$^{-1}$.
The dotted curve shows the 60~$\mu$Jy completeness limit of the
radio sample. The solid vertical line shows the minimum
luminosity of a ULIRG. The dashed horizontal line shows an upper
redshift cut-off of $z=2$ to provide as uniform a selection of
ULIRGs as possible, while still retaining a statistically
significant sample.
\label{figfirpower}
}
\end{inlinefigure}

In Figure~\ref{figcthist}a, we show histograms of the logarithmic
total FIR luminosities of the $z\le 2$ radio sources with either 
spectroscopic or photometric redshifts and ULIRG luminosities
divided into two categories: high X-ray luminosity sources 
(either $L_{0.5-2~{\rm keV}}$ or $L_{2-8~{\rm keV}}\ge
10^{42}$~ergs~s$^{-1}$) and low X-ray luminosity sources 
(both $L_{0.5-2~{\rm keV}}$ and 
$L_{2-8~{\rm keV}}<10^{42}$~ergs~s$^{-1}$). 
There are 38 sources in the low X-ray luminosity histogram 
and 20 sources in the high X-ray luminosity histogram. 
Excluding the bi-lobal source (the source with the highest FIR 
luminosity estimate in Fig.~\ref{figcthist}, which is likely a 
substantial overestimate for this radio-loud AGN) from being 
considered a candidate to contain a highly-obscured AGN,
we find a maximum ratio of highly-obscured AGN candidates 
to high-luminosity X-ray sources of 1.9 for radio sources in the quasar 
class. We stress again that this is an upper limit, since it includes 
star-forming ULIRGs as well as radio-loud AGNs, where the FIR-radio 
correlation is overestimating the FIR luminosity. Even if all 17 sources
without photometric redshifts (11 of which do not have any
obvious counterparts) were assumed to have ULIRG luminosities, 
to lie at $z\le 2$, and not to be X-ray luminous, the ratio would 
only increase to 2.7.

To show the classes of sources that are comprising the
low X-ray luminosity population, in Figure~\ref{figcthist}b,
we divide the population into absorbers, star formers, and
``other''. The latter category includes spectroscopically
unclassified sources and sources with only photometric redshifts.
In principle, we should be able to exclude the absorbers,
which are presumably radio-loud AGNs in the centers of elliptical
galaxies, from being considered candidates to contain highly-obscured
AGNs. However, in practice, apart from the bi-lobal source, we leave
the absorbers in because of the difficulties in
distinguishing between radio-loud and radio-quiet AGNs,
as discussed in \S\ref{secradloud}.

%
% FIGURE 18
%
\begin{inlinefigure}
\centerline{\psfig{figure=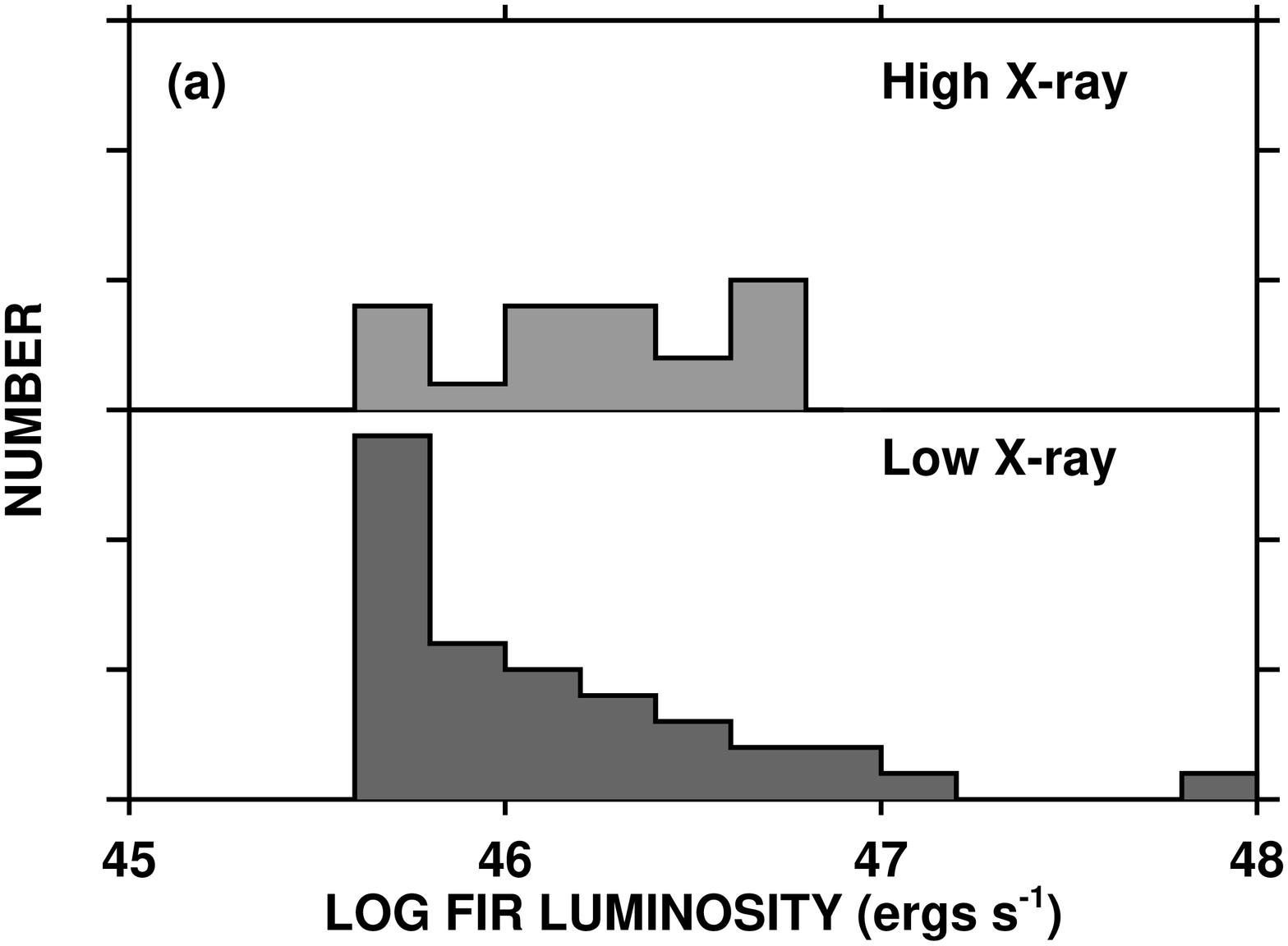,angle=90,width=3.5in}}
\centerline{\psfig{figure=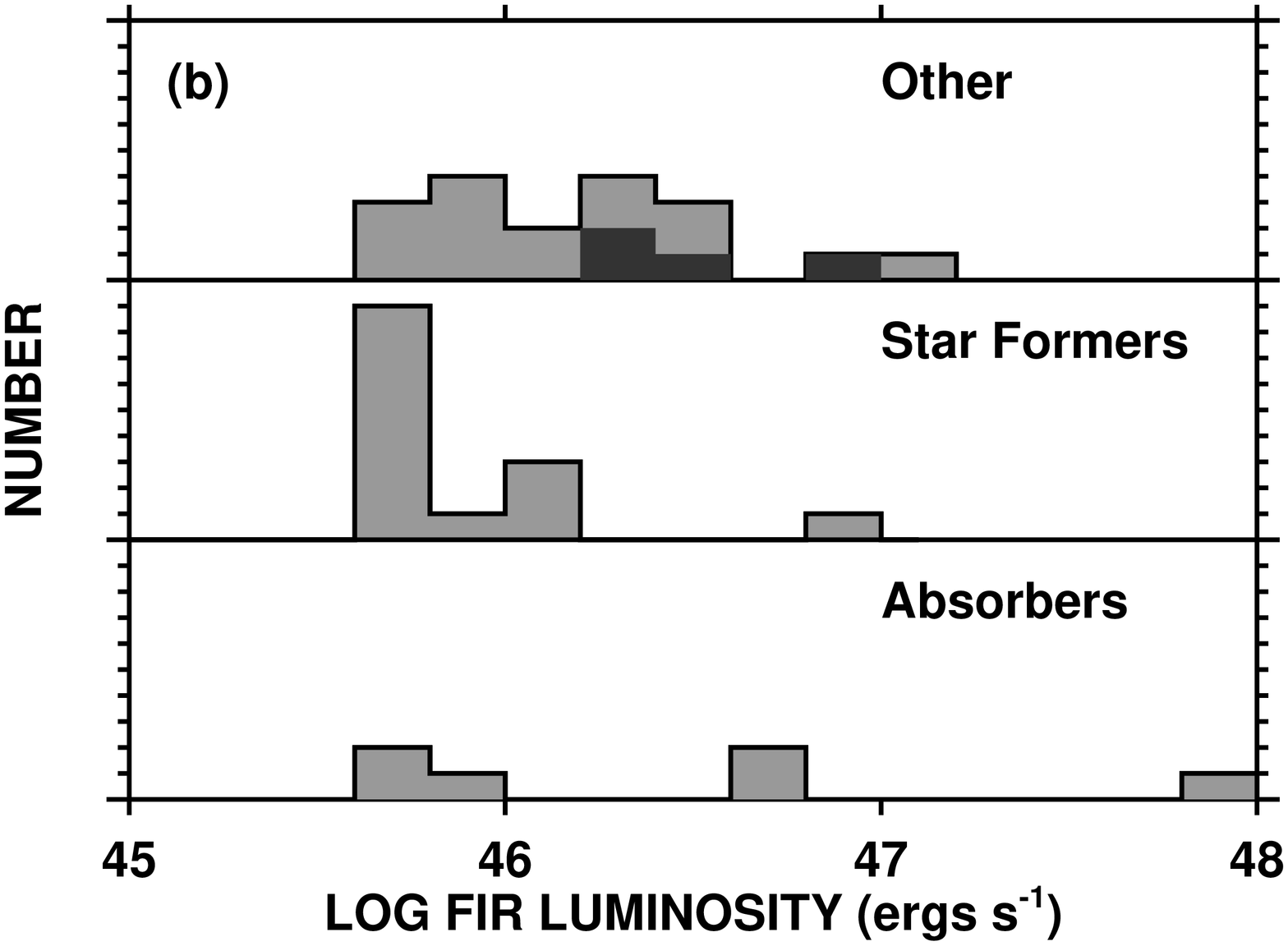,angle=90,width=3.5in}}
\figurenum{18}
\figcaption[]{
(a) Histograms of logarithmic FIR luminosity for the
$L_{FIR}\ge 4\times 10^{45}$~ergs~s$^{-1}$ and $z\le 2$ radio
sample with spectroscopic or photometric redshifts, divided
into two X-ray luminosity bins: high X-ray
(either $L_{0.5-2~{\rm keV}}$ or
$L_{2-8~{\rm keV}}\ge 10^{42}$~ergs~s$^{-1}$)
and low X-ray (both $L_{0.5-2~{\rm keV}}$ and
$L_{2-8~{\rm keV}}<10^{42}$~ergs~s$^{-1}$).
Each tickmark on the y-axis represents five sources.
(b) Histograms of the low X-ray luminosities from (a)
divided into three categories: absorbers, star formers, and
``other'' (spectroscopically unclassified sources and sources
with only photometric redshifts). The CS sources
are denoted by dark shading. Each tickmark on the y-axis
represents one source.
\label{figcthist}
}
\end{inlinefigure}

In Figure~\ref{fignumden}, we show in two redshift bins
the number densities of radio sources with either 
spectroscopic or photometric redshifts and
$L_{FIR}\ge 4\times 10^{45}$~ergs~s$^{-1}$,
divided according to X-ray luminosity. At these high FIR
luminosities, the ratio of highly-obscured AGN candidates 
to high X-ray luminosity sources 
does not appear to be changing much with redshift, with values 
of 2 and 1.8 for the lower and higher redshift bins.

We can get a rough check on whether the upper bound on the 
ratio continues to stay about the same at lower 
FIR luminosities, and hence lower redshifts, by redoing our 
analysis for LIRGs. Of course, there is much more contamination 
at these lower luminosities by star formation, since M82-type
sources will push into the sample and will not show up as 
luminous X-ray sources. In Figure~\ref{figfirpower2}, we again 
plot redshift versus total FIR luminosity, this time for the 
radio sources with LIRG luminosities
($4\times 10^{44}\le L_{FIR}<4\times 10^{45}$~ergs~s$^{-1}$;
{\em vertical lines\/}). The dashed 
horizontal line shows an adopted upper redshift cut-off of
$z=0.7$. The ratio of highly-obscured AGN candidates
to high X-ray luminosity sources for the LIRG 
population with $0.2 < z \le 0.7$ comes out to be 5.4, which
can be taken as an extreme upper limit.

%
% FIGURE 19
%
\begin{inlinefigure}
\centerline{\psfig{figure=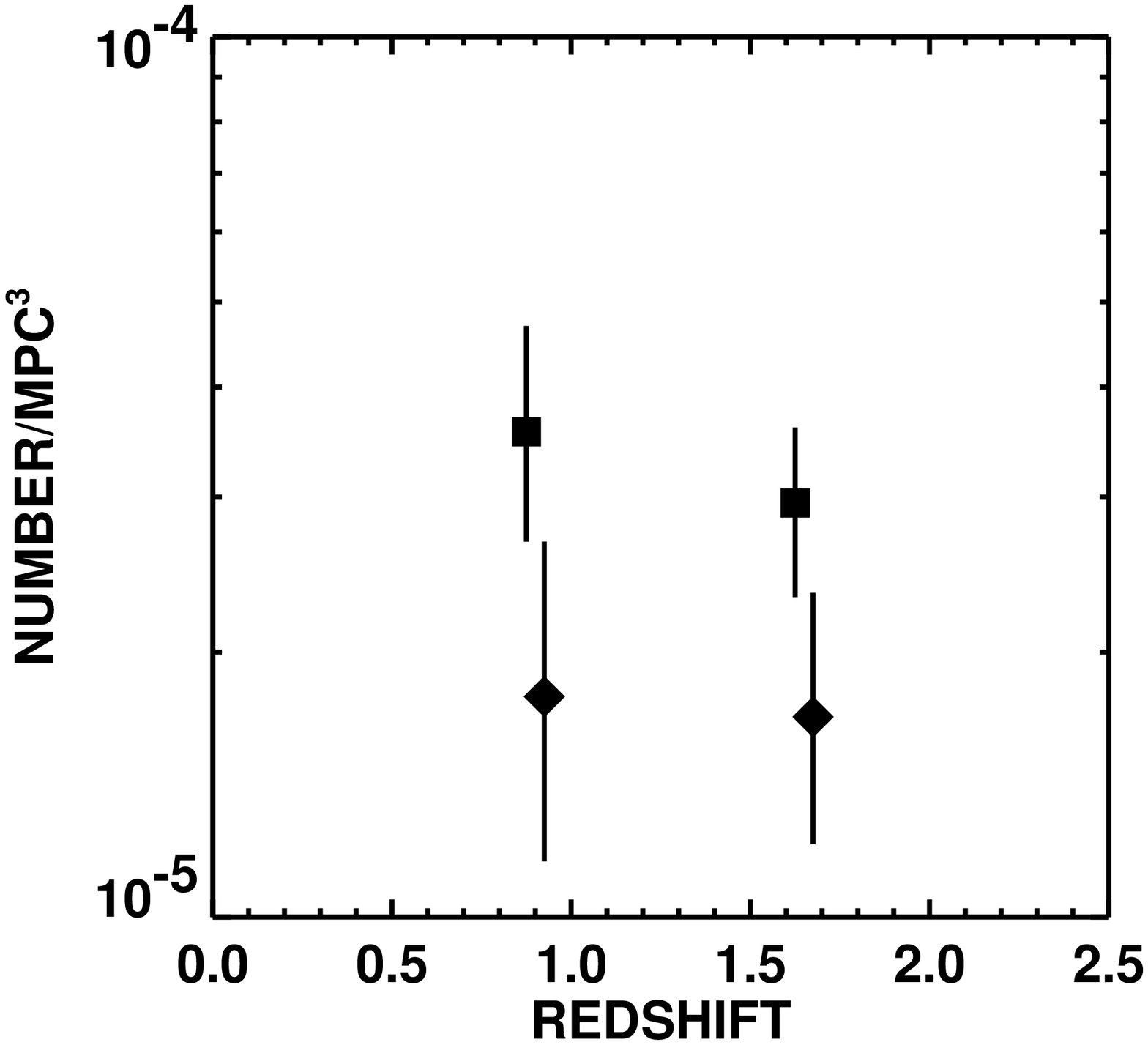,angle=90,width=3.5in}}
\figurenum{19}
\figcaption[]{
Number densities in two redshift bins
($0.5<z\le 1.25$ and $1.25<z\le 2$) of radio sources
with either spectroscopic or photometric redshifts
and $L_{FIR}\ge 4\times 10^{45}$~ergs~s$^{-1}$,
divided according to X-ray luminosity:
diamonds denote sources with either $L_{0.5-2~{\rm keV}}$ or
$L_{2-8~{\rm keV}}\ge 10^{42}$~ergs~s$^{-1}$, and
squares denote sources with both soft and hard X-ray
luminosities $<10^{42}$~ergs~s$^{-1}$.
Squares are slightly offset for clarity.
Poissonian $1\sigma$ uncertainties are based on the
number of sources in each redshift interval.
\label{fignumden}
}
\end{inlinefigure}

%
% FIGURE 20
%
\begin{inlinefigure}
\centerline{\psfig{figure=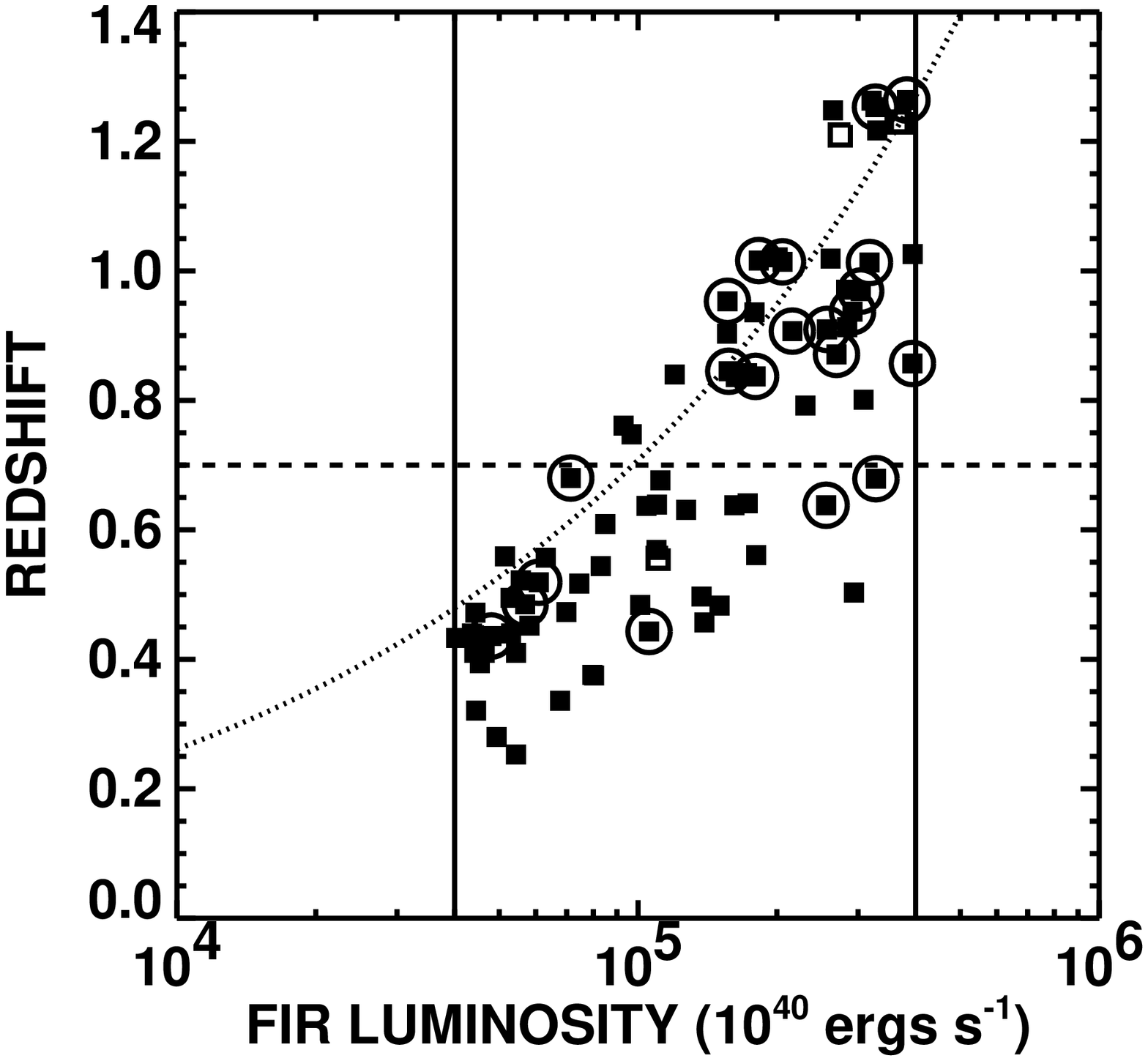,angle=90,width=3.5in}}
\figurenum{20}
\figcaption[]{
Redshift vs. total FIR luminosity for the radio sample with
LIRG luminosities. Solid squares and solid circles 
(the latter are CS sources)
denote spectroscopic redshifts, and open squares denote
photometric redshifts. Large, open circles denote sources that
have either $L_{0.5-2~{\rm keV}}$ or  
$L_{2-8~{\rm keV}}\ge 10^{42}$~ergs~s$^{-1}$.
The dotted curve shows the 60~$\mu$Jy completeness limit of the
radio sample. The two solid vertical lines show the minimum
luminosities of a LIRG and a ULIRG. The dashed horizontal line
shows an upper redshift cut-off of $z=0.7$ to provide as uniform
a selection of LIRGs as possible, while still retaining a
statistically significant sample.
\label{figfirpower2}
}
\end{inlinefigure}

Models of the XRB generally require a population 
of near--Compton-thick 
sources ($N_H\sim 10^{23-24}$~cm$^{-2}$). Typical estimates
are, on average, for $2-3$ times as many such sources 
as X-ray detected sources (e.g., Fabian \& Worsley 2004; 
Treister \& Urry 2005). However, such obscured sources may 
preferentially lie at luminosities lower than quasar luminosities 
(e.g., Ballantyne et al.\ 2006). Thus, our upper limit of
1.9 for the ratio of quasar-luminosity but not X-ray--luminous 
sources to quasar-luminosity, X-ray--luminous sources may be 
consistent with these results.

If we assume that a roughly factor of two ratio of 
highly-obscured AGN candidates to high X-ray luminosity
sources applies to the quasar luminosity population and
shows no redshift effect, then we estimate that the accreted
supermassive black hole mass density determined by 
Barger et al.\ (2005) for broad-line AGNs (their Eq.~9)
will increase to $3.6\times 10^5~M_\odot$~Mpc$^{-3}$. 
Although this is still within $2\sigma$ of the local supermassive 
black hole mass density found by Yu \& Tremaine (2002) of 
$(2.9\pm 0.5)\times 10^5~M_\odot$~Mpc$^{-3}$ for $h=0.7$,
it does not leave much room for the obscured accretion from
the optically narrow AGNs. Applying the same factor of
two corrections to Equations~6 and 7 from Barger et al.\ (2005),
which give the accreted supermassive black hole mass density
for all spectral types assuming bolometric corrections,
respectively, of 85 and 35 for the optically narrow AGNs,
would increase these numbers to $12\times 10^5$ and 
$6.3\times 10^5~M_\odot$~Mpc$^{-3}$.
This could be viewed as a consistency check that there are
not that many luminous obscured AGNs.

\section{Contributions to the X-ray Light}
\label{secxrb}

We would now like to understand what fraction of the X-ray light 
is coming from the different X-ray and radio populations, especially
from the radio-identified ULIRG population. 
To study this, we performed source-stacking analyses on the CDF-N 
2~Ms exposure, restricting to a $9.5'$ radius circle rather than the 
$10'$ radius circle used elsewhere in this paper, since a few 
of the radio sources in the $10'$ radius circle lie just outside 
the X-ray image. 

For each sample,
we first measured the X-ray fluxes of all of the sources in 
each of four passbands, $0.5-1$, $1-2$, $2-4$, and $4-8$~keV, using 
the images and exposure maps given in Alexander et al.\ (2003). 
For the X-ray sample, we measured the fluxes of all of the
sources in the Alexander et al.\ (2003) catalog that lie in the
$9.5'$ radius region. (Note that Worsley et al.\ 2005
found an increase in total resolved flux of about $2-5$\% in the
$>2$~keV bands when they applied the systematic flux correction
determined by Bauer et al.\ (2004) to account for Eddington bias
and some additional aperture/photometry effects that were not 
considered in Alexander et al.\ 2003, but we have not applied
that correction here.)

To calculate the contribution to the extragalactic background light 
(EBL) from a given sample, we summed the fluxes in the $9.5'$ radius 
circle and divided by the corresponding area.
We did not exclude any sources from the summations, even if a source
were not significantly detected in a particular band or had a
measured negative flux in the band. This simple averaging
is not optimal, since the errors in the X-ray fluxes are smaller
at smaller off-axis angles, but it avoids weighting the smaller
central region more highly.

To estimate the uncertainties on our surface brightness measurements,
we first generated a grid of 60,000 points in the $9.5'$ radius area
that are separated from each other by more than $4''$ (the aperture
diameter used to measure the counts in the X-ray sources) and
are located more than $6''$ away from any known X-ray source.
We next measured the fluxes in each of our four X-ray passbands
at these 60,000 positions. This serves as our base file. Then, for
each sample for which we want to determine the uncertainties, we
randomly generate a number (which matches the number of sources
in the sample) of integers between 1 and 60,000 and use the fluxes
corresponding to those positions in our base file. We do this
1000 times, which gives us 1000 flux values for each of the four
passbands. We then order those values from the most negative to the
most positive and use the 68\% confidence interval to determine our
uncertainties. (Most of our uncertainties are much smaller than the
data points.)

%
% FIGURE 21
%
\begin{inlinefigure}
\centerline{\psfig{figure=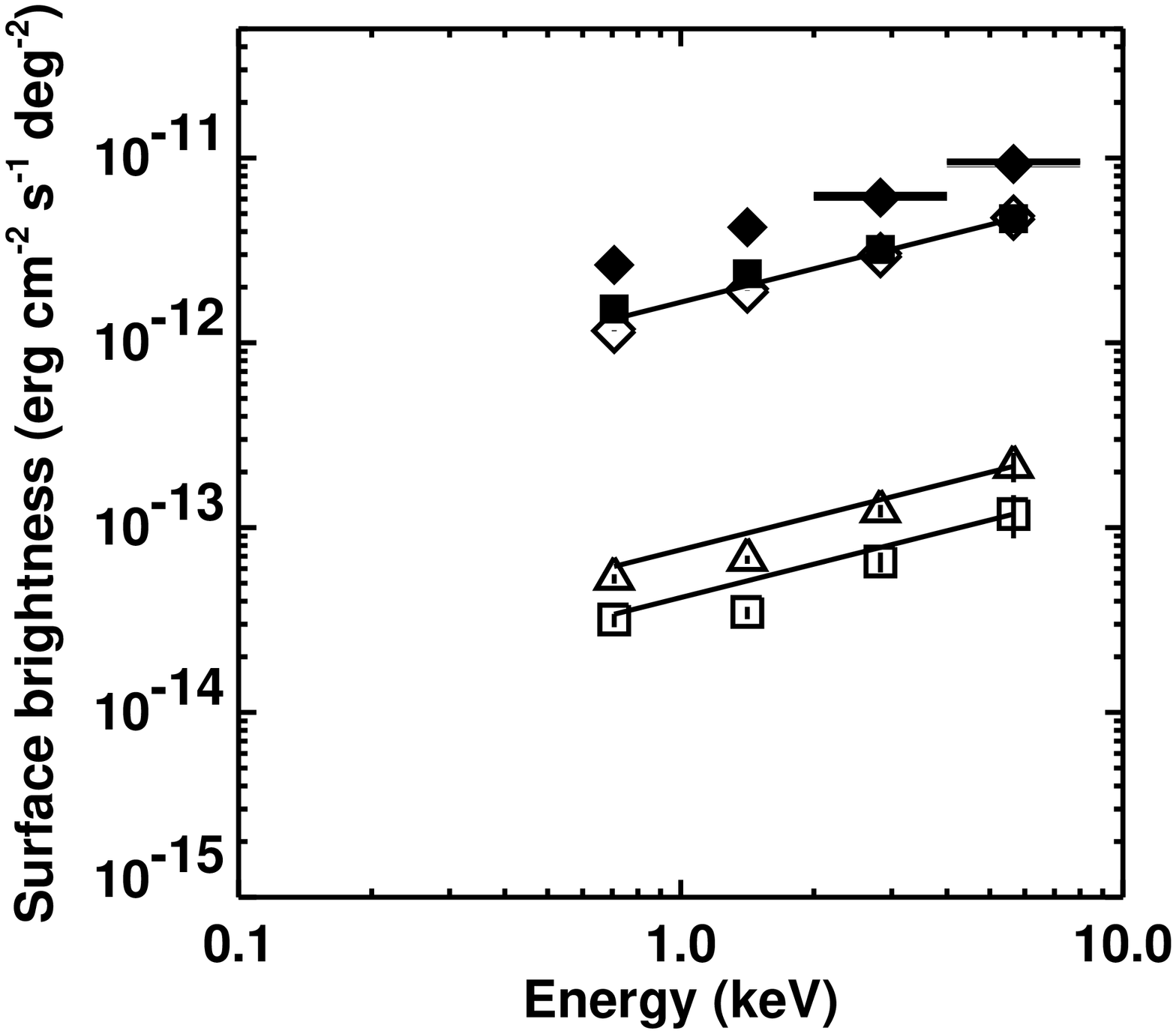,angle=90,width=3.5in}}
\figurenum{21}
\figcaption[]{
X-ray surface brightness vs. energy measured within a $9.5'$ radius
circle for the Alexander et al.\ (2003) X-ray sample
{\em (solid diamonds)\/}, for the X-ray sample without radio
counterparts {\em (open diamonds)\/}, for the present radio
sample {\em (solid squares)\/}, for the radio sample without
X-ray--luminous counterparts
(these sources have
$L_{0.5-2~{\rm keV}}$ or $L_{2-8~{\rm keV}}< 10^{42}$~ergs~s$^{-1}$)
{\em (open triangles)\/}, and for the radio sample without
X-ray counterparts at all {\em (open squares)\/}.
The uncertainties are 68\% confidence intervals, as described in the text.
For comparison purposes, the diagonal solid lines show the 1.4 photon
index of the XRB normalized to the highest energy bins of the X-ray
surface brightness measurements for the radio sample, for the
non--X-ray--luminous radio sample, and for the X-ray--undetected radio
sample. The horizontal solid lines show the $2-10$~keV XRB measurement
of Revnivtsev et al.\ (2005) converted into narrower energy bands by
adopting a photon index of 1.4. The thickness of the lines
represents the uncertainties on their measurement.
\label{figxraysb}
}
\end{inlinefigure}

In Figure~\ref{figxraysb}, we show the EBL from the X-ray sources in 
the Alexander et al.\ (2003) catalog with solid diamonds. 
Coincidentally, they match very well
the $2-10$~keV XRB measurement made by Revnivtsev et al.\ (2005) 
from a reanalysis of the HEAO1/A2 data {\em (thick horizontal 
lines; the thickness denotes the uncertainty range on their
measurement\/)}. Note that we converted the Revnivtsev et al.\ (2005)
$2-10$~keV measurement to $2-4$~keV, $4-8$~keV, and $8-10$~keV by 
adopting a photon index of 1.4. We chose this determination to
compare with since the A2 or Cosmic X-ray
Experiment instrument was especially designed for accurate
measurements of the XRB over a very wide sky solid angle
(e.g., Marshall et al.\ 1980; Boldt 1987; Gruber et al.\ 1999),
and the design of its detectors allows the internal instrumental 
background to be separated from the XRB with almost absolute 
accuracy (Rothschild et al.\ 1979; Boldt 1987). 

However, for this work, we are not interested in accurately determining 
the resolved fraction of the XRB. Such an analysis would require 
integrating the known $\log N-\log S$ distribution to take into account 
the rare, bright sources that are not sampled in this deep pencil-beam
survey, extrapolating the number counts to take into account
the sources with fainter fluxes, and deciding which of the XRB
measurements to compare with. Indeed, this was done 
by Moretti et al.\ (2003), and there is very good agreement between
their result and the Revnivtsev et al.\ (2005) measurement.
Rather, it is sufficient for us to measure the X-ray surface 
brightnesses from the known X-ray sources in the CDF-N and see 
that they are roughly consistent with the HEAO1/A2 
XRB measurement. In other words, we are not concerned with
uncertainties at the $10-20$\% level. Our aim is just to see roughly 
what fraction of the X-ray light comes from the various 
populations.

We begin by considering only the Alexander et al.\ (2003) X-ray 
sources that are not radio sources in the Richards (2000) catalog. 
In Figure~\ref{figxraysb}, we show the computed X-ray surface
brightnesses from these sources with open diamonds. We find that 
they contribute $50\pm 0.7$\% of the X-ray light in the $4-8$~keV band.

Next, we examine how much the total radio sample contributes to the
X-ray light. In Figure~\ref{figxraysb}, we show our computed 
X-ray surface brightnesses for this sample with solid squares. 
We see that the radio sample also contributes $50\pm 0.5$\%
of the X-ray light in the $4-8$~keV band. In other words, 
radio sources to the Richards (2000) 1.4~GHz flux limit of 40~$\mu$Jy 
are markers for half of the X-ray light at these high energies.
This contribution is the same as the contribution from the 
X-ray sources that are not detected at these radio fluxes. 

However, most of the light contributed by the radio sample 
comes from the X-ray--luminous radio sources. To show this, we 
next measure the X-ray surface brightnesses of the sample of 
radio sources without X-ray--luminous counterparts
(these sources have
$L_{0.5-2~{\rm keV}}$ or $L_{2-8~{\rm keV}}< 10^{42}$~ergs~s$^{-1}$;
hereafter, non--X-ray--luminous) 
in the Alexander et al.\ (2003) catalog 
{\em (open triangles)\/}, as well as of the sample of radio sources 
without X-ray counterparts at all (hereafter, X-ray--undetected)
{\em (open squares)\/}.

We find the contributions to the $4-8$~keV light from the 
non--X-ray--luminous population to be $2.3\pm 0.4$\% 
and from the X-ray--undetected population to be $1.2\pm 0.3$\%.
Thus, although some of the radio sources are 
contributing additional X-ray light that is not already in the X-ray 
sample, such contributions are very small. This immediately tells
us that the current radio source population cannot account for the 
background light that has been suggested may be missing at these energies.
Indeed, the percentages are about a factor of 5 lower than those
predicted for $N_H\sim 10^{24}$~cm$^{-2}$ sources at these energies
in typical XRB synthesis models (e.g., Comastri, Gilli, \& Hasinger 2006).

However, the inverse statement that some of these non--X-ray--luminous
or X-ray--undetected radio sources
contain highly-obscured AGNs is true. We can see this by
comparing the shapes of the X-ray surface brightness measurements 
for the various samples with the shape of the XRB. To make it 
easier to do the comparisons,
in Figure~\ref{figxraysb}, we normalize the 1.4 photon index of 
the XRB to the surface brightnesses in the highest 
energy bin for all but the X-ray sample and the X-ray sample 
without radio counterparts (to avoid cluttering up the figure too much).
We can see that the shape of the surface brightness 
measurements for the radio sample is slightly softer than that
of the XRB, while the shape of the surface brightness measurements 
for the X-ray sample without radio counterparts is slightly harder
than the XRB. However, the shapes of the surface brightness
measurements for the non--X-ray--luminous radio sample and 
the X-ray--undetected radio sample both appear much harder than that 
of the XRB. This result suggests that these two populations
are consistent with containing highly-obscured AGNs.

%
% FIGURE 22
%
\begin{inlinefigure}
\centerline{\psfig{figure=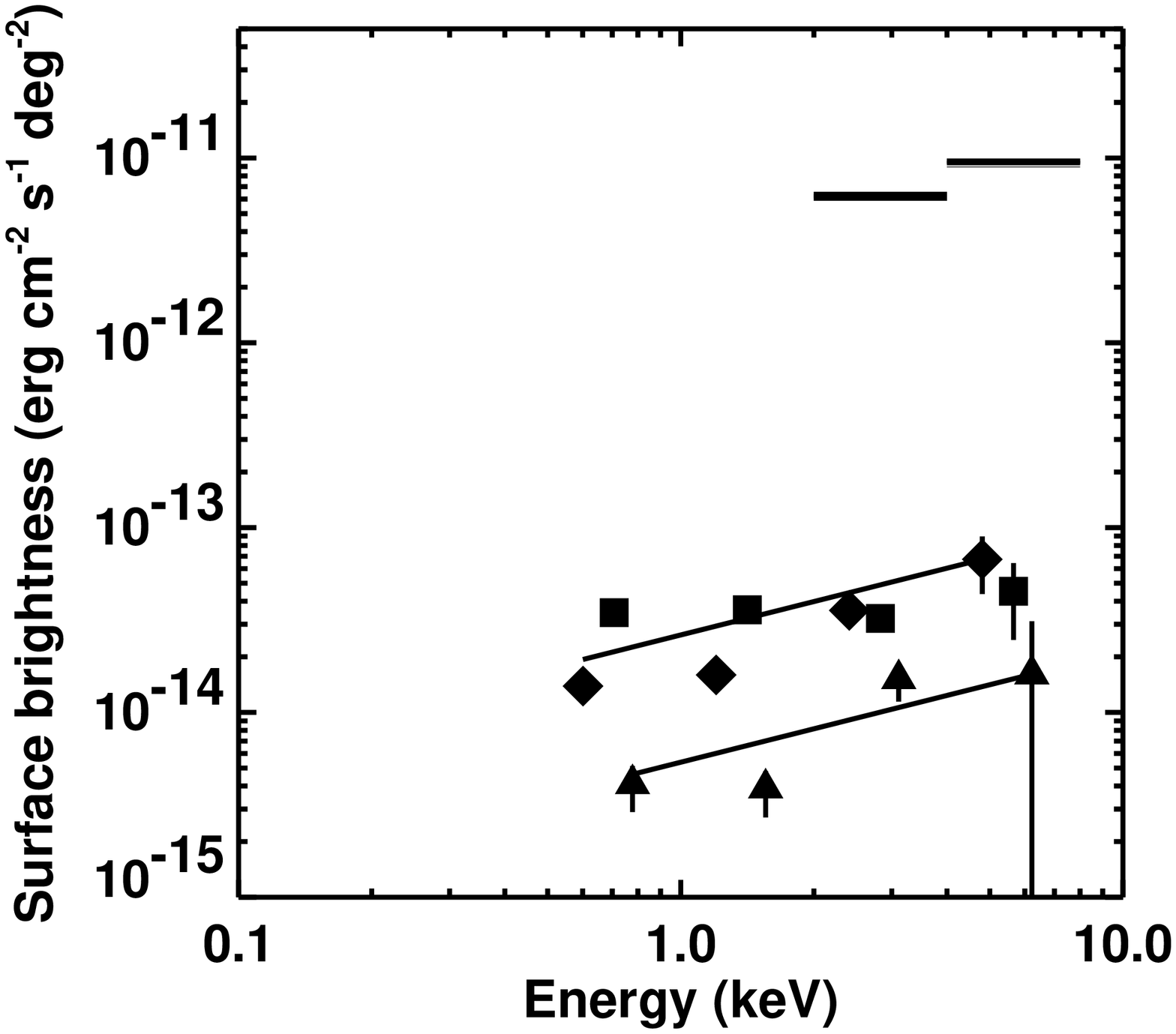,angle=90,width=3.5in}}
\centerline{\psfig{figure=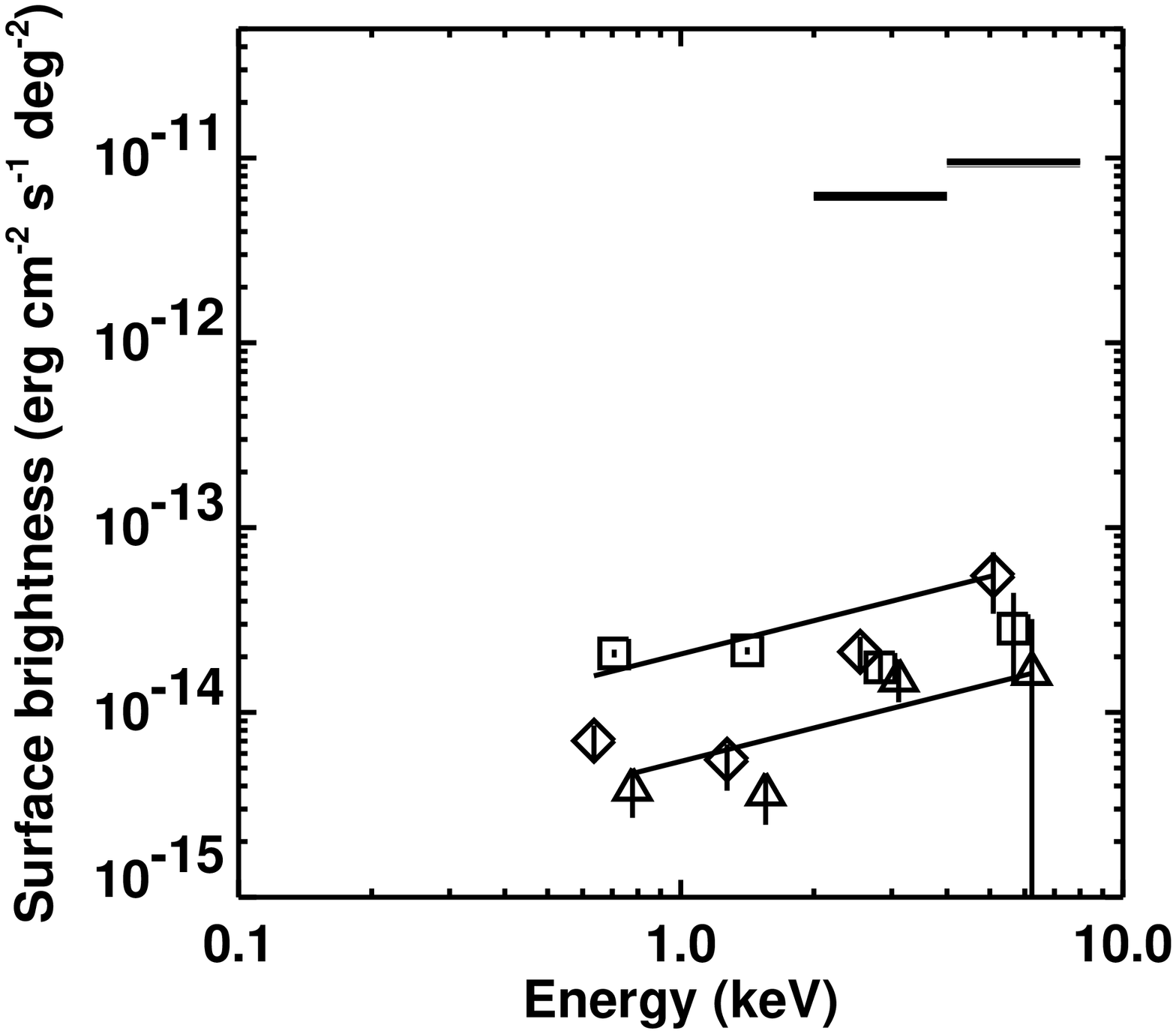,angle=90,width=3.5in}}
\figurenum{22}
\figcaption[]{
X-ray surface brightness vs. energy measured within a $9.5'$ radius
circle for three redshift intervals: $0<z\le 0.5$ {\em (squares)\/},
$0.5<z\le 1.5$ {\em (diamonds)\/}, and $1.5<z\le 3$ {\em (triangles)\/}
(a) for the radio sample without X-ray--luminous counterparts
(these sources have $L_{0.5-2~{\rm keV}}$ or
$L_{2-8~{\rm keV}}< 10^{42}$~ergs~s$^{-1}$)
in the Alexander et al.\ (2003) catalog {\em (solid symbols)\/}
and (b) for the radio sample without X-ray counterparts at all
{\em (open symbols)\/}.
Small offsets have been applied in the $x$-direction to the diamonds
and to the triangles for clarity. The uncertainties are 68\% confidence
intervals, as described in the text. For comparison purposes,
the diagonal solid lines show the 1.4 photon index of the XRB normalized
to the highest energy bins of the $0.5<z\le 1.5$ and $1.5<z\le 3$
samples. The horizontal solid lines show the
$2-10$~keV XRB measurement of Revnivtsev et al.\ (2005) converted
into narrower energy bands by adopting a photon index of 1.4.
The thickness of the lines represents the uncertainties on
their measurement.
\label{figxraysbz}
}
\end{inlinefigure}

One might wonder how the shapes of the X-ray surface brightness
measurements for the non--X-ray--luminous radio sample
and for the X-ray--undetected radio sample vary with
redshift. We show this in Figures~\ref{figxraysbz}a and b for 
the redshift intervals $z=0-0.5$ {\em (squares)\/}, 
$0.5-1.5$ {\em (diamonds)\/}, 
and $1.5-3$ {\em (triangles)\/}. The solid
symbols in Figure~\ref{figxraysbz}a denote the non--X-ray--luminous 
radio sample, and the open symbols in Figure~\ref{figxraysbz}b
denote the X-ray--undetected radio sample. Small offsets have
been applied in the $x$-direction to the diamonds and to the triangles 
for clarity. The uncertainties are again the 68\% confidence intervals.
We normalize the highest energy bins of the $z=0.5-1.5$ and $z=1.5-3$ 
samples to the 1.4 photon index of the XRB for comparison.
In both figures, we can see that the lowest redshift interval sample
{\em (squares)\/} is quite soft due to the contributions from
star-forming galaxies, while the higher redshift interval samples 
are hard due to the obscured nature of the sources.

%
% FIGURE 23
%
\begin{inlinefigure}
\centerline{\psfig{figure=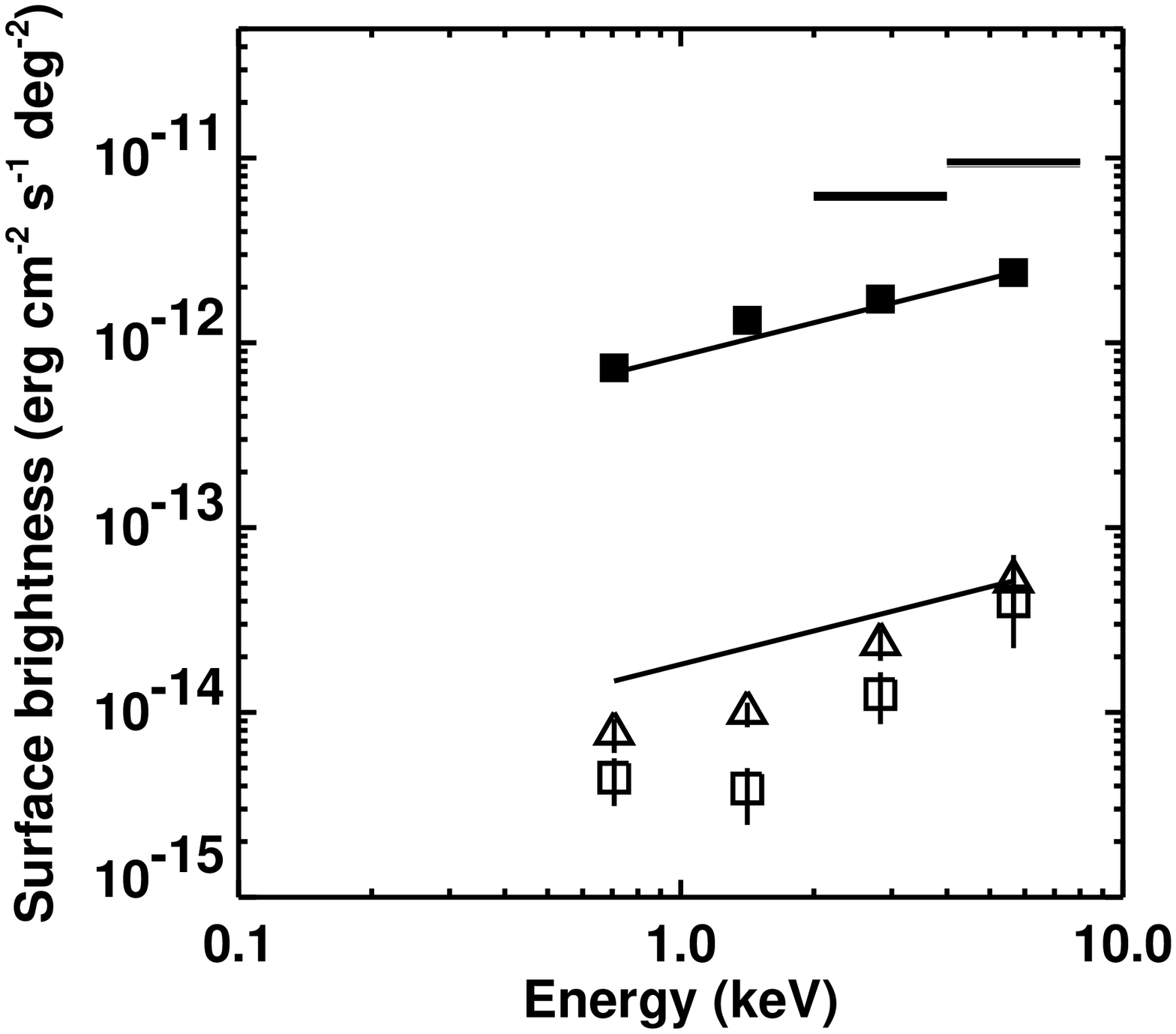,angle=90,width=3.5in}}
\figurenum{23}
\figcaption[]{
X-ray surface brightness vs. energy measured within a $9.5'$ radius
circle for the $z\le 2$ radio-identified ULIRG sample with X-ray--luminous
($L_{0.5-2~{\rm keV}}$ or $L_{2-8~{\rm keV}}\ge 10^{42}$~ergs~s$^{-1}$)
counterparts in the Alexander et al.\ (2003) catalog
{\em (solid squares)\/}, without X-ray luminous counterparts
{\em (open triangles)\/}, and without X-ray counterparts at all
{\em (open squares)\/}.
For comparison purposes, the diagonal solid lines show the 1.4 photon
index of the XRB normalized to the highest energy bins of the
X-ray--luminous and non--X-ray--luminous samples.
The horizontal solid lines show the $2-10$~keV XRB measurement of
Revnivtsev et al.\ (2005) converted into narrower energy bands by
adopting a photon index of 1.4. The thickness of the lines
represents the uncertainties on their measurement.
\label{figulirgsb}
}
\end{inlinefigure}

Since in \S\ref{secuplim} we focused on the $z\le 2$ radio-identified 
ULIRG sample, we now repeat our source-stacking analyses using only 
these data. Our results are shown in Figure~\ref{figulirgsb}.
We use solid squares to denote the X-ray surface brightnesses measured 
for the $z\le 2$ ULIRG sample with X-ray--luminous 
($L_{0.5-2~{\rm keV}}$ or $L_{2-8~{\rm keV}}\ge 10^{42}$~ergs~s$^{-1}$)
counterparts in the Alexander et al.\ (2003) catalog.
These sources contribute $25\pm 0.1$\% of the X-ray light in 
the $4-8$~keV band, which shows that sources with 
ULIRG radio powers are responsible for contributing a very
substantial fraction of the XRB. We use open triangles 
to denote our measurements for the $z\le 2$ non--X-ray--luminous ULIRGs. 
These contribute only $0.5\pm 0.2$\% of the X-ray light 
in the $4-8$~keV band. Finally, we use open squares to denote our
measurements for the $z\le 2$ ULIRGs with no X-ray 
detections at all in the Alexander et al.\ (2003) catalog. These 
contribute only $0.4\pm 0.2$\% to the $4-8$~keV light.
The uncertainties are again the 68\% confidence intervals.
We normalize the highest energy bins of the
X-ray--luminous ULIRGs and the non--X-ray--luminous 
ULIRGs to the 1.4 photon index of the XRB for comparison. 
We see that the shape of the X-ray--luminous ULIRGs is 
quite well matched to the XRB, while the shapes of the 
non--X-ray--luminous ULIRGs and the X-ray--undetected 
ULIRGs are considerably harder, rising steeply towards 
the higher energies.
Again, this is consistent with these samples being highly obscured. 
However, even if we interpret them as such, their contributions to 
the $4-8$~keV light are very small, and they are unlikely to 
contribute substantially to the XRB at even higher energies.

\section{Summary}
\label{secsummary}

In this paper, we analyzed the nature and evolution of microJansky 
radio sources using a highly spectroscopically complete, 
deep VLA survey of the HDF-N region. We supplemented the 
spectroscopic identifications with photometric redshifts measured
from the rest-frame ultraviolet to MIR spectral energy distributions.
Our results are as follows.

$\bullet$ We found that the fraction of radio sources that can be 
optically spectroscopically identified is fairly independent of 
radio flux, with about $60-80$\% identified at all fluxes.
We spectrally classified the galaxies into four spectral types: 
absorbers, star formers, Seyfert galaxies, and broad-line AGNs. 

$\bullet$ We did not confirm the existence of an X-ray--radio 
correlation for star-forming galaxies in either the hard or 
soft X-ray bands. Previous claims of such a correlation appear 
to be the result of selection effects.

$\bullet$ We did not observe any correlation between 1.4~GHz flux
and $R$ magnitude, which helps to explain the uniformity of
the spectroscopically identified fraction with radio flux.
Previous claims of the existence of such a correlation were 
likely due to the much shallower optical observations that were 
used in those studies.

$\bullet$ We also did not observe any correlation between 1.4~GHz
flux and redshift for the spectroscopically and photometrically
identified sources. The redshift distribution is consistent
with being invariant with radio flux, though we may be missing
the higher redshift tail of the redshift distribution function
due to optical spectroscopic bias.

$\bullet$ The observed invariance in the rest-frame colors
and only small amount of change in the absolute magnitudes 
of the host galaxies suggests that radio
observations sample the same type of host galaxies at all 
redshifts and that the host galaxy properties are not evolving
much with redshift.

$\bullet$ There is dramatic evolution in the radio powers of
the host galaxies with increasing redshift, and similar optical
galaxies are hosting more powerful radio galaxies at higher
redshifts. A quantitative description the evolution of the 1.4~GHz 
luminosity function can be found in Cowie et al.\ (2004a).

$\bullet$ Assuming that the locally determined FIR-radio 
correlation holds at high redshifts, we estimated total FIR
luminosities for the radio sources. We note that these will be
overestimates in the case of radio-loud AGNs, but we conservatively 
did not try to remove the radio-loud AGNs from the sample, since
we are only determining upper limits. For the radio sources with
total FIR luminosities comparable to quasar-like bolometric 
luminosities, we obtained an upper limit of 1.9 for the 
ratio of candidate highly-obscured AGNs to X-ray--luminous sources.

$\bullet$ We measured the X-ray surface brightnesses for various
X-ray and radio populations within a $9.5'$ radius circle to 
determine their contributions to the X-ray light. 
For the known X-ray sources in the CDF-N, we found our 
measurements to be (coincidentally) consistent
with the HEAO1/A2 XRB measurement made by Revnivtsev et al.\ (2005). 
We found that the Richards (2000) radio sample contributes half 
of the $4-8$~keV light, which is also how much the X-ray sources 
without radio counterparts contribute. However, most of the light
from the radio sample comes from the X-ray--luminous radio sources.
The non--X-ray--luminous and X-ray--undetected subsamples
contribute only $2.3\pm 0.4$\% and 
$1.2\pm 0.3$\%, respectively. This tells us that the
current radio source population cannot account for the
background light that one might think is missing at these energies.
However, the shapes of the surface brightness measurements for
these two samples both appear much harder than that of the XRB,
which suggests that these two populations are consistent
with containing highly-obscured AGNs. For 
the $z\le 2$ radio-identified ULIRG sample, we found that the
X-ray--luminous subsample contributes a very substantial
one-quarter of the $4-8$~keV light,
while the non--X-ray--luminous and X-ray--undetected
subsamples each contribute less than a percent. Again, the 
shapes of the latter two subsamples are considerably harder 
than the XRB and hence are consistent with containing 
highly-obscured AGNs.

\acknowledgments
We thank the referee for helpful comments that improved the manuscript. 
We thank L. Silva for providing the template spectral energy distributions.
This research has made use of the NASA/IPAC Extragalactic 
Database (NED) which is operated by the Jet Propulsion Laboratory, 
California Institute of Technology, under contract with the 
National Aeronautics and Space Administration.
We gratefully acknowledge support from NSF grants
AST 02-39425 (A.~J.~B.) and AST 04-07374 (L.~L.~C.), 
the University of Wisconsin Research 
Committee with funds granted by the Wisconsin Alumni Research 
Foundation, the Alfred P. Sloan Foundation, and the David and
Lucile Packard Foundation (A.~J.~B.).

%\newpage

\clearpage
%
% TABLE 1
%
\begin{deluxetable}{cccccccc}
\tablecaption{\label{tab1}Bi-lobal Source and 11 Sources
Without Obvious Counterparts}
\tablewidth{0pt}
\tablehead{\multicolumn{6}{c}{Radio (J2000.0)} &
           \colhead{Radio Flux} &
           \colhead{Redshift} \cr
           \multicolumn{3}{c}{R.A.} &
           \multicolumn{3}{c}{Decl.} &
           \colhead{($\mu$Jy)} &
           \colhead{} \\
           }
\startdata
12 & 37 & 25.73 & 62 & 11 &  28.50  & 5960\tablenotemark{a} & 1.641\tablenotemark{b}    \\
12 & 37 & 02.29 & 62 & 23 &  31.50  & 78.7 & 0.642\tablenotemark{c}    \\
12 & 35 & 40.39 & 62 & 16 &  23.80  & 60.1 & \nodata  \\
12 & 35 & 41.00 & 62 & 18 &  28.20  & 89.0 & \nodata  \\
12 & 35 & 57.71 & 62 & 08 &  08.41  & 68.6 & \nodata  \\
12 & 36 & 08.24 & 62 & 15 &  53.00  & 59.3\tablenotemark{a} & \nodata  \\
12 & 36 & 24.28 & 62 & 10 &  17.00  & 54.2\tablenotemark{a} & \nodata  \\
12 & 36 & 43.88 & 62 & 05 &  59.20  & 62.5 & \nodata  \\
12 & 36 & 46.70 & 62 & 12 &  26.49  & 72.0 & \nodata  \\
12 & 36 & 51.72 & 62 & 05 &  02.49  & 57.6\tablenotemark{a} & \nodata  \\
12 & 36 & 54.70 & 62 & 10 &  39.60  & 48.2 & \nodata  \\
12 & 37 & 23.05 & 62 & 05 &  39.59  & 78.5\tablenotemark{a} & \nodata  \\
\enddata
\tablecomments{Units of right ascension are hours,
minutes, and seconds, and units of declination are degrees,
arcminutes, and arcseconds.}
\tablenotetext{a}{Also present in the Biggs \& Ivison (2006)
catalog.}
\tablenotetext{b}{Bi-lobal source centered
on a small red galaxy with optical coordinates
(12:37:25.9, 62:11:29.0).}
\tablenotetext{c}{Source appears to be associated
with a pair of interacting galaxies at $z=0.642$. However,
because the association is uncertain, we assume this source
is unidentified.}
\end{deluxetable}

%
% TABLE 2
%

\begin{deluxetable}{cccccccc}
\tablecaption{\label{tab2}Omitted Redshifts Due to Uncertain Identifications}
%\tabletypesize{\scriptsize}
%\rotate
%\tablewidth{0pt}
\tablehead{\multicolumn{6}{c}{Radio (J2000.0)} &
           \colhead{Radio Flux} &
           \colhead{Literature} \cr
           \multicolumn{3}{c}{R.A.} &
           \multicolumn{3}{c}{Decl.} &
           \colhead{($\mu$Jy)} &
           \colhead{Redshift} \cr
           }
\startdata
12 & 35 & 53.26 & 62 & 13 & 37.70 & 58.4 & 2.098\tablenotemark{a} \cr %chapman
12 & 36 & 08.24 & 62 & 15 & 53.00 & 59.3 & 0.459\tablenotemark{b} \cr
12 & 36 & 11.41 & 62 & 21 & 49.79 & 111.0 & 0.294\tablenotemark{c} \cr
12 & 36 & 21.27 & 62 & 17 & 08.40 & 148.0 & 1.992\tablenotemark{a} \cr 
12 & 36 & 36.91 & 62 & 13 & 20.41 &  50.0 & 0.680\tablenotemark{d} \cr
12 & 36 & 42.10 & 62 & 13 & 31.41 & 467.0 & 4.420\tablenotemark{e} \cr 
12 & 36 & 46.70 & 62 & 12 & 26.49 &  72.0 & 2.970\tablenotemark{f} \cr
12 & 36 & 51.76 & 62 & 12 & 21.30 &  49.3 & 0.401\tablenotemark{g} \cr
12 & 36 & 56.60 & 62 & 12 & 07.60 &  46.2 & 0.321\tablenotemark{d} \cr
12 & 37 & 01.57 & 62 & 11 & 46.60 & 128.0 & 0.884\tablenotemark{b} \cr
12 & 37 & 06.77 & 62 & 07 & 22.50 &  72.8 & 0.518\tablenotemark{d} \cr
12 & 37 & 50.27 & 62 & 13 & 59.00 &  90.5 & 0.231\tablenotemark{d} \cr
\enddata
\vskip -1cm
\tablecomments{Units of R.A. are hours, minutes, and seconds, 
and units of Decl. are degrees, arcminutes, and arcseconds.}
\tablenotetext{a}{Chapman et al.\ (2005) redshift for
SMM J$123553.26+621337.7$ and Swinbank et al.\ (2004)
redshift for SMM J$123621.27+621708.4$. 
Each radio source has an $8~\mu$m counterpart, but there is 
no blue light at those positions. The redshifts appear to be 
of neighbor objects, which may or may not be associated 
with the radio sources.}
\tablenotetext{b}{Wirth et al.\ (2004) redshifts for
GOODS J$123608.02+621554.0$ and GOODS J$123701.80+621144.2$.
The first radio source is far off-axis. The second radio
source has an $8~\mu$m counterpart, but the redshift
is of a neighbor galaxy.}
\tablenotetext{c}{Barger et al.\ (2002) redshift for
a neighbor galaxy, CXOHDFN J$123611.40+622149.9$.}
\tablenotetext{d}{Barger et al.\ (2000) redshifts for 
neighbor galaxies.}
\tablenotetext{e}{Waddington et al.\ (1999) redshift. The radio 
source has an $8~\mu$m counterpart, but the identification is offset.}
\tablenotetext{f}{Steidel et al.\ (2003) redshift. The radio source 
has a weak $8~\mu$m counterpart, but the redshift is of a neighbor 
galaxy.}
\tablenotetext{g}{Lanzetta et al.\ (1996) redshift.
The radio source has an $8~\mu$m counterpart, but the redshift 
is of a bright neighbor.}
\end{deluxetable}

\end{document}